\documentclass[a4paper,11pt]{book}
 \usepackage{amsmath,epsfig,amssymb,amsfonts,verbatim}
 \usepackage{fancyhdr}
 \usepackage{makeidx}
  \usepackage{hyperref}
 \makeindex

 \hypersetup{colorlinks,%
 filecolor=magenta,%
 citecolor=red,%
 linkcolor=blue,%
 urlcolor=cyan,%
 pdftex}

 \renewcommand{\theequation}{\arabic{chapter}.\arabic{equation}}

 \newcommand{\beq}{\begin{equation}}
 \newcommand{\eeq}{\end{equation}}
 \newcommand{\beqa}{\begin{eqnarray}}
 \newcommand{\eeqa}{\end{eqnarray}}
 \newcommand{\colsep}{\setlength\arraycolsep{1pt}}

 \newcommand{\nn}{\nonumber}
 \newcommand{\F}{\Phi}
 
 \newcommand{\x}{\vec{x}}
 \newcommand{\bx}{\bar{x}}
 \newcommand{\bz}{\bar{z}}
 \newcommand{\by}{\bar{y}}
 \newcommand{\bw}{\bar{w}}
 \newcommand{\tl}{\tilde{l}}
 
 \newcommand{\pd}{\partial}
 \newcommand{\bm}{\bar m}
 \newcommand{\bn}{\bar n}

 \newcommand{\g}{\gamma}
 \newcommand{\mathD}{\mathcal{D}}
 \newcommand{\mathT}{\mathcal{T}}
 \newcommand{\hj}{\hat{\jmath}}
  \newcommand{\al}{\alpha}
  \newcommand{\bt}{\beta}
  \newcommand{\half}{\frac{1}{2}}

 \def\h{\eta}
 \def\ka{\kappa}
 \def\z{\zeta}
 \def\m{\mu}
 \def\n{\nu}
 \def\r{\rho}
 \def\s{\sigma}
 \def\t{\theta}

 \def\p{\partial}

 \def\f{\varphi}
 \def\a{\alpha}
 \def\b{\beta}
 \def\d{\delta}
 
 \def\g{\gamma}
 \def\G{\Gamma}
 
 \def\ba{\begin{eqnarray}}
 \def\ea{\end{eqnarray}}
 \def\bea{\begin{eqnarray}}
 \def\eea{\end{eqnarray}}
 \def\nb{\nonumber}
 \def\jh{{\hat{\jmath}}}
 
 \def\ie{{\it i.e.},}
 
 \def\be{\begin{equation}}
 \def\ee{\end{equation}}

\begin{document}

 % OOOOOOOOOOOOOOOOOOOOOOOOOOOOOOOOO     TITLE     OOOOOOOOOOOOOOOOOOOOOOOOOOOOOOOOO

 % OOOOOOOOOOOOOOOOOOOOOOOOOOOOOOOOO     TITLE     OOOOOOOOOOOOOOOOOOOOOOOOOOOOOOOOO

 % OOOOOOOOOOOOOOOOOOOOOOOOOOOOOOOOO     TITLE     OOOOOOOOOOOOOOOOOOOOOOOOOOOOOOOOO

 % OOOOOOOOOOOOOOOOOOOOOOOOOOOOOOOOO     TITLE     OOOOOOOOOOOOOOOOOOOOOOOOOOOOOOOOO

 % OOOOOOOOOOOOOOOOOOOOOOOOOOOOOOOOO     TITLE     OOOOOOOOOOOOOOOOOOOOOOOOOOOOOOOOO

\thispagestyle{empty}

\vspace*{-1.5truecm}

\begin{center}
\begin{flushright}
 \texttt{ROM2F/2005/22}
 \end{flushright}
 \vspace{.7cm}
\makebox[\textwidth]{\raisebox{0.68cm}{
\begin{minipage}[h]{13truecm}
        \hspace*{0.2cm}
        {\sc \LARGE UNIVERSIT\`A DEGLI STUDI DI ROMA\\
        \hspace*{3.36cm}
        ``TOR \rule{0pt}{24pt}VERGATA''}\\
    \end{minipage}}}
%\hspace*{-0.6cm}
\end{center}

\vspace{0.2cm}

\begin{center}
    \includegraphics[height=1.45truecm]{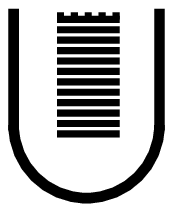}
\end{center}

\begin{center}
FACOLT\`A DI SCIENZE MATEMATICHE, FISICHE E NATURALI \\
Dipartimento di Fisica \\
\end{center}

\vspace{2cm}

\begin{center}
{\LARGE {\bf Strings on AdS$_3$$\times$S$^3$ and the Plane-Wave
Limit}}
\end{center}
\begin{center}
{\Large {Issues on PP-Wave/CFT Holography}}
\end{center}

\vspace{1.5cm}

\begin{center}
{\large A PhD thesis presented by} \\\vspace{.3cm}
\rule{0pt}{22pt}{\large {\sl Oswaldo Zapata Mar\'{\i}n}} \rule{0pt}{22pt}
\end{center}

\vspace{1.5cm}  \noindent
Advisor: \\ 
{\large {\sl Prof. Massimo Bianchi (Univ. Tor Vergata)} \\
\\
Evaluating Commmittee: \rule{0pt}{30pt} \\
{\large {\sl Prof. Ignatios Antoniadis (CERN)}} \\
{\large {\sl Prof. Edi Gava (ICTP)}}\\

\vspace{1.0cm}
\begin{center}
Rome, March 2005
\end{center}

 \newpage\thispagestyle{empty}

 \tableofcontents

% OOOOOOOOOOOOOOOOOOOOOOOOOOOOOO       Acknowledgements    OOOOOOOOOOOOOOOOOOOOOOOO

% OOOOOOOOOOOOOOOOOOOOOOOOOOOOOO       Acknowledgements    OOOOOOOOOOOOOOOOOOOOOOOO

% OOOOOOOOOOOOOOOOOOOOOOOOOOOOOO       Acknowledgements    OOOOOOOOOOOOOOOOOOOOOOOO

% OOOOOOOOOOOOOOOOOOOOOOOOOOOOOO       Acknowledgements    OOOOOOOOOOOOOOOOOOOOOOOO

 \newpage\thispagestyle{empty}

 \begin{center}
 $\star$
 \end{center}
 \vskip2cm

 \begin{flushright}
 \large\emph{ A mis seres m\'as queridos:\\
 Lili, mi pap\'a, mi mam\'a,\\
 Emiliano, Paco y Serenita}
 \end{flushright}
 \vskip3cm

 \begin{center}
 {\huge \bf Acknowledgements}
 \end{center}
 \vskip1cm

 I am grateful to Massimo Bianchi for his assistance and friendship
 during my studies. I am also indebted to F.~Fucito,
 G.~Pradisi, Ya.~Stanev and, specially, A.~Sagnotti, for patiently teaching me theoretical physics. \\

 I enjoyed many discussions with my collaborators: M.~Bianchi, G.~D'Appollonio, E.~Kiritsis and
 K.~L.~Panigrahi. \\

 I would like to recognize and thank the Istituto Nazionale di Fisica Nucleare and the ICSC World-Laboratory,
 in particular its President
 Prof. A.~Zichichi, for financial support during part of my PhD. My thanks to
 the Abdus Salam International Center for Theoretical Physics in Trieste,
 where I enjoyed a peaceful time to write this thesis. I am also grateful to the Max Planck 
 Institute for Gravitational Physics in Berlin, and its director H. Nicolai, for hospitality during completion of this work. \\

 I am deeply acknowledged to all the people who made possible, and sometimes impossible, my stay in Rome.
 Special thanks to
 Alfredo, Bruno, Fabrice, Jovanka, Kamal, {\bf Lea}, Marco, Maurizio, Paco and Vladimir.\\

 A big thank you to all the graduate students of the group. \\

 Dejo aqu\'{\i} constancia de la deuda contra\'{\i}da con mi amigo 
 y mentor Cmdte. Luigi:
 {\sl `gracias por haberme contagiado parte de tu entusiasmo por la f\'{\i}sica te\'orica'}.\\

 \newpage\thispagestyle{empty}

% OOOOOOOOOOOOOOOOOOOOOOOOOOOOO     Introduction   OOOOOOOOOOOOOOOOOOOOOOOOOOOOOOOOOOOOO

% OOOOOOOOOOOOOOOOOOOOOOOOOOOOO     Introduction   OOOOOOOOOOOOOOOOOOOOOOOOOOOOOOOOOOOOO

% OOOOOOOOOOOOOOOOOOOOOOOOOOOOO     Introduction   OOOOOOOOOOOOOOOOOOOOOOOOOOOOOOOOOOOOO

% OOOOOOOOOOOOOOOOOOOOOOOOOOOOO     Introduction   OOOOOOOOOOOOOOOOOOOOOOOOOOOOOOOOOOOOO

% OOOOOOOOOOOOOOOOOOOOOOOOOOOOO     Introduction   OOOOOOOOOOOOOOOOOOOOOOOOOOOOOOOOOOOOO

 \chapter{Introduction and Conclusions}
 \label{int}\setcounter{chapter}{1}
 \noindent

 It is strongly believed that the Standard Model of particle
 physics is just an excellent low energy limit of
 a more fundamental theory that must account for physics at very high
 energies. For more than twenty years the SM has
 successfully described data from experiments probing
 energies up to the order of 1 TeV. It has also predicted new
 particles and phenomena subsequently verified, and it continues to provide new insights
 into our physical world. In spite of this, at higher
 energies new physics could appear and new theoretical frameworks
 should explain it.
 A viable framework to deal with is String Theory, a theory
 that assumes that the fundamental building blocks of nature are
 one-dimensional objects, \emph{i.e.}, strings.

 String Theory is a quantized theory of gravity that is expected to be also a unified model of all the
 fundamental interactions.
 More recently there has been a renewal of interest in understanding the link
 between string theory and particle physics, this in order to find an
 exact description of confining gauge theories in terms of
 strings. The idea is not new and goes back
 to the early '70 when 't\,Hooft proposed
 a dual model of strings for describing gauge theories with a large number of colors.

 In order to clarify this duality it is convenient to introduce the concept of Black Hole,
 a very massive object originated in a gravitational collapse,
 inside of which all the forces of nature are in action.
 For our purposes, a Black Hole (BH) simply can be regarded as a region of
 spacetime from where no information can
 escape beyond its boundary,
 \emph{i.e.}, the information inside the BH is inaccessible to
 distant observers. Moreover, Black Holes are very simple objects since
 their properties do not depend
 on the kinds of constituents they are made of,
 but instead in some basic properties as the mass, charge and angular
 momentum.

 The simplest black hole, the Schwarzschild BH, has
 a Event Horizon which is a sphere of area $A=4\pi G^2M^2/c^4$.
 It can be proved that
 this area cannot decrease in any classical process.
 On the other hand, gravitational collapsing objects
 which give rise to black holes seem to violate the
 second law of thermodynamics. This is easy to see: since the initial
 collapsing object has
 a non-vanishing entropy and the final BH cannot radiate, hence, the
 entropy of the system has decreased. The problem is solved by
 providing an entropy to the Black Hole.
 For a Schwarzschild BH it was proposed by Bekenstein that
 the entropy is proportional to the Event Horizon area,
 a quantity that can only increase as the entropy does in classical
 thermodynamics,
 \be
 S_{BH}=\frac{1}{4}\frac{A}{l_p^2}~.
 \ee
 The Generalized Second Law (GSL) of thermodynamics
 extends the usual Second Law to
 include the entropy of black holes in a composite system, counting the
 entropy of the standard matter system and also that of the black hole:
 $S_{TOT}=S_{MATT}+S_{RAD}+S_{BH}$. This is the entropy that
 always increases.
 Starting with
 a collapsing object of entropy $S$, the GSL of thermodynamics imposes
 that $S\leq S_{BH}$.
 This is the Holographic Bound,
 and it states that the entropy of a matter system entirely
 contained inside a surface of area $A$,
 cannot exceed that of a black hole of the same size.
  Alternatively, the Holographic Bound can be rephrase saying that the information of
 a system is completely stocked in its boundary surface.

 This statement is generalized by the Holographic Principle. It
 claims that any physical process occurring in
 $D+1$ spacetime dimensions,
 as described by a quantum theory of gravity, can be equivalently described by
 another theory, without gravity, defined on its $D$-dimensional boundary.
 Some authors believe that this statement is universal and
 a fundamental principle of nature.
 Nevertheless, the principle has been tested only in few
 cases. An example is the AdS/CFT correspondence,
 that exactly relates superstring theory in
 a $D$-dimensional space with a superconformal field theory on
 the boundary.

 Finally, we would like to comment on the Black Hole
 Information Paradox and see how the
 Holographic Principle resolves it. The paradox can be posed
 in the following terms. Since a collapsing object is in a definite
 quantum state before it starts to contract, we expect
 the final object to be in exactly the same configuration. However, the
 thermal radiation of the BH comes as mixed states and so the information
 we get from the inside of the BH does not
 reproduce the information booked in the original object. We can say
 that the initial information is lost inside the Black Hole. This
 paradox is solved by the Holographic Principle, since the
 full dynamics of the gravitational theory is now described by a
 standard, though complex, quantum system with unitary
 evolution.

 So far the most accurate holographic proposal relating
 gauge theories to strings is the novel AdS/CFT correspondence.
 In two words, it says that ST defined in a negatively curved anti-de Sitter
 space (AdS) is equivalent to a certain Conformal Field Theory (CFT)
 living on its boundary.
 One concrete example is AdS$_5$/CFT$_4$: it states that
 type IIB superstring theory in
 AdS$_5$ is
 equivalently described  by an extended ${\cal N}=4$ super-CFT
 in four dimensions. The other five
 dimensions of the bulk are
 compactified on S$^5$. The five-sphere with isometry group
 $SO(6)$ is chosen in order to match with
 the $SU(4)$ R-symmetry of the super Yang-Mills theory.
 The AdS/CFT correspondence is a weak/strong
 coupling duality, allowing us to probe the strong coupling regime in the gauge theory
 from perturbative means in the string side, and viceversa.
  This can be seen from the fundamental relation between the superstring side of the
 correspondence and the super-Yang-Mills theory
 \be\label{adscftfund}
 \left( \frac{R}{l_s}  \right)^4=4\pi g_s N ~\leftrightarrow~ g_{YM}^2N\equiv \lambda~,
 \ee
 where $R$ is the curvature radius of the anti-de Sitter space,
 $N$ is the number of colors (considered very large) in the gauge group and $\lambda$ is
 the 't\,Hooft coupling.

 In the Supergravity limit the string length is
 much smaller than
 the radius of the AdS space, given
 \be
 1\ll  \left( \frac{R}{l_s}  \right)^4 ~\leftrightarrow~ \lambda~.
 \ee
 In this limit the bulk theory is manageable, being ${\mathcal N}=8$ Gauged Supergravity, but in the
 boundary side it turns out that the gauge theory is in a strongly couple regime, where a
 perturbative analysis is senseless. This establishes the weak/strong coupling nature of the duality.
 This is an advantage is we want to
 study the strongly coupled regime of one of the theories, since we can
 always use perturbative results in the dual theory. However,
 the difficulties in finding a common perturbative sector where to
 test the correspondence makes it very hard to prove its full validity.
 The strong formulation of the
 AdS/CFT correspondence claims its validity at the string quantum level,
 nevertheless,
 so far no one has been able to quantitatively prove it beyond the classical 
 Supergravity approximation.

 The main challenges of AdS/CFT are two-fold: i)~
 to shed light in the strongly regime of
 non-abelian theories, as a step further in
 the understanding of more realistic QCD-like
 theories;
 ii)~to provide a full proof of the correspondence. The latter is a non-trivial task since we do not have an independent
 non-perturbative definition of string theory in order to compare it with
 the boundary theory in the strongly coupled regime. Pointing in this direction, a couple
 of years ago
 a new proposal was suggested, that goes under the name of BMN Conjecture, and opened the
 possibility to test the correspondence beyond the SUGRA limit.\\

 Since the two-dimensional sigma model for superstrings on AdS$_5$$\times$S$^5$
 supported by R-R flux
 is far from been manageable, even at the
 non-interacting level $g_s=0$, the proof of the stringy regime
 of the AdS/CFT correspondence is hard to carry off. Nevertheless, under some
 conditions, it was
 found by Berenstein, Maldacena and Nastase (BMN) that the
 two sides of the correspondence have an overlapping perturbative
 regime that allows for a test of the duality at the stringy regime.

 Long before the AdS/CFT duality,
 it was known that type IIB superstring theory had two
 maximally supersymmetric spaces:
 flat ten-dimensional spacetime and AdS$_5$$\times$S$^5$.
 More recently it was discovered that in the Penrose limit
 AdS$_5$$\times$S$^5$ reduces to a gravitational
 plane-wave, PP-Wave for short, that gives the third and
 last maximally supersymmetric space of type IIB superstring.
 In this background the theory was shown to be
 described by a free, massive, two-dimensional worldsheet sigma model
 easily quantizable in the light-cone gauge. With these exact string results
 at hand, it remained to associate the limit in the bulk to a process in the boundary theory.
 This last step was fulfilled by BMN.

 In the bulk, the energy of a state
 is given
 by the generator of translations in time $E=i\,\pd_{t}$, and the
 angular momentum around a
 great circle of S$^5$ is associated to $J=-i\,\pd_{\phi}$. The light-cone
 hamiltonian and momentum can be taken to be
 \be\label{bmnHPlc}
 H_{lc}\equiv\,\pd_{x^+}\propto E-J~, \qquad P_{lc}\equiv\,\pd_{x^-}\propto
 \frac{E+J}{R^2}~,
 \ee
 where $x^{\pm}$ are the spacetime light-cone coordinates,
 and an explicit formula for $H_{lc}$ is known.
 From the second equation, we note that the condition of
 non-vanishing light-cone momenta selects those angular
 momenta going with the  radius $R$ as $J\propto R^2$. Moreover,
 the limit $N\to \infty$ imposes $J^2\sim N$.
 It turns out that this limit is different from the large $N$ limit we comment above
 since in that case the expansion in $\lambda\equiv g_s N$ implied
 $g_s$ small, something not required by the present approach.

 On the
 other side of the correspondence, the energy $E$ is identified with the
 scaling dimension $\Delta$ of the operator on the boundary. The
 angular momentum $J$ is associated to an $U(1)$ subgroup of the
 $SU(4)$ R-symmetry of ${\cal N}=4$ super Yang-Mills. With these
 identifications, the
 fundamental relation of the BMN correspondence is settled to be
 \be\label{bmnfund}
 H_{lc}\propto \Delta-J~.
 \ee
 As we said before, according to the AdS/CFT duality the limit we carry out in the
 bulk should have certain consequences in the boundary.
 BMN conjecture states that the limit has its counterpart in a
 truncation of the class of operators defined in the superconformal
 theory. Only operators with large R-charge and $\Delta\sim J$ survive the limit,
 including string excitations.

 The BMN conjecture opens the possibility to test the
 AdS/CFT correspondence beyond the
 supergravity regime, nevertheless, not all the ideas involved are
 conceptually well established.
 One of these is the fate of holography in the BMN limit. It
 seems that the beautiful holographic picture of the AdS/CFT duality is
 completely lost in the plane-wave background. But maybe a remnant of it can survive.\\

 \begin{center}
 {\large\bf This Thesis}\\
 \end{center}

 The main obstacle towards a complete understanding of the holographic duality beyond
 the supergravity approximation, that captures the strongly coupled
 regime of the boundary conformal field theory, is
 our limited comprehension of the quantization of superstring theory in
 R-R backgrounds.
 One possible
 exception is the background AdS$_3$$\times$S$^3$$\times$$\mathcal{M}$
 supported by a NS-NS three-form flux\footnote{S-duality relates NS-NS three-form flux to R-R flux
 and one may in principle resort to the hybrid formalism
 of Berkovits, Vafa and Witten to make part of the space-time
 supersymmetries manifest and compute three-point
 amplitudes.}, which is the near horizon
 geometry of a bound-state of fundamental strings (F1) and
 five-branes (NS5). Since the dynamics on the world-sheet is governed by an
 $\widehat{SL}(2,{\mathbb{R}})\times\widehat{SU}(2)$ Wess-Zumino-Witten model, string amplitudes can be computed
 using CFT techniques.
 The dual two-dimensional superconformal field theory is expected
 to be the non-linear sigma model with target space the
 symmetric orbifold ${\mathcal M}^N/{\mathcal S}_N$, where $\mathcal M$ can be either
 T$^4$ or $K3$.

  In this thesis we give explicit results for bosonic
  string amplitudes
  on AdS$_3$$\times$S$^3$ and the corresponding plane-wave limit. We also
  analyze the consequences of our approach for understanding holography in this set up,
  as well as its possible generalization to other models. The
  original materials appeared (or are
  to appear) in a
  series of publications by the author and collaborators \cite{oz,bdkz,bpz,bdkz2}.\\

 \begin{itemize}
 \item \underbar {Chapter \ref{adscft}}: After reviewing the
 physics of the two sides of the AdS$_3$/CFT$_2$ correspondence, we perform the plane-wave
 limit in an heuristic way in order to set the basis of the more technical material
 that follows in this thesis. In general,
 a precise correspondence between bulk and boundary dynamics has
 been a longstanding challenge in AdS/CFT.
 A discussion on the AdS$_3$/CFT$_2$ discrepancy at
 the supergravity level can be found in the last part of this chapter.
 \item \underbar {Chapter \ref{str-ads3s3}}: We recall the necessary tools
 for dealing with Wess-Zumino-Witten (WZW) models and
 display the full spectrum of strings on AdS$_3$$\times$S$^3$.
 We then compute bosonic string amplitudes on
 this background
 and  determine their Penrose limit.
 A crucial role is played by the charge variables that, from a group
 theoretical point of view, are
 introduced in order to encode the content of the (in)\,finite dimensional
 irreducible representations. More physically, they are seen as coordinates
 in the holographic boundary.
 In this approach, plane-wave chiral primaries are
 obtained by rescaling the charge variables with the level of the
 algebras $k_1,k_2\to \infty$.
 \item \underbar {Chapter \ref{str-pp}}: We compute tree-level bosonic
 string amplitudes in the Hpp-wave limit of
 AdS$_3$$\times$S$^3$ supported by NS-NS three-form
 flux. The corresponding WZW model is obtained contracting
 the algebras $\widehat{SL}(2,\mathbb{R})_{k_1}$ and
 $\widehat{SU}(2)_{k_2}$
 according to $k_1,k_2\to \infty$ with $\mu_1^2 k_1 = \mu_2^2 k_2$.
 We examine the irreducible representations of the model and
 define vertex operators. We then compute two,
 three and four-point functions.
 As a starting point for more realistic models, we
 consider only scalar tachyon vertex operators with no excitation in the internal worldsheet CFT.
 The computation of
 such string scattering amplitudes heavily relies on current
 algebra techniques on the worldsheet,
 generalizing in some sense the results for the Nappi-Witten model developed in
 \cite{dak}.
 We show that these amplitudes exactly match the ones
 computed in the third chapter. It is worth stressing that these
 amplitudes are
 well defined even for $p=0$ states, which are difficult
 (if not impossible) to analyze in the light-cone gauge.
 We have thus provided further evidence for the consistency of the
 BMN limit in this setting.
 \item \underbar {Chapter \ref{holography}}: We propose an extension
 of the previous procedure and the holographic interpretation
 of the charge variables to more interesting and realistic
 models.
 While for the $\widehat{SL}(2,\mathbb{R})_L\times
 \widehat{SL}(2,\mathbb{R})_R$
 algebra, underlying
 AdS$_3$, the charge variables $x$ and $\bar{x}$ can be viewed as coordinates on the
 two-dimensional boundary, in the case of the $\widehat{\mathcal
 H}_{2+2n}$ algebra, underlying a pp-wave geometry,
 the corresponding charge variables
 $x^\alpha$ and $x_\alpha$ become coordinates on a $2n$-dimensional
 holographic screen \cite{kirpio}. This picture replaces the
 one-dimensional null boundary representing the
 geometric boundary in the Penrose limit.
 The approach with charge variables suggests that correlation functions in
 the BMN limit of ${\mathcal N}=4$ super-CFT are indeed well defined and computable.
 \item \underbar {Chapter \ref{outlook}}: Emphasizing on the holographic charge variables we tackle the
 full superstring model.
 We construct vertex operators and give instructions on how  to compute
 scattering amplitudes in the Hpp-wave limit and interpret their
 structure. In principle, one would like
 to address important issues as the spectrum, trilinear couplings
 and operator mixing in a more quantitative
 way.
 In the last part we propose a precise correspondence between
 states in the symmetric product and superstring vertex operators.
 \item \underbar {Appendix \ref{app A}}: The alternative Wakimoto free
 field representation is given and string amplitudes computed.
 These results are shown to coincide with the amplitudes of chapter
 3 and 4.

 \end{itemize}

 Alternatively, one may consider turning on R-R fluxes. The hybrid formalism of Berkovits, Vafa
 and Witten \cite{bvw} seems particularly suited to
 this purpose as it allows the computation of string amplitudes, at least for the massless modes \cite{bobdol},
 and the study of the Penrose limit in a covariant way \cite{berkpp}. The mismatch for 3-point
 functions of chiral primaries (or rather their superpartners) and the consequent lack of a
 non-renormalization theorem for these
 couplings calls for additional investigation in this direction and a careful comparison with
 the boundary CFT results. Once again, the BMN limit may shed some light on this
 issue as well as on the short-distance logarithmic behavior, found in \cite{mo3} for AdS$_3$ and in \cite{dak} for
 NW,
 that could require a resolution of the operator mixing or a scattering matrix interpretation.

 In this thesis we have argued that the BMN limit of physically sensible correlation functions are well
 defined and perfectly consistent, at least for the CFT dual to AdS$_3$$\times$S$^3$. In particular it should not
 lead to any of the difficulties encountered in the case of ${\cal N}=4$ SYM as a result of the use of
 perturbative schemes or of the light-cone gauge.
 In conclusion, we hope we have presented enough arguments
 in order to consider (super)\,strings on the
 plane-wave limit of AdS$_3$$\times$S$^3$ supported by NS-NS fluxes as a source of extremely
 useful insights in holography and the duality between string theories and field theories.

 % OOOOOOOOOOOOOOOOOOOOOOOOOOOOOOOOOOO  AdS3/CFT2  OOOOOOOOOOOOOOOOOOOOOOOOOOOOOOOOOOOOOOOOOOOOOO

 % OOOOOOOOOOOOOOOOOOOOOOOOOOOOOOOOOOO  AdS3/CFT2  OOOOOOOOOOOOOOOOOOOOOOOOOOOOOOOOOOOOOOOOOOOOOO

 % OOOOOOOOOOOOOOOOOOOOOOOOOOOOOOOOOOO  AdS3/CFT2 OOOOOOOOOOOOOOOOOOOOOOOOOOOOOOOOOOOOOOOOOOOOOO

 % OOOOOOOOOOOOOOOOOOOOOOOOOOOOOOOOOOO  AdS3/CFT2  OOOOOOOOOOOOOOOOOOOOOOOOOOOOOOOOOOOOOOOOOOOOOO

 % OOOOOOOOOOOOOOOOOOOOOOOOOOOOOOOOOOO  AdS3/CFT2  OOOOOOOOOOOOOOOOOOOOOOOOOOOOOOOOOOOOOOOOOOOOOO

 \renewcommand{\theequation}{\arabic{chapter}.\arabic{equation}}
 \chapter{Overview of AdS$_3$/CFT$_2$}
 \label{adscft}\setcounter{chapter}{2}
 \noindent

 In this chapter we present the state of the holographic
 correspondence holding between
 string theory in AdS$_3$ and the two-dimensional CFT living on its boundary
 \cite{malda}\footnote{Due to the introductory
 nature of this chapter, we mostly
 refer to some review articles the author found useful
 in preparing this thesis. For a more complete list of references see next chapters.}.
 This chapter is organized as follows. In section \ref{adsbh} we
 first introduce the anti-de Sitter spaces,
 the geometry relevant for the correspondence,
 and then we review the microscopic physics of some
 generalized black holes. In the last part of this section we comment on the core idea of the correspondence,
 \emph{i.e.}, its holographic property. In
 section \ref{stads3} we describe
 the dynamics of strings moving on general group manifolds, in particular on AdS$_3$. From a semi-classical analysis we get a
 valuable geometrical picture of the quantum spectrum. In
 section \ref{ads3cft2} we introduce the
 AdS$_3$/CFT$_2$ correspondence. Finally, in section \ref{ppwaves} we discuss
 the possibility of testing the AdS/CFT correspondence beyond the
 supergravity approximation, in concrete, in the plane-wave limit.

 % OOOOOOOOOOOOOOOOOOOOOOOOOO      AdS-BH       OOOOOOOOOOOOOOOOOOOOOOOOOOOOOOOOOOOOOOOO

 % OOOOOOOOOOOOOOOOOOOOOOOOOO      AdS-BH       OOOOOOOOOOOOOOOOOOOOOOOOOOOOOOOOOOOOOOOO

 \section{Anti-de Sitter spaces and black holes}
 \label{adsbh}
 \noindent

 \emph{Anti-de Sitter}\index{Anti-de Sitter spaces|textbf} geometries are solutions of
 Einstein equations of
 motion with negative cosmological constant
 $\Lambda$ =$-(D-1)(D-2)/2R^2$, where $D$ is the number of spacetime dimensions of the AdS space and
 $R$ its curvature radius. These spaces
 have the important property of being maximally
 symmetric, or in other words, to have a maximal number of Killing
 vectors, in all
 $D(D+1)/2$.
 The AdS$_{p+2}$/CFT$_{p+1}$
 correspondence proposes the identification of the
 isometry group of AdS$_{p+2}$ with the conformal symmetry of the
 flat Minkowski space ${\mathbb R}^{1,p}$.\\

 An AdS$_{p+2}$ space of curvature radius $R$ can be
 defined by the connected hyperboloid\,\cite{agmoo}
 \beq\label{adshyper}
 X_{-1}^2+X_0^2-\sum_{i=1}^{p+1}X_i^2=R^2~,
 \eeq
 where $X_M$ $(M=-1,0,\dots\,p+1)$ are
 the coordinates on the embedding space. The latter is a (p+3)-dimensional
 flat space ${\mathbb R}^{2,p+1}$ with metric
 \beq\label{metric}
 ds^2=-dX_{-1}^2-dX_0^2+\sum_{i=1}^{p+1}dX_i^2~.
 \eeq
 From here we see that the AdS$_{p+2}$ space has isometry
 group SO$(2,p+1)$ and by construction is
 homogeneous and isotropic. Even if by definition the ambient space has two
 time directions, notice that one of them is orthogonal to the hypersurface defining the AdS
 space, leaving us with only one time direction as desired.
 As we will see below, the closed timelike curves
 arising in this picture can be avoided
 by choosing an appropriate
 set of coordinates (global coordinates, see below) and then unwrapping the time direction.\\

 We can now define global coordinates on the hyperboloid (\ref{adshyper})
 {\colsep
 \beqa\label{glcoord}
 X_{-1}&=&R\cosh \rho \,\sin \tau~, \qquad X_0=R\cosh\rho\,\cos\tau~,\nn\\
 X_i&=&R\sinh \rho\,\Omega_i~,
 \eeqa}
 with  $i=1, \dots, p+1$ and $\sum_i\Omega_i^2=1\,$. Inserting
 (\ref{glcoord})
 in the metric (\ref{metric}), this takes the form
 \beq\label{metricgc}
 ds^2=R^2\,(-\cosh^2\rho\,d\tau^2+d\rho^2+\sinh^2\rho\,d\Omega^2)~.
 \eeq
 It is worth remarking that the  global coordinates cover the whole AdS
 space. We can get rid of
 closed timelike curves by considering
 a non-compact coordinate system where time is non-compact, $-\infty<\tau<\infty$.
 When talking about the AdS/CFT correspondence we will always refer to this covering space. \\

 Another useful parametrization is given by the Poincar\'e coordinates,
 {\colsep
 \beqa\label{poincoord}
 X_{-1}&=&Ru\,t,\qquad\qquad X_0=\frac{1}{2u}\,[1+u^2(R^2+\x\,^2-t^2)],\nn\\
 X_i&=&Ru\,x_i,\nn\\
 X_{p+1}&=&\frac{1}{2u}\,[1-u^2(R^2-\x\,^2+t^2)],
 \eeqa}
 where (this time) $i=1, \dots, p$\,. With this change of variables, the
 induced metric on  AdS$_{p+2}$ becomes
 \beq\label{poinmet}
 ds^2=R^2\left[ u^2(-dt^2+d\x\,^2)+ \frac{du^2}{u^2} \right].
 \eeq

 Since in this thesis
 we will be particularly interested in AdS$_3$ spaces, $\ie$ $p=1$
 in our notation, it will be helpful to have the
 metric in global coordinates for this special case
 \index{AdS$_3$ metric|textbf}
 \beq \label{3glmet}
 ds^2=R^2(-\cosh^2\rho\,d\tau^2+d\rho^2+\sinh^2\rho\,d\phi^2),
 \eeq
 where the unit vector $\Omega_i\,$ in (\ref{metricgc})
 reduces to a single parameter $\phi$. In terms of this coordinates
 and choosing the universal covering range already mentioned,
 $0\leq\rho<\infty\,$, $-\infty<\tau<\infty$ and $0\leq\phi<2\pi$, the AdS$_3$
 space can be seen as a solid cylinder (see Figure \ref{fig1}).

 \begin{figure}[!htb]
 \begin{center}
 \includegraphics[width=0.2\textwidth]{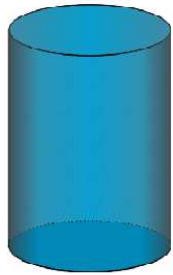}
 \end{center}
 \caption{Representation of an AdS$_3$ space in global coordinates.}
 \label{fig1}
 \end{figure}

 For later use, let us also write down the full
 AdS$_3$$\times$S$^3$ metric\index{AdS$_3$$\times$S$^3$ metric|textbf} in
 global coordinates
 \be \label{ads3s3}
 ds^2=R^2(-\cosh^2\rho\,d\tau^2+d\rho^2+\sinh^2\rho\,d\phi^2)
 +R^2(\cos^2\theta\,d\psi^2+d\theta^2+\sin^2\theta\,d\varphi^2).
 \ee
 Both spaces have radius $R$. The variables
 $0\leq\theta,\psi,\varphi<2\pi$ are angles in the
 three-sphere.\\

 The boundary of AdS$_{p+2}$ is the $p+1$-dimensional surface
 at $\rho\to \infty$, a region that cannot be reached by a massive
 particle while a light ray can go and get back in a finite
 time. The AdS/CFT correspondence states that the dual description of the gravity
 theory lives precisely in this boundary space. We will come back to this later on.
 After this overview of Anti-de Sitter spaces now we would like to understand
 how these backgrounds can be generated from string theory principles and what is their exact role
 in the correspondence.\\

 In the middle nineties it was shown that there are only five
 consistent superstring theories (types IIA, IIB, HE and HO
 with closed strings and type I
 that unavoidably contains both open strings and closed strings),
 all of them living in ten spacetime dimensions.
 This was supported by the idea that some of the theories were
 related between them by fundamental symmetries, and by the fact,
 unexpectedly, that
 non-perturbative dualities of type-II theories allowed for R-R $p$-brane
 solutions
 in \emph{supergravity}\index{Supergravity|textbf} -- the low energy effective theory of the
 corresponding superstring\footnote{An Introduction to these subjects can
 be found in \cite{pbranes}.}. Moreover, $p$-branes were shown to have an
 equivalent description in terms of hyperplanes in spacetime, a sort of R-R charged
 membranes --
 in fact BPS solitons, which in turn can be a source of closed strings.
 These last objects are called \emph{Dp-branes}\index{Dp-branes|textbf}. They are
 non-rigid
 hyperplanes with $p$ space dimensions and $p+1$ dimensional
 world-volume.
 Due to the open-closed duality of the string models, the $Dp$-branes are also
 supposed to be the
 slices of space where the ends of an open string can sit (see \cite{strings} for
 an introduction to string theory). This dual vision of
 $Dp$-branes is at the heart of the most exciting
 developments in string theory, including the celebrated AdS/CFT
 correspondence.\\

 D$p$-branes are solutions of supergravity equations,
 and it can be proved that for type IIB theory the only admissible BPS D$p$-branes
 are for $p=-1, \,1,\,3,\,5,\,7,\,9$, each of them been a particular
 solution. But not
 all imaginable configurations of branes produces new supergravity
 solutions, actually a D$p$-D$p'$ configuration needs to satisfy some
 conditions in order to  generate
 a stable solution, compensating the R-R charge repulsion with some attractive potential.
 For the D1-D5 system we will be concerned with, the condition $p-p'=4$ insures that we have a
 stable configuration, or in other terms a bound
 state at threshold.
 By this procedure the supergravity solution we will get at first hand is not the
 AdS$_3$$\times$S$^3$ space we are looking for, but instead a more complex
 space that in a certain limiting region reduces to it. To present our analysis
 in the more convenient way, we choose to begin generating
 this more general space and then going to the
 suitable limiting case.\\

 The relevant set-up for AdS$_3$$\times$S$^3$ consists in
 D1-branes
 living on D5-branes. To visualize it let us draw a table
 showing where the branes are located.\\
 \nopagebreak
 \begin{center}
 \begin{tabular}{|c|c|c|c|c|c|c|c|c|c|c|}
 \hline&$x_0$&$x_1$&$x_2$&$x_3$&$x_4$&$x_5$&$x_6$&$x_7$&$x_8$&$x_9$\\
 \hline {\bf \small{D1}}&--&--&$\perp$&$\perp$&$\perp$&$\perp$&$\perp$&$\perp$&$\perp$&$\perp$\\
 \hline {\bf \small{D5}}&--&--&$\perp$&$\perp$&$\perp$&$\perp$&--&--&--&--\\
 \hline
 \end{tabular}
 \end{center}
 \nopagebreak
 \begin{center}
 {\bf Table 1.} The D1-D5 brane configuration.
 \end{center}

 A dash under the coordinate $x_{M}$ ($M=0,\,\dots ,\,9$) means that the brane is extended in
 that direction. On the other hand, a $\perp$ symbol means that the brane is
 perpendicular to $x_M$, or in other words that the brane looks like a point particle
 along that direction. It can be seen that due to the presence of
 the D-branes the ten dimensional Lorentz symmetry
 is broken from the original $SO(1,9)$ to $SO(1,1)_{01}\times SO(4)_{2345}\times SO(4)_{6789}$,
 where the indices stand for the spacetime directions shown in Table 1. Truly
 speaking, in string theory the $SO(4)_{6789}$ symmetry is broken by wrapping the directions on the
 four dimensional manifolds T$^4$ (or $K3$), but at low energies the compactified dimensions
 are too small, restoring it as a symmetry
 of the supergravity solution, $U(1)^4\to SO(4)_{6789}$. \\

 The \emph{D1-D5 brane configuration}\index{D1-D5 system|textbf} described above is a IIB supergravity solution
 with black hole type metric
 \beq\label{d1d5}
 ds^2=f_1^{-1/2}f_5^{-1/2}(-dx_0^2+dx_1^2)+
 f_1^{1/2}f_5^{1/2}(dr^2+r^2d\Omega_3^2)+f_1^{1/2}f_5^{-1/2}dx_Adx_A,
 \eeq
 where $d\Omega_3$ stands for the metric on the three-sphere, $A=6,\,7,\,8,\,9$
 are the directions along the four torus
 and $r^2=x_2^2+x_3^2+x_4^2+x_5^2$ measures the transverse
 direction to the D1 and D5-branes. The harmonic functions of the
 transverse directions are
 \beq\label{f1f5}
 f_1=1+\frac{g_s\alpha'Q_1}{vr^2},\qquad
 f_5=1+\frac{g_s\alpha'Q_5}{r^2}.
 \eeq
 The volume of the four-torus is given by
 $V_{T^4}=(4\pi^2\alpha')^2v$ and
 $Q_{1}$ ($Q_5$) is the number of D1(D5)-branes. In addition to the metric
 there are other fields in the supergravity solution (see for example \cite{wadia}).
 For the time been, it is enough for us to remark on the presence of a non-zero R-R
 three-form flux $F_3$.\\

 Since it is not well known how to quantize the superstring
 in the presence of generic R-R backgrounds \footnote{At present,
 the \emph{Pure Spinors Formalism}\index{Pure spinors|textbf} is the most promising approach to tackle
 this outstanding problem \cite{berk}.}, it is convenient to
 consider the S-dual configuration of $Q_1$ fundamental strings living
 on $Q_5$ NS5-branes, a system that is supported by a NS-NS
 3-form flux $H_3$. S-duality is a weak$\to$strong transformation
 that applied to the D1-D5 system
 essentially transforms the fields according to $ds^2\to e^{-\phi}\,ds^2$, $\phi\to -\phi$
 and $F_3\to H_3$, obtaining
 \beq\label{f1ns5}
 ds^2=f_1^{-1}(-dx_0^2+dx_1^2)+f_5(dr^2+r^2d\Omega_3^2)+dx_Adx_A.
 \eeq
 For the \emph{F1-NS5 bound state}\index{F1-NS5 bound state|textbf} the relevant harmonic functions are
 \beq\label{f1f5f1ns5}
 f_1=1+\frac{g_s'^2\alpha'Q_1}{v'r^2},\qquad
 f_5=1+\frac{\alpha'Q_5}{r^2}, \eeq
 where $v'$ is defined as for D1-D5, but now measured
 in the rescaled coordinates, and the string coupling becomes $g_s'=1/g_s$.
 Clearly, this does not look like the solution we are looking for,
 indeed it
 is more like a black hole geometry of the Reissner-Nordstrom type.
 As we will see next, it is only near the horizon of such a black object/brane that we
 will get the desired  AdS$_3$$\times$S$^3$ space.

 The \emph{near horizon limit}\index{Near horizon limit|textbf} of the F1-NS5 system
 is simply obtained by going close
 enough to the branes, that is, taking $r\to 0$ in (\ref{f1ns5}),
 while keeping fixed $v'$ and $g_6'=g_s'/\sqrt{v'}$.
 With these prescriptions the metric takes the new form
 \beq\label{f1ns5nhl}
 ds^2=\frac{r^2}{\alpha'Q_5}(-dx_0^2+dx_1^2)+\frac{\alpha'Q_5}{r^2}dr^2
 +\alpha'Q_5 d\Omega_3^2+dx_Adx_A.
 \eeq
 Above we have also used the fact that in the near horizon geometry the string coupling constant
 is $g_s^2=v\frac{Q_5}{Q_1}$. It is not hard to see that
 (\ref{f1ns5nhl}) is just another way to write the
 AdS$_3$$\times$S$^3$$\times$T$^4$ metric. By simply changing
 variables $r=\frac{R^2}{u}$ and writing the common curvature radius
 in terms of the NS5-branes as $R^2=Q_5\alpha'$, we get
 \beq\label{ads3pointmet}
 ds^2=R^2\left[\frac{1}{u^2}(-dx_0^2+dx_1^2)+\frac{du^2}{u^2}\right]+
 R^2\,d\Omega_3^2+dx_Adx_A~,
 \eeq
 that is AdS$_3$$\times$S$^3$$\times$T$^4$ in Poincar\'e coordinates, see
 (\ref{poinmet}). \\

 These geometries are closely related to the \emph{BTZ black hole}\index{BTZ black holes|textbf},
 a background suitable to study the microscopic
 physics of black holes (see \cite{townsend} for an introduction).
 We can
 think of BTZ black holes as five-dimensional
 near-horizon geometries coming from the Kaluza-Klein reduction of the
 six-dimensional D1-D5 black string (\ref{d1d5}).
 In this limit,
 and after changing variables,
 the BTZ metric can be written as
 \beq\label{btz}
 ds_{BTZ}^2=-\frac{(r^2-r_+^2)(r^2-r_-^2)}{R^2\,r^2}dt^2+
 \frac{R^2\,r^2}{(r^2-r_+^2)(r^2-r_-^2)}dr^2+r^2\left(d\phi+\frac{r_+r_-}{Rr^2}dt\right)^2~,
 \eeq
 where $r_+(r_-)$ is the outer(inner) horizon and $\phi$ is
 periodic, $\phi=\phi+2\pi$. Notice that in the simplest case
 $r_+=r_-=0$, the BTZ geometry reduces locally to AdS$_3$. In
 spite of this, the
 global identification of $\phi$ suggests a crucial difference:
 the fermions of the two spaces have opposite boundary conditions.
 This has significant consequences for the quantum analysis of
 black holes. However we will not dwell on such topics (the reader interested in
 an accurate presentation can look at reference \cite{wadia}).\\

 The \emph{no-hair theorem}\index{No-hair theorem|textbf} establishes that any
 stationary black hole
 can be completely characterized with only three quantities -- nothing
 more than the
 three physical quantities that
 describe the object before it collapses: its mass, angular
 momentum and charge.
 There are also the \emph{Black
 Hole Laws}\index{Black holes laws|textbf}. The zero law says that the surface
 gravity $\kappa$ - this quantity plays the role of the temperature - is uniform over the whole
 horizon. The first law relates the change of the horizon area with
 the three fundamental properties associated with a black hole
 \beq\label{bh1law}
 dA=\frac{8\pi}{\kappa}\left[\,dM-\omega\, dJ-\Phi\, dQ \,\right],
 \eeq
 where $\omega$ is the angular momentum and $\Phi$ the
 electrostatic potential. The second law states that the horizon
 area, a measure of the entropy, can never decrease. Any physical
 process will give rise to an
 increase of the total entropy (ordinary matter plus black
 holes). That the black hole can not completely cool down is the statement of
 the fourth law.\\

 One of the most important results in black hole physics,
 found by Hawking, states
 that these objects radiate with the spectrum of a
 black body (thermal radiation) at certain temperature $T_H$.
 Moreover, an entropy $S$ is also associated to the black hole.
 For a D-dimensional black hole the entropy and temperature are
 given by
 \beq \label{btzst}
 S=\frac{A_d}{4\hslash G_D},\qquad\qquad
 T_H=\frac{\hslash\kappa}{2\pi}\,.
 \eeq
 This entropy is a  generalization to higher
 dimensions of the \emph{Bekenstein-Hawking
 entropy formula}\index{Bekenstein-Hawking formula|textbf}.

 For the BTZ black hole introduced above, the role of the electrical charge
 of the standard Reissner-Nordstrom black hole is played by the
 R-R charges.
 The formulas for the entropy and the temperature are
 \beq \label{btzst}
 S=\frac{2\pi}{4G_3}~,\qquad\qquad T_H=\frac{(r_+^2-r_-^2)}{2\pi
 r_+R^2}~.
 \eeq
 BTZ Black Holes have played an
 important role in recent
 developments in string theory. In part because the computation of Hawking radiation
 in the full supergravity theory was shown to coincide with the semiclassical analysis.
 Moreover, the agreement found between
 Hawking radiation, and also the temperature, calculated in the D1-D5 black hole and in a
 superconformal
 field theory on two-dimensions, led to propose the AdS/CFT
 correspondence. There was also found a
 three-dimensional analogue of
 the Hawking-Page transition, where the process was naturally interpreted as
 the fluctuation of the partition function from AdS$_3$ geometry, dominating at low
 energies, to the Euclidean BTZ black hole that prevail at high energies.

 In an attempt to solve the puzzle arising from the information loss
 paradox, \emph{i.e.,} the fact that a black hole absorbs everything without any emission,
 it was
 proposed that the entropy of a matter system confined in a volume
 with boundary area $A$ should be upper bounded by
 the entropy of a black hole with same horizon area.
 Pointing in the same direction, the \emph{Holographic
 Principle}\index{Holographic principle|textbf}
 \cite{holpr} claims that any physically sensible formulation of a fundamental
 quantum theory of gravity, such as string theory, defined in a region with boundary of area $A$
 is fully described by $A/4$ number of degrees of freedom per Planck area.
 This principle has been formulated at a great level of
 generality and for any theory that quantizes gravity. Nevertheless, only
 string theory, thanks to the
 AdS/CFT correspondence, really fulfills the requirements of the
 principle in an
 accurate manner. Let us close this section showing this last
 statement for the more standard AdS$_5$/CFT$_4$.

 We begin by introducing an infrared cutoff $\delta$ in order to regularize the
 bulk spacetime, thus the area of the S$^3$$\times$S$^5$ boundary
 is roughly given by $A=R^8/\delta^3$. In the boundary side we
 introduce the ultraviolet cutoff $\delta$ (there is an UV/IR relation),
 and after using the bulk-boundary formula $R^4=4\pi g_{YM}^2N\alpha'^2$, we find
 that the total number of degrees of freedom of the $U(N)$
 boundary gauge theory is roughly given by $n\sim N^2\delta^3\sim A$. In this way the
 correspondence saturates the
 holographic bound, or, equivalently, the number of degrees of
 freedom of the boundary CFT agrees with the number of degrees of
 freedom contained in the bulk S$^3$$\times$S$^5$. Therefore, we have
 heuristically prove that the
 AdS/CFT duality is a holographic proposal. There is a slice by slice holographic
 correspondence between bulk physics and boundary theory, and the latter
 in the form of a conformal field theory generates
 the unitary evolution of the boundary data.

% OOOOOOOOOOOOOOOOOOOOOOOOOO      Strings on AdS3 X N     OOOOOOOOOOOOOOOOOOOOOOOOOOOOOOOOOOOOOOOO

% OOOOOOOOOOOOOOOOOOOOOOOOOO      Strings on AdS3 X N    OOOOOOOOOOOOOOOOOOOOOOOOOOOOOOOOOOOOOOOO

 \section{Strings on AdS$_3$$\times$$\mathcal N$}
 \label{stads3}
 \noindent

 In this section we introduce some basic facts about affine conformal
 field theories and present a helpful picture of strings moving on
 AdS$_3$. This section is based on \cite{mo1}.\\

 AdS$_3$ has isometry group $SO(2,2)$, that is isomorphic to
 the product $SL(2,{\mathbb R})\times SL(2,{\mathbb R})$, one for each left and
 right movers on the worldsheet $\Sigma$.
 On general ground, strings on group manifolds,
 such as AdS$_3$, can be
 described by
 \emph{Wess-Zumino-Witten (WZW) models}\index{WZW models|textbf}\footnote{Sometimes
 also called WZNW
 models
 to emphasize the contribution due to Novikov.}. These are
 conformal invariant theories whose basic object is a group
 element $g$ that takes values in the Lie group G, $g:\,(\tau_{ws},\sigma)\to
 G\,$.\\

 These theories are nicely formulated in terms of
 a non-linear sigma model action plus a \emph{Wess-Zumino term}\index{Wess-Zumino term|textbf},
 \beq\label{swzw}
 S_{\mathrm{WZW}}=\frac{k}{16\pi\alpha'}\int_{\Sigma}d^2\sigma\,\textrm{Tr}
 (g^{-1}\pd g\,g^{-1}\pd g) + \frac{ik}{8\pi}\int_{M}d^3\sigma\,\textrm{Tr}
 [\epsilon_{\alpha\beta\gamma}
 (g^{-1}\pd^{\alpha}g)\,(g^{-1}\pd^{\beta}g)\,(g^{-1}\pd^{\gamma}g)],
 \eeq
 where the second term, a total derivative, is an integral over a three dimensional
 manifold $M$ with boundary $\Sigma$. We will see that for AdS$_3$
 the constant $k$, the level of the affine algebra, is equal to
 $Q_5$
 the number of D5-branes generating the background.

 The equations of
 motion $\pd_-(\pd_+gg^{-1})=0$ derived from (\ref{swzw})
 admit a general solution constructed with purely right and left moving
 contributions, $g=g_+(x^+)g_-(x^-)$.
 Here we have defined the light-cone coordinates on the
 worldsheet $x^{\pm}=\tau_{ws}\pm \sigma$.
 In fact, what makes these models so particular is
 the presence of left and right independent conserved currents
 \beq\label{chcurr}
 J_R(x^+)=k \,\mathrm{Tr}(g\pd_+ g^{-1}),\qquad    J_L(x^-)=k \,\mathrm{Tr}(\pd_- g g^{-1})      ,
 \eeq
 where $T^a$ are the generators of the Lie algebra of $G$.

 This property allows for the much larger \emph{affine symmetry}\index{Affine symmetry|textbf}
 \beq \label{currentsym}
 g(x^+,x^-)\to \Omega(x^+)g(x^+,x^-)\bar\Omega^{-1}(x^-),
 \eeq
 where $\Omega$ and $\bar\Omega$ are two arbitrary matrices valued in $G$.

 For these models the metric can be written in terms of the field $g$ according to the
 Maurer-Cartan formula\index{Maurer-Cartan formula|textbf}
 \beq\label{mcform}
 {\mathrm g}_{\mu\nu}=\frac{1}{2}\textrm{Tr}(g^{-1}\pd_{\mu}g\,g^{-1}\pd_{\nu}g).
 \eeq

 We will leave for the next chapter the quantum formulation
 of the conformal field theory supported with affine algebras, for the moment we will
 just concentrate  in realizing classically the WZW model for the
 matrix representation of the group $SL(2,{\mathbb R})$.\\

 An element of $SL(2,{\mathbb R})$ can be parameterized as
 \beq\label{gsl2X}
 g=
 \left(
 \begin{array}{cc}
   X_{-1}+X_1 & X_0-X_2 \\
   -X_0-X_2 & X_{-1}-X_1 \
 \end{array}
 \right)\,\,.
 \eeq

 It is straightforward to prove that the metric we get using the
 formula (\ref{mcform}) gives the AdS$_3$ element in the form (\ref{metric}). Of course, we
 can also parameterize the $SL(2,{\mathbb R})$ element $g$ in terms of  the global
 variables $\rho$, $\tau$ and $\phi$. We can take
 \beq\label{gsl2gc}
 g=e^{\frac{i}{2}(\tau+\phi)\sigma_2}e^{\rho\sigma_3}e^{\frac{i}{2}(\tau-\phi)\sigma_{2}},
 \eeq
 with $\sigma_i$ ($i=1,\,2,\,3$) the Pauli matrices. With $g$ given in this
 form, we obtain correctly the metric (\ref{3glmet}). From now on we
 will mostly use the standard normalization $R=1$.\\

 The authors of \cite{mo1} proved that the most general solution
 that this model admits consists on two independent contributions of the
 form
 \beq \label{g+g-}
 g_+=U\,e^{v_+(x^+)\sigma_2},\qquad g_-=e^{u_-(x^-)\sigma_2}\,V,
 \eeq
 where $U$ and $V$ are constant elements of $SL(2,{\mathbb R})$.
 Appropriately choosing values of the parameters in (\ref{g+g-})
 they also showed that two different physical solutions arise.
 We will not give the algebraic expressions here, for our purposes a
 picture of the solution will be more than enough. The
 two kinds of solutions, timelike (A) and spacelike (B), are shown in Figure \ref{figs2}.

 \nopagebreak
 \begin{figure}[!h]
 \hspace{2.7cm}
 \includegraphics[width=0.2\textwidth]{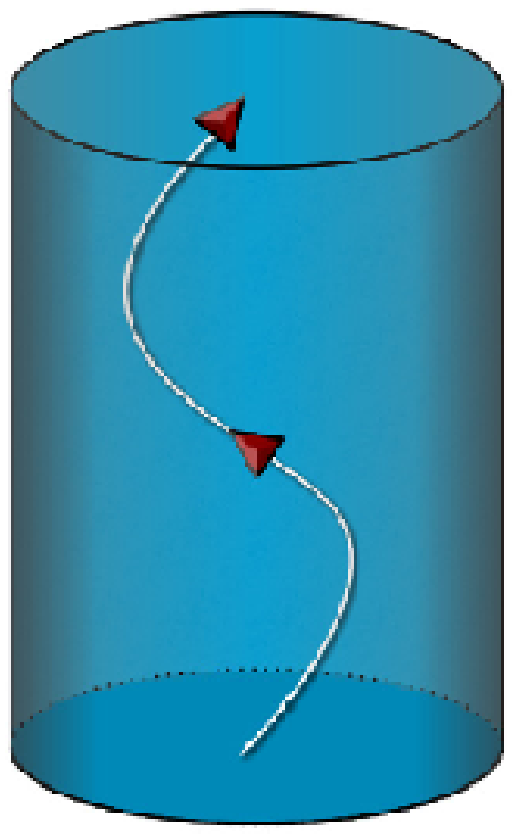}\hspace{4cm}
  \includegraphics[width=0.2\textwidth]{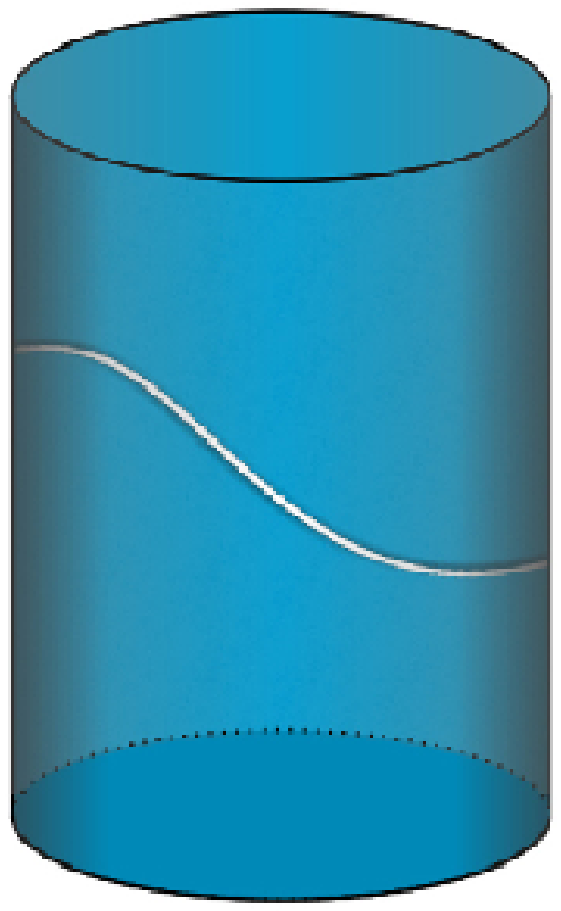}
 \begin{center}
 ~\quad(A)\hspace{6.8cm}(B)
 \end{center}
 \caption{Timelike and spacelike geodesics. Time goes
 upward.}
 \label{figs2}
 \end{figure}

 Moreover, we can generate new solutions by shifting two of the
 parameters according to
 \beq\label{spflow}
 \tau\to \tau+\omega\tau_{ws},\qquad\qquad\qquad
 \phi\to\phi+\omega\sigma,
 \eeq
 where $\omega$ is an integer that to some extent can be
 interpreted as a \emph{winding number}\index{Winding number $\omega$|textbf}. The last
 interpretation comes from the fact that the \emph{spectral flow} transformation (\ref{spflow})
 gives the two different solutions drawn in Figure \ref{figs3},
 interpreted as short and long strings wrapping on AdS$_3$.

 \nopagebreak
 \begin{figure}[!h]
 \hspace{2.7cm}
 \includegraphics[width=0.2\textwidth]{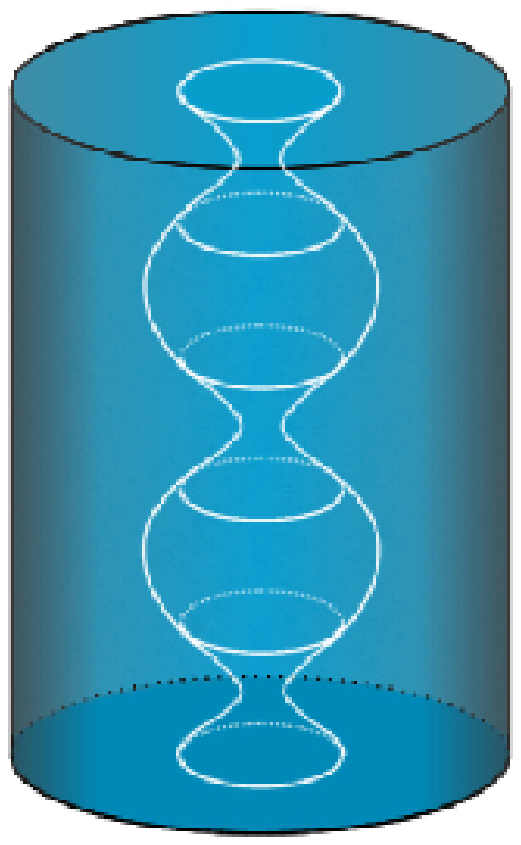}\hspace{4cm}
  \includegraphics[width=0.2\textwidth]{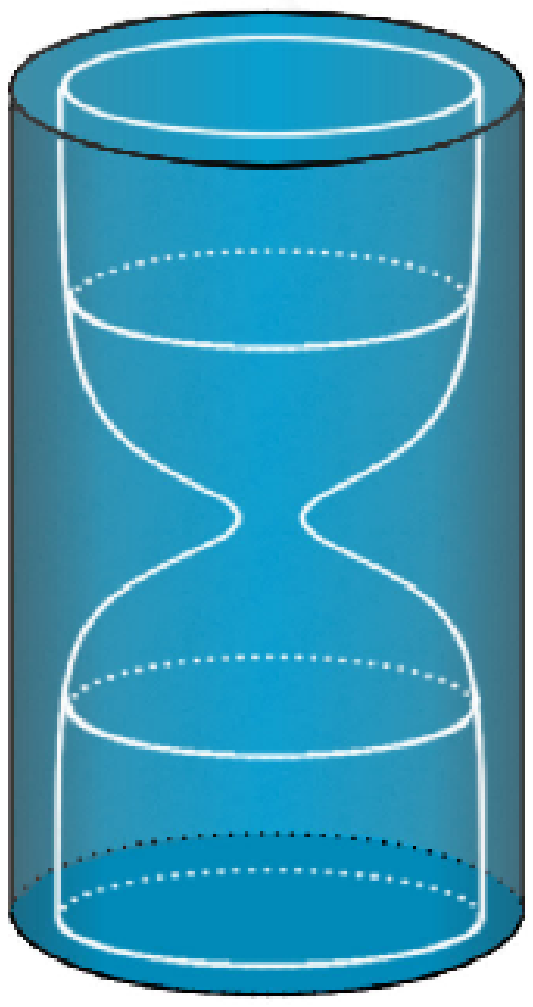}
 \begin{center}
 \qquad(A)\hspace{6.6cm}(B)
 \end{center}
 \caption{Short and long strings propagating in AdS$_3$.}
 \label{figs3}
 \end{figure}

 From Figure \ref{figs3} (A) we see that timelike spectral flowed geodesics behave as
 strings contracting and expanding continuously
 around the axis of
 the AdS$_3$ cylinder without ever approaching the boundary.
 This behavior is due to the opposite forces coming from
 the tension of the string and the NS-NS $B$ field repulsion.
 These objects are called \emph{short
 strings}\index{Short strings AdS$_3$|textbf}. We also have the
 strings shown in Figure \ref{figs3} (B). They come in from the boundary, shrink
 around the axis and expand away reaching again the boundary, this are the
 \emph{long strings}\index{Long strings AdS$_3$|textbf}. The name
 winding number for the parameter $\omega$ can be misleading,
 since during the process of collapsing and expanding its value
 can change.\\

 This concludes our classical analysis of bosonic strings
 moving on AdS$_3$. A full quantum approach will be carried off in
 the next chapters and we will read the previous interpretation in the spectrum
 of the theory. In chapter 6 we shall see that the fermionic
 string is still described by a  model of this kind.

% OOOOOOOOOOOOOOOOOOOOOOOOOO      AdS3/CFT2      OOOOOOOOOOOOOOOOOOOOOOOOOOOOOOOOOOOOOOOO

% OOOOOOOOOOOOOOOOOOOOOOOOOO      AdS3/CFT2       OOOOOOOOOOOOOOOOOOOOOOOOOOOOOOOOOOOOOOOO

 \section{The AdS$_3$/CFT$_2$ holographic duality}
 \label{ads3cft2}
 \noindent

 AdS$_3$/CFT$_2$ conjecture proposes that superstring theory on
 AdS$_3$$\times$S$^3$$\times$T$^4$ is dual to a conformal theory
 leaving on the boundary of AdS$_3$. This particular realization
 of the
 correspondence is especially interesting since, unlike AdS$_5$/CFT$_4$,
 it can be checked beyond the
 supergravity approximation. This is due to two major facts:
 $a)$ as we saw, strings on AdS$_3$
 can be described by an exactly solvable conformal theory, $\ie$ a WZW
 model,
 $b)$ the boundary of the three-dimensional
 anti-de Sitter space is two-dimensional, giving rise to an infinite number
 of generators of the conformal group.

 In order to make the presentation
 as clear as possible, we will first restrict to the supergravity
 regime. Then, for the moment we consider the volume of the compactification
 manifold of the order of the string
 length, so we can ignore the Kaluza-Klein modes around these
 directions and superstring theory reduces to supergravity
 on AdS$_3$$\times$S$^3$$\times$${\mathcal M}$: (2,0) and (1,1)
 supergravity for
 ${\mathcal M}=K3$ and T$^4$ respectively.

 It is known that the low energy
 dynamics of the D1-D5 system is described by an
 $U(Q_1)\times U(Q_5)$ gauge theory in two dimensions with $\mathcal
 N=(4,4)$ supersymmetry. Moreover, the Higgs branch of this gauge
 theory description\footnote{The sector
 where the scalars have trivial
 expectation value.}
 flows in the infrared, $\ie$ near the horizon, to an $\mathcal N=(4,4)$ super-conformal
 field
 theory with central charge $c=6\,Q_1Q_5$ on certain  manifold $\widetilde{\mathcal M}$.
 On the other side, we can think the D1-branes
 as solitonic strings of
 the D5-brane theory. These arguments (see \cite{wadia} for details)
 led to propose that the conformal theory
 is a non-linear sigma model on the \emph{symmetric
 orbifold}\index{Symmetric orbifold ${\mathcal S}^N(\mathcal M)$|textbf}
 \beq\label{T4symorb}
 \mathcal S^{Q_1Q_5}({\widetilde{\mathrm T}}^4)=
 \frac{(\widetilde{\mathrm T}^4)^{Q_1Q_5}}{\mathcal
 S_{Q_1Q_5}},
 \eeq
 where we choose $\widetilde{\mathcal M}={\widetilde{\mathrm T}}^4$, and
 the tilde on the torus indicates that the four dimensional
 torus of this theory is not necessarily the same as the one in
 the bulk theory. $\mathcal S_{Q_1Q_5}$ is the permutation group
 of $Q_1Q_5$ variables.

 The first evidence for this correspondence comes as usual
 from the analysis of the symmetries of
 the bulk theory and the superconformal boundary theory. In the next
 table we show how the symmetries match.\\

 \begin{center}\label{ads3isom}
 \begin{tabular}{|c|c|}
 \hline {\bf \footnotesize{Bulk IIB SUGRA}}&{\bf \footnotesize{Boundary $\mathcal N$=(4,4) SCFT}}\\
 \hline \hspace{5mm}\footnotesize{AdS$_3$ isometry} \footnotesize{$SO(2,2)\,\,\,\,\,\,$}&\hspace{5mm}\footnotesize{global Virasoro $SL(2,{\mathbb R})\times SL(2,{\mathbb R})\,\,\,\,\,\,$}\\
 \hline \hspace{5mm}\footnotesize{S$^3$ isometry} \footnotesize{$SO(4)$}&\footnotesize{R-symmetry $SU(2)\times SU(2)$}\\
 \hline \hspace{5mm}\footnotesize{T$^4$ isometry $SO(4)$}&\scriptsize{
 $\widetilde{\mathrm T}^4$} \footnotesize{isometry $SO(4)$}\\
 \hline \hspace{5mm}\footnotesize{16 supersymmetries}&global supercharges\\
 \hline
 \end{tabular}
 \end{center}
 \nopagebreak
 {\bf Table 2.} Correspondence between the symmetries of the bulk
 and boundary theory.\\

 Even if a first evidence comes from the matching of the global
 symmetries on the two sides of the correspondence, the duality
 should also say something about the more interesting interacting
 regime. The following formula does the job for AdS$_5$/CFT$_4$
 \beq\label{wittenamp}
 \mathcal{Z}_{\mathrm{string}}\left[\,   \phi
 |_{\mathrm{boundary}}  =\phi_0(x)\, \right]
 = \left\langle e^{\int\,d^4x\,\phi_0(x)\mathcal{O}(x)}
 \right\rangle_{\mathrm{CFT}}~.
 \eeq
 Two remarks are in order. The first one is that there is a
 one to one correspondence between operators $\mathcal{O}(x)$ in the boundary gauge
 theory and fields $\phi(x_0,x)$ propagating
 in the bulk of the AdS$_5$ space. Secondly, note that in the left hand side of
 (\ref{wittenamp}) the bulk field
 $\phi(x_0,x)$ is evaluated in the boundary, $\phi(x_0,x)\to
 \phi(x)$. We need to keep in mind these properties while
 constructing the AdS$_3$/CFT$_2$ duality.\\

 Instead of working directly with AdS$_3$ we will deal with its Euclidean
 version, the non-compact manifold
 $H_3^+=SL(2,C)/SU(2)$\index{Euclidean AdS$_3\equiv H_3^+$|textbf}, see
 Section \ref{pl-ads3s3}.
 We parameterize $H_3^+$ with the
 set of coordinates ($\phi, \gamma, \bar{\gamma}$). It is true that this two
 spaces are related by analytic continuation, nevertheless,
 some attention should be paid since the
 WZW model constructed on $SL(2,{\mathbb R})$ and on the coset $SL(2,C)/SU(2)$ are not the same.
 We will point on these differences
 whenever necessary.

 Because $SL(2,C)$ has infinite dimensional
 representations, it is convenient to introduce a complex variable
 $x$ (and its conjugate $\bx$), in such a way to
 encode the components of a given representation in a
 compact form, specifically, in a function
 $\Phi_{h,\bar h}(\phi,\gamma,\bar{\gamma};x,\bx)$. Using these auxiliary
 variables, it was shown that the
 $SL(2,C)/SU(2)$ classical theory has the most general solution given by
 \beq\label{Phi}
 \Phi_{h,\bar{h}}(\phi,\gamma,\bar{\gamma};x,\bx)
 =\frac{1-2h}{\pi}\,\left( e^{-\phi}+(\gamma-x)(\bar{\gamma}-\bx)
 e^{\phi} \right)^{-2h}.
 \eeq
 It can be shown that near the boundary of AdS$_3$, $\ie$ at $\phi\to\infty$,
 this function has the same form as the bulk
 to boundary propagator used in supergravity computations. This
 identification
 gives a strong motivation for interpreting $(x,\bx)$ as coordinates on
 the boundary. The \emph{charge variables}\index{Charge variables|textbf}
 $(x,\bx)$ we introduced are fundamental in the
 approach we use throughout this thesis, so this will be further
 discussed in the next chapters.\\

 In Table 2 we matched the symmetries on both sides of the AdS$_3$/CFT$_2$ correspondence,
 now
 we are ready to see the analogue of (\ref{wittenamp}). In this case, N-point
 functions in the Euclidean worldsheet and in the Euclidean boundary
 conformal theory are believed
 to be related by
 \beq\label{wittenamp3}
 \left\langle \prod_{i=1}^N \int d^2z_i \, \Phi_{h_i} (z_i,\bz_i;x_i,\bx_i)
 \right\rangle_{\Sigma}=
 \left\langle  \prod_{i=1}^N \mathcal{O}_{h_i}(x_i,\bx_i)
 \right\rangle_{\mathrm{CFT}},
 \eeq
 where $N$ is the number of insertions and $(x_i,\bx_i)$ is
 identified as the location of the operator $\Phi_{h}$ in the dual boundary
 theory\footnote{To concrete computations of correlation
 functions in the string side
 of the correspondence we will dedicate an important part of this thesis.}.
 In (\ref{wittenamp3}) there is  a one to one correspondence
 between vertex operators  in the worldsheet theory  and
 vertex operators in the boundary theory, $\int d^2z\,\Phi_{h_i}(z_i,\bz_i;x_i,\bx_i)
 \longleftrightarrow \mathcal{O}_{h_i}(x_i,\bx_i)$. For example, the vertex
 operator for the graviton
 corresponds to the energy-momentum tensor of the CFT.

 Finally we would like to comment
 on the fact that in the real AdS$_3$ with Lorenzian metric, as in other non-compact sigma models, the vertex
 operators belong to non-unitary representations. This non-unitarity of the model and
 the analysis of the singularities in the
 correlation functions show that there is no state/operator
 correspondence, neither IR/UV relation, unless we extend somehow the concept of state.

% OOOOOOOOOOOOOOOOOOOOOOOOOO      BMN       OOOOOOOOOOOOOOOOOOOOOOOOOOOOOOOOOOOOOOOO

% OOOOOOOOOOOOOOOOOOOOOOOOOO      BMN       OOOOOOOOOOOOOOOOOOOOOOOOOOOOOOOOOOOOOOOO

 \section{Plane wave limit and BMN conjecture}
 \label{ppwaves}
 \noindent

 Originally proposed for the AdS$_5$/CFT$_4$ correspondence, the
 \emph{BMN conjecture}\index{BMN conjecture|textbf} has attracted a lot of attention, including
 remarkable developments in the analysis of the
 \linebreak AdS$_3$/CFT$_2$ case. In few words, the idea
 of BMN is to take the pp-wave limit on both sides of the duality
 and see how this affects and possibly reformulate the
 correspondence. The BMN proposal is a limiting case of the AdS/CFT
 correspondence, but this should not make us think that the
 new correspondence is then trivial. As we will see, things
 are a bit more complicated. This will involve establishing a new
 operator map and matching the Hilbert spaces on both
 sides. In the first part of this section we
 focus on the bulk side and
 only in the end we sketch the PP-Wave$_3$/CFT$_2$ correspondence itself.  \\

 Plane-fronted gravitational waves with parallel rays,
 \emph{pp-waves}\index{Parallel Plane waves|textbf}, are
 defined as spacetime solutions of Einstein equations of motion
 having a globally constant null Killing vector field $v^{\mu}$, $\ie$
 it satisfies
 $\nabla_{\mu}v_{\nu}=0$ and $v_{\mu}v^{\mu}=0$.
 We will not deal with the most general form of pp-wave backgrounds but instead
 with a very special type, namely, the
 ones which
 have $D$-dimensional metric given by
 \beq \label{ppwavegen}
 ds^2=-2\,du\,dv-\frac{1}{4}\,du^2\sum_{I=1}^{D-N-2}\,\mu_I^2x^Ix^I+
 \sum_{I=1}^{D-N-2}\,dx^I\,dx^I+
 \sum_{i,j=1}^{N}g_{ij}\,dx^i\,dx^j~,
 \eeq
 where $u$ and $v$ are the light-cone directions in spacetime, $\mu$ is certain
 constant and $g_{IJ}$ is the metric on the transverse directions.
 The main subject of the discussion will not be affected if
 we suppose that the internal
 manifold is the $N$-dimensional torus T$^N$, so from now on in
 most of the equations we consider
 $g_{ij}=\delta_{ij}$, or we simply drop this term.
 Also notice that in the limit $\mu\to 0$ the pp-wave reduces to the
 flat space metric.

 Another form for the metric can be obtained by changing variables 
 \beq\label{ppwave}
 ds^2=-2du\,dv-\frac14\,
 du^2\sum_{I=2}^{D}\mu_I^2\,y_I\,\by_I+\sum_{I=2}^{D}dy_Id\by_I\,.
 \eeq

 The most interesting property of plane wave geometries is that any
 spacetime reduces to it in a certain limit. The process 
 is called
 \emph{Penrose limit}\index{Penrose limit|textbf}. The idea is to choose a null geodesic on the
 given spacetime and zoom
 into a region very close to it, the spacetime
 around the chosen point is a plane-wave. Using this procedure we can generate new supergravity
 backgrounds starting from already known solutions.
 In particular, we can apply the Penrose modus operandi to the ten-dimensional
 AdS$_5$$\times$S$^5$ in order to get plane-wave solutions,
 affecting the AdS/CFT correspondence in a certain way to be clarified. This is
 exactly what BMN conjecture does. Nevertheless, the fate of holography in this limit is
 not yet well understood. We hope that with the work done in this thesis
 we will
 contribute to understand a little bit better this point (see chapter \ref{holography}).\\

 Before entering deeper in the BMN proposal, two words on the
 motivations that brought to it.
 For a long time it was known that some pp-waves, of the kind used here,
 were $\alpha'$ exact solutions of supergravity theories, $\ie$ the
 pp-wave geometries are supergravity solutions that do not get
 $\alpha'$ corrections. This result led to study strings on
 pp-waves,
 supported either by NS-NS or R-R charges, and finally revealing the spectrum in the light-cone gauge.
 On the other hand, it was shown that in addition to the flat space
 and AdS$_5$$\times$S$^5$, the Penrose limit of the latter was
 the only additional maximally supersymmetric solution of type IIB
 supergravity. Hence, BMN got the main ideas to formulate the
 PP-Waves/CFT correspondence. Unfortunately, in the light-cone
 gauge the pp-wave string interactions
 and the spectrum at $p^+=0$ are much harder to determine than
 it was at first expected (for a review see \cite{jabbari}).

 But unlike AdS$_5$$\times$S$^5$, the sigma model for strings
 moving on
 AdS$_3$$\times$S$^3$ with a background NS-NS $B$ field is a well known conformal theory, so it is
 by itself
 a natural ground where to test the correspondence beyond the supergravity
 regime. Furthermore, we do not need
 to go to the light-cone gauge because the model can be solved in a
 fully covariant way (see chapter \ref{str-ads3s3}). Even more important
 is that these two properties are
 preserved in the pp-wave limit. These are the main reasons why in this thesis we will
 concentrate on this model and thus try to
 get new insights for the holographic duality, results that then we would like to extend to the more
 realistic AdS$_5$/CFT$_4$. \\

 The Penrose limit of AdS$_3$$\times$S$^3$  can be carried off
 choosing a lightlike trajectory moving along the axis of the AdS$_3$
 cylinder and turning around a great circle of S$^3$. This can be
 performed changing variables in the following way
 \be \label{ads3s3topp}
 t=\frac{\mu u}{2}+\frac{v}{\mu R^2}, \qquad \psi=\frac{\mu
 u}{2}-\frac{v}{\mu R^2},\qquad \rho=\frac{r}{R}, \qquad
 \theta=\frac{r_2}{R},
 \ee
 and taking the limit $R\to \infty$.  In the plane
 wave background the transverse directions are compact with size
 $\sqrt{\mu p^+}$.

 We expect that the
 Penrose limit of a WZW model should give rise to another conformal model. This
 topic is very important for us and will be discussed widely in the next chapters. What we
 would like
 to stress here is that such theories are all generalizations of the more basic
 \emph{Nappi-Witten (NW) model}\index{Nappi-Witten models|textbf}, associated with the central
 extension of the group $T_2\wedge SO(2)$,
 consisting of translations and
 rotations in the  two-dimensional Euclidean plane.

 Appropriately parameterizing the group
 element and utilizing the Maurer-Cartan formula
 (\ref{mcform}) we
 recover the pp-wave metric in the form (\ref{ppwave}).
 Moreover, it can be shown that the Nappi-Witten Lie algebra,
 $\ie$ the local behavior of the NW group,
 is determined by an ${\mathcal H}_4$ \emph{Heisenberg
 algebra}\index{Heisenberg algebra ${\mathcal H}_4$|textbf} given by \cite{dak}
 \beq\label{h4}
 [P^+,P^-]=-2i\mu K,\qquad [J,P^+]=-i\mu P^+, \qquad [J,P^-]=i\mu
 P^-\,.
 \eeq
 The generators $P^\pm$ stand for the translations in $\mathbb R^2$, $J$ for the
 rotation symmetry and $K$ is the central extension element. If we
 take into account the holomorphic and the anti-holomorphic
 contributions, it can be shown that the central
 generator $K=\bar K$ is common to both algebras.

 This is the simplest of the Heisenberg algebras we can associate to
 pp-waves backgrounds, in fact, for the most general case (\ref{ppwavegen})
 the ${\mathcal H}_{2+2n}$ Heisenberg algebra\index{Heisenberg algebra ${\mathcal H}_{2+2n}$|textbf} reads
 \beq\label{h2+2n}
 [P^+_i,P^-_j]=-2i\mu_i \delta_{ij}K,\qquad [J,P^+_i]=-i\mu_i P^+_i, \qquad
 [J,P^-_i]=i\mu_i P^-_i\,,
 \eeq
 where $i,j=1,\dots\, 2n$. Considering the holomorphic and anti-holomorphic part
 the total number of generators is $2\,(2+2n)-1$, since as we said
 above they share the same central element.\vskip.2cm

 Since in the AdS/CFT duality time translation is associated to
 the dilatation operator, it follows that null geodesics along the axis of the
 AdS$_3$ cylinder are related to operators
 on the boundary with large conformal dimension $\Delta$. On the
 other hand, fast rotations around the three-sphere imposes large
 R-charge $J$. Formally, the BMN proposal states that
 \be\label{bmn1}
 H_{lc}\equiv\pd_u=\frac{1}{\mu}\,p^-\equiv \Delta-J=\mathrm{fixed}~,
 \ee
 \be
 P_{lc}\equiv\pd_v=p^+\equiv\frac{(\Delta+J)}{\mu R^2}=\mathrm{fixed}~.
 \ee
 Operators on the boundary that
 survive the limit $R\to\infty$ are called
 \emph{BMN operators}\index{BMN operators|textbf}.
 Strings propagating in the plane-wave limit of AdS$_3$$\times$S$^3$ are described by a two-dimensional effective field theory
 with coupling $g_2^2=g_6^2\,(\mu p\,\alpha')^2$. The
 coupling of such theory can be written as $J^2/N$, that in the double scaling limit
 $N\to \infty$ and $J\to \infty$ ($J^2/N$ kept fixed)
 differs from the $J^4/N^2$ dependence of the
 standard BMN proposal (Penrose limit of AdS$_5$$\times$S$^5$), see \cite{bmn,ppreviews}.
 %Starting from the two and three-point functions given
 %in \cite{lunmat}, the authors of  \cite{gms} have shown that in fact
 %the orbifold theory present a $J^2/N$ expansion. \\

 As mentioned earlier, the boundary CFT dual to superstrings on
 AdS$_3$$\times$S$^3$$\times$$\mathcal{M}$ is a non-linear $\sigma$-model
 on the symmetric orbifold
 $\mathit{Sym}^{N}(\mathcal{M})=(\mathcal{M})^{N}/{\mathcal
 S}_{N}$, where $\mathcal S_{N}$ is the symmetric group of
 $N=Q_1Q_5$ elements. Following the BMN conjecture,  that
 relates light-cone momenta in the pp-wave background to conformal
 dimensions and R-charges of the operators in the dual theory, the
 spectrum of the SCFT was shown to be given by \cite{tseytlin}
 {\colsep\label{pp3cft2}
 \beqa
 \mathrm{R\!-\!R}: \quad\quad \Delta-J&=&\sum_n N_n\,
 \sqrt{1+\left(\frac{n\,g_s\,Q_5^{{}^R}}{J} \right)^2}
 +g_s\,Q_5^{{}^R}\,\frac{L_0^{{}^\mathcal M}+\bar L_0^{{}^\mathcal
 M}}{J}\, ,\\
 \mathrm{NS\!-\!NS}: \quad\quad \Delta-J&=&\sum_n N_n\,
 \left(1+\frac{n\,Q_5^{{}^{NS}}}{J}
 \right)+Q_5^{{}^{NS}}\,\frac{L_0^{{}^\mathcal M}+\bar
 L_0^{{}^\mathcal M}}{J}\, .
 \eeqa}
 \noindent
 where R-R and NS-NS stand for the nature of the 3-form flux.\\

 At a first scrutiny, it seemed
 that the BMN correspondence failed to correctly  match the spectra
 on the two sides, this even for states corresponding to operators with
 large R-charge\footnote{\cite{hik} is a good introduction to the subject.}.
 Nevertheless, it has been suggested
 that this might be due to the fact that the boundary CFT$_2$ is
 sitting at the orbifold point, which is not the case for the bulk
 description. In principle one can dispose of this mismatch by a
 marginal deformation along the moduli space of the CFT$_2$.
 Alternatively one may in principle be able to
 extrapolate the string spectrum to the symmetric orbifold point and find precise agreement \cite{gms,hiksug,gavnar}.

 \section{The state of the AdS$_3$/CFT$_2$ discrepancy }
 \noindent

  Using standard techniques
  of super--CFTs on symmetric products $\mathcal S^N(\mathcal
  M)=(\mathcal M)^N/\mathcal S_N$,
  the authors of \cite{jevicki}
  have computed three-point
  functions of chiral primary operators on the
  symmetric orbifold (T$^4)^N/\mathcal S_N$.
  On the other hand, in \cite{mihail}
  the dual cubic couplings for scalar primaries of
  type IIB supergravity compactified on T$^4$ were found. In this last case, three-point
  interactions were derived solving the linear equations of motion
  after a non-trivial redefinition of the fields.
  In open contradiction with the AdS/CFT proposal, the results found in the two sides
  disagree. In  \cite{arpanth}, alternative results from
  those of
  \cite{mihail} are given, but leaving the discrepancy still unsolved.
  In \cite{arpanth} it is pointed out that maybe there is a problem with the procedure
  used in \cite{mihail}, specifically, the field redefinition in order to cancel the derivative
  terms.

  In an attempt to clarify this disagreement Lunin and Mathur \cite{lunmat}
  have developed
  a novel formulation for computing correlators on the
  boundary conformal theory. They work directly with the non-abelian permutation group
  ${\mathcal S}_N$, instead of the more studied ${\mathbb Z}_N$.  In this new approach, supersymmetric
  three-point functions reduce to a bosonic contribution times a factor that can be
  calculated using bosonization representation. It should be stressed that this
  formalism makes no reference to the specific form of the compact manifold, it only
  uses the fact that the
  theory has $\mathcal N=(4,4)$ supersymmetry. For the three-point functions they find,
  remarkably,
  a result that behaves as the one of
  \cite{mihail}. It is also underlined that there is no agreement with \cite{arpanth}.

 \subsection{Supergravity in D=6}
 \noindent

  Supersymmtry in six-dimensions has
  pseudo Majorana-Weyl spinor supercharges. They can have either
  positive chirality $Q_+^i$ $(i=0, 1,\dots\, N_+)$ or negative
  chirality $Q_-^i$ $(i=0, 1,\dots\,N_-)$, where $N_{\pm}$ are even numbers.
  The automorphism group is the symplectic $USp(N_+)\times
  USp(N_-)$.

  The supercharges satisfy the anticommutators
  {\colsep
  \bea
  \{Q_+^i,Q_+^{jT}\}&=&\frac12\,(1+\bar{\gamma})\,\gamma^M C_-P_M\Omega^{ij}_+  ~,   \nn\\
  \{Q_-^i,Q_-^{jT}\}&=&\frac12\,(1-\bar{\gamma})\,\gamma^M C_-P_M\Omega^{ij}_-  ~,  \\
  \{Q_+^i,Q_-^{jT}\}&=&\frac12\,(1+\bar{\gamma})\, C_-Z^{ij}  ~,   \nn
  \eea}
  where $P_M$ are the six-dimensional Poincar\'e generators, $C_-$ is the charge
  conjugation matrix and $Z^{ij}$ are the central charges.

  For the theory with $N_+=2$ and $N_-=0$, the chiral supercharges $Q_+^{1,2}$ transform under $USp(2)$.
  We can double the number of them simply adding another doublet of chiral supercharges,
  with the composite spinor now transforming under $USp(4)$. If  supersymmetry
  is related to the number of chiral spinors by
  ${\cal N}=(N_+/2,N_-/2)$, what we just did can be reread as the extension of chiral ${\cal N}=(1,0)$
  to ${\cal N}=(2,0)$.  ${\cal N}=(2,0)$ is sometimes called  ${\cal N}=4b$ to emphasize that there are
  four chiral symplectic Majorana-Weyl supercharges.

  We are interested in this particular theory since ten-dimensional type IIB supergravity
  compactified on a four-dimensional manifold  $K3$, gives rise
  to ${\cal N}=(2,0)$  supergravity in six dimensions coupled to $21$ matter
  tensor multiplets, by definition the latters containing only particles of spin $\leq
  1$.
  Looking at the irreducible representations of the super-Poincar\'e algebra of ${\cal N}=(2,0)$,
  we find that there is a graviton $g_{MN}$, four chiral
  gravitini $\psi_M$ and five self-dual two-form fields $B^i_{MN}$. On the other
  hand, each  matter multiplet contains an anti-self dual tensor field, four
  fermions and five scalars. The field content is shown in the
  next table.
  {\small
  \begin{center}
  \begin{tabular}{|ccc|ccc|}
    \hline
    {}& \footnotesize{\bf SUGRA}  & && \footnotesize{\bf MATTER} & \\
    \hline\hline
    \, \,&&&&&\vspace{-3mm}\\$g_{MN}$ & $\psi_M$ &
    \footnotesize{$B^i_{MN}$}\, \,& \,\,\footnotesize{$B^r_{MN}$} & $\chi^r$ & \footnotesize{$\phi^{r}$} \,\,\\
     1 & 4 & 5 \,\,\,& 1 & 4 & 5\,\,\,\, \\
    \hline
  \end{tabular}
  \end{center}}
  \begin{quote}
  {\bf Table 3.} Field content of pure ${\cal N}=(2,0)$ 6D supergravity and matter multiplets.
  Indices $i=1,\dots\,5$ transform under the group $SO(5)_R$ of the R-symmetry and $r=1,\dots\,n$
  is the $SO(n)$ vector index of the rotating tensor multiplet\footnote{The presentation below is independent of the number
  $n$ of matter multiplets, but for type IIB superstring compactified on $K3$ we must consider
  $n=21$.}.
  \end{quote}
  \vskip.2cm

  In general, chiral superalgebras defined in $D=4k+2$ \,($k=0,1,2$)\, dimensions
  possess antisymmetric tensor fields with either \,self-dual or anti-self-dual field strengths,
  $F_{\mu_1\dots\,\mu_{2k+1}}=\pm\widetilde{F}_{\mu_1\dots\,\mu_{2k+1}}$.
  The main difficulty raised when constructing a consistent interacting formulation of such theories,
  in addition to the absence of an explicit action principle,
  is the mixing of self-dual and anti-self-dual tensor fields by the
  global $SO(5,n)$. This was solved in \cite{romans}
  noting that in our case the field strengths do not need to have definite duality
  properties under the $SO(5,n)$ group but only under the local composite
  $SO(5)\times SO(n)$. This is seen studying the scalar
  sector of the theory.

  It is known that scalar fields in supergravity theories can be described by
  non-linear sigma models on  cosets ${\bf G/H}$,
  where ${\bf G}$ is the non-compact group of the isometry
  transformations and ${\bf H}$ is the maximal compact subgroup of ${\bf
  G}$, called the isotropy group. In the present case,
  the $5n$ scalars of the theory, five for each matter multiplet, can compactly be encoded in a field $\phi^{ir}$,
  with $i=1,\dots\,5$ and  $r=1,\dots\,n$. These scalars parameterize the  $SO(5,n)/\big(SO(5)_R\times
  SO(n)\big)$
  non-linear sigma model.
  We can regroup them in a $(5+n)\times (5+n)$
  vielbein matrix
  \be
  V=\left(%
  \begin{array}{cc}
  u^i_I & v^i_A \\
  w^r_I & x^r_A \\
  \end{array}%
  \right)~,
  \ee
  where we have split the $SO(5,n)$ index in $I=1,\dots\, 5$ and $A=1,\dots\, n$ for
  later convinience. The
  invariant metric of the $SO(5,n)$ manifold
  is $\eta_{IJ}=\textrm{diag}\,({\mathbb I}_{5\times 5},-{\mathbb I}_{n\times
  n})$. As usual the sechsbein $V$ converts  curved indices transforming under $SO(5,n)$
  into flat indices $i,r$ transforming under the composite local
  symmetry $SO(5)_R\times SO(n)$.

  The vacuum has  $u^i_I=\delta^i_I,\, x^r_A=\delta^r_A, v^i_A=w^r_I=0$ and
  the fluctuations of it away from the identity is given by
  {\colsep
  \bea\label{dV}
  V^i_I&=&\delta^i_I+\phi^{ir}\delta_I^r+\frac12
  \phi^{ir}\phi^{jr}\delta_I^j~,\nn\\
  V^r_I&=&\delta^r_I+\phi^{ir}\delta_I^i+\frac12
  \phi^{ir}\phi^{is}\delta_I^s~.
  \eea}
  Here we have included second order corrections in order to consider
  later on cubic couplings of chiral primaries.

  We can also define
  \be
  dV\,V^{-1}=
  \left(%
  \begin{array}{cc}
  Q^{ij} & \sqrt2\,P^{is} \\
  \sqrt2\,P^{jr} & Q^{rs} \\
  \end{array}%
  \right)~,
  \ee
  and it can be seen that $Q^{ij}$ and $Q^{rs}$ are the connections of the $SO(5)_R$ and
  $SO(n)$ respectively.

  Supersymmetry transformations and field
  equations for pure ${\cal N}=(2,0)$ supergravity in six dimensions coupled to matter
  multiplets were derived in \cite{romans}. For completeness, we recall a couple of these results.
  The bosonic field equations are the six-dimensional Einstein equations for the
  metric
  \be
  R_{MN}=H^i_{MPQ}\,{H^i_N}^{PQ}+H^r_{MPQ}\,{H^r_N}^{PQ}+2{P_M}^{ir}\,{P_N}^{ir}~,
  \ee
  and the equations for the scalars
  \be
  D^MP_M^{ir}-\frac{\sqrt2}{3}\,H^{i\,MNP}\,H^r_{MNP}=0~,
  \ee
  this in addition to the self and anti-self duality conditions for the field strengths
  $B_{MN}^i$ and $B_{MN}^r$, respectively.
  In the fermionic sector
  we have field equations mixing $\psi_M$ and $\chi^r$, see for example \cite{sezgin}.

  Imposing $\langle\psi_M\rangle=\langle\chi^r\rangle=0\,$ and asking the supersymmetry
  transformations to vanish in the vacuum, we obtain the Killing spinor equations
  $$
  D_{\mu}\epsilon+\frac12\gamma_{\mu}\Gamma^5\epsilon=0~,\qquad
  D_{a}\epsilon-\frac{i}{2}\gamma_{a}\Gamma^5\epsilon=0~.
   $$
  In order to determine the spectrum of the theory,
  fluctuation are linearized according to
  \be
  g_{MN}=\bar{g}_{MN}+h_{MN}~,\qquad
  P^{ir}_M=\frac{1}{\sqrt2}\,\pd_M\phi^{ir}~,\qquad
  G^I=\,dB^I=\bar G^I +g^I~,
  \ee
  and first order corrections to the vielbein are considered.\\

  States are classified according to the
  symmetries of the theory and they are generically written as $D^{[l_1,l_2]}[\Delta_0,s_0][R,S]$.
  In this notation,
  $[l_1,\,l_2]$ labels the highest weigh representation of the S$^3$ isometry group
  $SO(4)_{\rm gauge}$; ($\Delta_0$,\,$s_0$),
  with $\Delta_0$ the energy and $s_0$ the spin of the lowest energy state,
  labels the representations of the isometry
  group $SO(2,2)$ of AdS$_3$; and, finally, $R$ indicates a representation
  of the R-symmetry group $SO(4)_R$\footnote{\,Global $SO(5)_R$
  symmetry group is broken to $SO(4)_R$ due to the non-vanishing
  expectation value of the Freund-Rubin parameter of
  the field strengths.} while $S$ is a generic representation of
  $SO(n)$.

  Since $SO(4)$ is isomorphic to $SU(2)\times SU(2)$ we
  can rewrite the quantum numbers $l_1$ and $l_2$ in terms of the spin
  $SU(2)$s representations
  \be\label{l12j12}
  [l_1,l_2]~\longrightarrow~ [\,j_1=\frac12\,(l_1+l_2)\,,\,j_2=\frac12\,(l_1-l_2)\,]~.
  \ee
  In the tables 2 and 3 we use the quantum numbers $j_1$ and $j_2$.

  Once we identify the highest spin representation of a multiplet,
  all the other states follow
  repeatedly applying the supercharge generators. In \cite{sezgin} it was found that
  the spin-2 supermultiplet has
  $D^{[l+1,0]}[l+3,2][0,1]$ as its highest spin state and the spin-1 supermultiplet
  is generated from $D^{[l+1,-1]}[l+3,1][0,n]$.
  In order to construct them, the relevant commutators are
  \be
  \left[E,\,Q_{\pm}  \right]=\mp\frac12\,Q_{\pm}~, \qquad
  \left[J,\,Q_{\pm}  \right]=-\frac12\,Q_{\pm}~, \qquad
  \Gamma^5Q_{\pm}=\pm Q_{\pm}~,
  \ee
  where $E$ and $J$ are respectively the energy and spin operators of $SO(2,2)$. The former
  commutation relations tell us that $Q_-$ raises the
  energy and decreases the spin by $\frac12$, while $Q_+$ decreases both the
  energy and the spin.

  The explicit form of the supercharges is obtained from the Killing spinor equations
  \be
  Q_{\pm}^{[\pm1/2,-1/2]}\,\left[\mp1/2,-1/2\right][2_{\pm},0]~,\qquad
  ~~\bar Q_{\pm}^{[1/2,\pm1/2]}\,\left[\pm1/2,1/2\right][2_{\pm},0]~.
  \ee
  Analogous formulas can be found for
  the right part. It can be proved that these supercharges in fact generate the $SU(1,1|2)_L\oplus SU(1,1|2)_R$
  superalgebra.\\
%  $$
%  [l_1,l_2]\otimes
%  [1/2,\pm1/2]=\left[l_1+1/2,l_2\pm1/2\right]\oplus [l_1-1/2,l_2\mp1/2]
%  $$

 In the next diagrams and tables we show explicitly how the supercharges, acting on the
 highest spin representations, generate the full supermultiplets. In table 6 we extract from tables 4 and
 5 the lowest energy states. \\
 \newpage.
 \vspace{2cm}

 \begin{center}
 \setlength{\unitlength}{.65mm}
 \begin{picture}(130,100)(-15,-20)
 \put(50,100){\makebox(0,0){{\small $D^{\left[l+1,0\right]}\left[l+3,2\right]\left[0,1\right]$}}}
 \put(35,95){\vector(-1,-1){15}} \put(65,95){\vector(1,-1){15}}
 \put(15,75){\makebox(0,0){{\small $D^{\left[l+3/2,1/2\right]}\left[l+\frac52,\frac32\right]\left[2_{+},1\right]$}}}
 \put(85,75){\makebox(0,0){{\small $D^{\left[l+1/2,1/2\right]}\left[l+\frac72,\frac32\right] \left[2_{-},1\right]$}}}
 \put(10,70){\vector(-1,-1){15}} \put(30,70){\vector(1,-1){15}}
 \put(70,70){\vector(-1,-1){15}} \put(90,70){\vector(1,-1){15}}
 \put(-10,50){\makebox(0,0){{\small $D^{\left[l+2,1\right]}\left[l+2,1\right]\left[0,1\right]$}}}
 \put(50,50){\makebox(0,0){{\small $D^{\left[l+1,1\right]}\left[l+3,1\right]\left[4,1\right]$}}}
 \put(110,50){\makebox(0,0){{\small $D^{\left[l,1\right]}\left[l+4,1\right]\left[0,1\right]$}}}
 \put(-5,45){\vector(1,-1){15}} \put(45,45){\vector(-1,-1){15}}
 \put(55,45){\vector(1,-1){15}} \put(105,45){\vector(-1,-1){15}}
 \put(15,25){\makebox(0,0){{\small $D^{\left[l+3/2,3/2\right]}\left[l+\frac52,\frac12\right] \left[2_{-},1\right]$}}}
 \put(85,25){\makebox(0,0){{\small $D^{\left[l+1/2,3/2\right]}\left[l+\frac72,\frac12\right] \left[2_{+},1\right]$}}}
 \put(18,20){\vector(1,-1){16}} \put(82,20){\vector(-1,-1){16}}
 \put(50,0){\makebox(0,0){{\small $D^{\left[l+1,2\right]}\left[l+3,0\right] \left[0,1\right]$}}}
 \end{picture}
 \vspace{-1cm}
 \begin{quote}
 {\bf Diagram 1.} Generation of the spin-2 supermultiplet for $l\geq 0$.
 The supercharge $Q_+$ acts down-left in the diagram and $Q_-$ do it in the down-right direction.
 \end{quote}
 \vskip.6cm
 {\small
 \begin{tabular}{ccccc}
 \hline \vspace{-3mm}\\ \vspace{.5mm}
 $\Delta$ &$s_{0}$
 & $SO(4)_{\mathrm{gauge}}$  & $SO(4)_R\times SO(n)$& $\left
 [h,\bar{h} \right ]$
 \\
  \hline\\ \vspace{-2mm}
 $l+3$ &$\pm 2$ & ${\left
 [\frac{l+1}{2},\frac{l+1}{2} \right ]}$  &$\left [0,1 \right ]$ &
  $\frac{\left [l+3\pm2,l+3\mp2 \right ]}{2}$
 \\&&&&\\\vspace{-2mm}
 $l+\frac52$ &$\pm 3/2$& $\left
 [\frac{2l+3\pm1}{4},\frac{2l+3\mp1}{4}
 \right ]$ & $\left
 [2_{\pm},1 \right ]$ &  $\frac{\left
 [2l+5\pm3,2l+5\mp3 \right ]}{4}$
 \\ &&&&\\\vspace{-2mm}
 $l+\frac72$ &$\pm 3/2$& $\left
 [\frac{2l+1\pm1}{4},\frac{2l+1\mp1}{4}
 \right ]$ & $\left
 [2_{\mp},1 \right ]$ &  $\frac{\left
 [2l+7\pm3,2l+7\mp3 \right ]}{4}$
 \\ &&&&\\\vspace{-2mm}
 $l+2$ & $\pm 1$& $\left
 [\frac{2l+2\pm1}{2},\frac{l+2\mp1}{2}
 \right ]$ &$\left [0,1 \right ]$ &
 $\frac{\left [l+2\pm1,l+2\mp1 \right ]}{2}$
 \\ &&&&\\\vspace{-2mm}
 $l+3$ & $\pm 1$& $\left
 [\frac{l+1\pm1}{2},\frac{l+1\mp1}{2}
 \right ]$  & $\left [4,1 \right ]$ &
 $\frac{\left [l+3\pm1,l+3\mp1 \right ]}{2}$
 \\ &&&&\\\vspace{-2mm}
 $l+4$ &$\pm 1$& $\left
 [\frac{l\pm1}{2},\frac{l\mp1}{2}
 \right ]$ & $\left [0,1 \right ]$ &
  $\frac{\left [l+4\pm1,l+4\mp1 \right ]}{2}$
 \\ &&&&\\\vspace{-2mm}
 $l+\frac52$ &$\pm 1/2$& $\left
 [\frac{2l+3\pm3}{4},\frac{2l+3\mp3}{4}
 \right ]$ & $\left
 [2_{\mp},1 \right ]$ &  $\frac{\left
 [2l+5\pm1,2l+5\mp1 \right ]}{4}$
 \\ &&&&\\\vspace{-2mm}
 $l+\frac72$ &$\pm 1/2$& $\left
 [\frac{2l+1\pm3}{4},\frac{2l+1\mp3}{4}
 \right ]$ & $\left
 [2_{\pm},1 \right ]$ &  $\frac{\left
 [2l+7\pm1,2l+7\mp1 \right ]}{4}$
 \\ &&&&\\\vspace{-2mm}
 $l+3$ & $\,\,\,\,0$& $\left
 [\frac{l+1\pm2}{2},\frac{l+1\mp2}{2}
 \right ]$   &$\left [0,1 \right
 ]$ & $\frac{\left [l+3,l+3 \right ]}{2}$
 \\ &&&&\\ \hline
 \end{tabular}}
 \end{center}
 \begin{quote}
 {\bf Table 4.} Spin-2 supermultiplet tower. The multiplet is organized
 from the highest spin state to the lowest one. The $SO(4)_{\rm gauge}$
 quantum numbers are $[j_1,j_2]$ as defined in (\ref{l12j12}). For each non-scalar field,
 there is a state with negative spin $s=-s_0$, it has same energy $\Delta$ and transforms in
  the inverse representation of both $SO(4)_{\rm gauge}$ symmetry and $SO(n)$ global
  group.
 \end{quote}

 \newpage.
 \vspace{2cm}

 \begin{center}
 \setlength{\unitlength}{.65mm}
 \begin{picture}(130,100)(-15,-20)
 \put(50,100){\makebox(0,0){{\small $D^{\left[l+1,-1\right]}\left[l+3,1\right]\left[0,n\right]$}}}
 \put(35,95){\vector(-1,-1){15}} \put(65,95){\vector(1,-1){15}}
 \put(15,75){\makebox(0,0){{\small
 $D^{\left[l+3/2,-1/2\right]}\left[l+\frac52,\frac12\right]\left[2_{+},n\right]\quad$}}}
 \put(85,75){\makebox(0,0){{\small $\quad D^{\left[l+1/2,-1/2\right]}\left[l+\frac72,\frac12\right]
 \left[2_{-},n\right]$}}}
 \put(10,70){\vector(-1,-1){15}} \put(30,70){\vector(1,-1){15}}
 \put(70,70){\vector(-1,-1){15}} \put(90,70){\vector(1,-1){15}}
 \put(-10,50){\makebox(0,0){{\small $D^{\left[l+2,0\right]}\left[l+2,0\right]\left[0,n\right]$}}}
 \put(50,50){\makebox(0,0){{\small $D^{\left[l+1,0\right]}\left[l+3,0\right]\left[4,n\right]$}}}
 \put(110,50){\makebox(0,0){{\small $D^{\left[l,0\right]}\left[l+4,0\right]\left[0,n\right]$}}}
 \put(-5,45){\vector(1,-1){15}} \put(45,45){\vector(-1,-1){15}}
 \put(55,45){\vector(1,-1){15}} \put(105,45){\vector(-1,-1){15}}
 \put(15,25){\makebox(0,0){{\small $D^{\left[l+3/2,1/2\right]}\left[l+\frac52,-\frac12\right]
 \left[2_{-},n\right]\quad$}}}
 \put(85,25){\makebox(0,0){{\small $\quad D^{\left[l+1/2,1/2\right]}\left[l+\frac72,-\frac12\right]
 \left[2_{+},n\right]$}}}
 \put(18,20){\vector(1,-1){16}} \put(82,20){\vector(-1,-1){16}}
 \put(50,0){\makebox(0,0){{\small $D^{\left[l+1,1\right]}\left[l+3,-1\right] \left[0,n\right]$}}}
 \end{picture}
 \vspace{-1cm}
 \begin{quote}
 {\bf Diagram 2.} Generation of the spin-1 supermultiplet. For the
 $SO(n)$ singlet $l\geq 0$, whereas the supermultiplet in the vector
 representation of $SO(n)$ has $l\geq -1$.
 \end{quote}
 \vskip.6cm
 {\small
 \begin{tabular}{ccccc}
 \hline \vspace{-3mm}\\ \vspace{.5mm}
 $\Delta$ &$s_{0}$
 & $SO(4)_{\mathrm{gauge}}$  & $SO(4)_R\times SO(n)$& $\left
 [h,\bar{h} \right ]$
 \\
  \hline\\ \vspace{-2mm}
 $l+3$ &$\pm 1$ & ${\left
 [\frac{l+1\mp1}{2},\frac{l+1\pm1}{2} \right ]}$  &$\left [0,n \right ]$ &
  $\frac{\left [l+3\pm2,l+3\mp2 \right ]}{2}$
 \\&&&&\\\vspace{-2mm}
 $l+\frac52$ &$\pm 1/2$& $\left
 [\frac{2l+3\mp1}{4},\frac{2l+3\pm1}{4}
 \right ]$ & $\left
 [2_{\pm},n \right ]$ &  $\frac{\left
 [2l+5\pm3,2l+5\mp3 \right ]}{4}$
 \\ &&&&\\\vspace{-2mm}
 $l+\frac72$ &$\pm 1/2$& $\left
 [\frac{2l+1\mp1}{4},\frac{2l+1\pm1}{4}
 \right ]$ & $\left
 [2_{\mp},n \right ]$ &  $\frac{\left
 [2l+7\pm3,2l+7\mp3 \right ]}{4}$
 \\ &&&&\\\vspace{-2mm}
 $l+2$ & $0$& $\left
 [\frac{l+2}{2},\frac{l+2}{2}
 \right ]$ &$\left [0,n \right ]$ &
 $\frac{\left [l+2\pm1,l+2\mp1 \right ]}{2}$
 \\ &&&&\\\vspace{-2mm}
 $l+3$ & 0& $\left
 [\frac{l+1}{2},\frac{l+1}{2}
 \right ]$  & $\left [4,n \right ]$ &
 $\frac{\left [l+3\pm1,l+3\mp1 \right ]}{2}$
 \\ &&&&\\\vspace{-2mm}
 $l+4$ &0& $\left
 [\frac{l}{2},\frac{l}{2}
 \right ]$ & $\left [0,n \right ]$ &
  $\frac{\left [l+4\pm1,l+4\mp1 \right ]}{2}$
 \\ &&&&\\ \hline
 \end{tabular}}
 \end{center}
 \begin{quote}
 {\bf Table 5.} Spin-1 supermultiplet tower. In the fourth
 column, the value of $n$ in $[R,n]$ can be $n=1$ or $n=21$ depending if the
 multiplet is a singlet or a vector of the global $SO(n)$.
 \end{quote}

 \newpage

  \begin{center}
 {\small
 \begin{tabular}{crcccrc}
  \hline \vspace{-3mm}\\ \vspace{.5mm}
 $\Delta$ & $s_0$ & $SO(4)_{\rm gauge} $ & $ SO(4)_R$ & $SO(n)$ & \# dof & 6D origin
 \\ \hline \hline
 \multicolumn{7}{c}{{\footnotesize Non-propagating gravity multiplet $({\bf
 3,1})_S+({\bf 1,3})_S$}}
 \\
  \hline \vspace{-3mm}\\ \vspace{.5mm}
 $2$ & $2$  & $\left[0,0\right]$ & $\left[0,0\right]$ & $1$
 & $0$ & $g_{\mu\nu}$ \\ $\vspace{1mm}$
 $\frac32$& $\frac32$  & $\left[0,\frac12\right]$ &
 $\left[0,\frac12\right]$ & $1$ & $0$
 & $\psi_\mu$ \\ $\vspace{1mm}$
 $1$ & $1$  & $\left[0,1\right]$ & $\left[0,0\right]$ & $1$ & $0$
 & $g_{\mu m}, B^5_{\mu m}$ \\
 \hline
 \multicolumn{7}{c}{{\footnotesize Spin-{\tiny $\frac12$} hypermultiplet $({\bf 2,2})_S$}} \\
  \hline \vspace{-3mm}\\ \vspace{.5mm}
 $1$ & $0$  & {\small $\left[\frac12,\frac12\right]$} &
 $\left[0,0\right]$ & $n$ & $4n$
 & $\phi^{5r}, B^r_{mn}$ \\ $\vspace{1.5mm}$
 $\frac32$&$\frac12$   & {\small $\left[\frac12,0\right]$} &
 $\left[0,\frac12\right]$ & $n$ & $4n$
 & $\chi^r$ \\ $\vspace{1.5mm}$
 $2$ & $0$ & $\left[0,0\right]$ & $\left[\frac12,\frac12\right]$ &
 $n$ &
 $4n$  & $\phi^{ir}$  \\
 \hline
 \multicolumn{7}{c}{{\footnotesize Spin-$1$ multiplet $({\bf 3,3})_S$}} \\
  \hline \vspace{-3mm}\\ \vspace{.5mm}
 $2$ & $0$  & $\left[1,1\right]$ & $\left[0,0\right]$ & $1$
 & $9$
 & $B^5_{mn}, g_m{}^m, g_\mu{}^\mu$ \\ $\vspace{1.5mm}$
 $\frac52$ &$\frac12$ & $\left[1,\frac12\right]$ &
 $\left[0,\frac12\right]$ & $1$ & $12$
 & $\psi_m$  \\ $\vspace{1.5mm}$
 $3$ &
 $0$&$\left[\frac12,\frac12\right]$&$\left[\frac12,\frac12\right]$
 & $1$ & $16$
 &$ B^{r}_{mn}$ \\ $\vspace{1.5mm}$
 $3$ & $1$ & $\left[1,0\right]$ & $\left[0,0\right]$ & $1$ & $3$
 & $g_{\mu m}, B^5_{\mu m}$ \\ $\vspace{1.5mm}$
 $\frac72$ & $\frac12$  & $\left[\frac12,0\right]$ &
 $\left[\frac12,0\right]$ & $1$ & $4$
 & $\psi_m$   \\ $\vspace{1.5mm}$
 $4$ &  $0$  & $\left[0,0\right]$ & $\left[0,0\right]$ & $1$ & $1$
 & $g_m{}^m, g_\mu{}^\mu$ \\
 \hline \\
 \end{tabular}}
 \end{center}
  \begin{quote}
  {\bf Table 6.~}Lowest mass spectrum of six-dimensional $N=(2,0)$
  supergravity on AdS$_3$$\times$S$^3$ \cite{nicolai}.
  From the formula $\Delta(\Delta-2)=M^2$ for scalar fields we note
  the presence of $9+4n$ massless scalars.
  \end{quote}

  \subsection{The boundary CFT$_2$}
  \noindent

  Type IIB superstring on AdS$_3$ has been proposed to be dual to a CFT theory living on
  the
  two-dimensional boundary with
  $\mathcal N =(4,4)$ supersymmetry. For AdS$_3$$\times$S$^3$$\times$$\mathcal M$ the
  boundary theory
  is a
  sigma model whose target space is the
  symmetric orbifold $(\mathcal M)^N/\mathcal S_N$. We denote by ${\mathcal S}_N$ the
  permutation group
  of $N$ variables and
  $N=Q_1Q_5$, with $Q_1$ and $Q_5$ respectively the number of fundamental strings
  and NS5-branes
  generating the background.
  The action of the permutation group ${\mathcal S}_N$ should be understood as
  the identification of
  the set of points ($X_1,\dots,X_N$) of the $N$ copies of $\mathcal M$ with the
  other sets of points obtained by
  permuting in all different ways the $X_i'\,$s. Similarly for the
  fermionic coordinates $\psi_i$.\\

 The Hilbert space of the  orbifold theory $(\mathcal M)^N/\mathcal S_N$ is obtained as follows \cite{dmv2}.
 Starting from the
 Hilbert space $\textsf{H}$ of the CFT  on ${\cal M}^N$,
 the action of the discrete group $G\equiv {\cal S}_N$ leaves us with the reduced  twisted sector
 $\textsf{H}_g$. In addition to this, among all these states we need to keep only
 those states
 invariant under the centralizer group $C_g$. After these identifications the final Hilbert
 space of each conjugacy class of $G$ is denoted $\textsf{H}_g^{C_g}$. Thus, the
 full Hilbert space of the orbifold theory is
 \be
 \textsf{H}\,\left(\frac{{\cal
 M}^N}{G}\right)=\bigoplus_{[g]}\textsf{H\,}_g^{C_g}~,
 \ee
 where the sum is over all the different conjugacy classes
 $[g]$ of $G$.

 For the theory here under consideration the discrete symmetry is the
 permutation group ${\cal S}_N$, whose conjugacy classes are
 characterized by decompositions of the form
 \be
 [g]=(1)^{N_1}\dots\,(s)^{N_s}~,
 \ee
 where $(n)$ is the number of elements that are permuted and $N_n$ is the number
 of times this operation is carried off. For a system with $N$ elements it is clear that $\sum_n\,nN_n=N$. This
 decomposition of the conjugacy classes is useful since we can identify the
 centralizer group as
 \be
 C_g=S_{N_1}\times (S_{N_2}\times \mathbb Z_2^{N_2}) \times \cdots \times (S_{N_s}\times \mathbb
 Z_s^{N_s})~.
 \ee

 In this notation, a generic $S_{N_n}$ permutes the sets of $(n)$ elements,
 while $Z_n$ acts on the internal elements of $(n)$.
 Now we can decompose the twist sector in smaller
 Hilbert spaces $\textsf{H\,}^{\mathbb Z_n}_{(n)}$ invariant under the action of $\mathbb Z_n$,
 such that
 \be\label{symhilbert}
 \textsf{H}\,({\cal S}^N{\cal M})=\bigoplus_{\sum nN_n}~
 \bigotimes_{n>0}\,{\cal S}^{N_n}\,\textsf{H\,}_{(n)}^{\mathbb Z_n}~.
 \ee

 Hence, the spectrum of CFTs on symmetric products ${\cal M}^N/{\cal S}_N$ is
 build by the action of $\mathbb Z_n$--twist fields ($n=1,\dots\,N$) on the vacuum of the
 theory \cite{dfms}. \\

  From the point of view of the CFT$_2$ the action of the permutation group
  is realized by twist fields $\sigma_n$, for permutations of length $n$, acting on the $N$ copies of $c=6$
  theories.
  If we suppose the theory defined on a cylinder, we can see the twist operator $\sigma_n$ as linking
  $n$ conformal field theories defined on each $\mathcal M$ in
  such a way that the final CFT lives in a circle that is $n$
  times larger than the initial one. The vacuum energy difference
  tells us that the twist field $\sigma_n$ has conformal
  dimension $\Delta_n={\frac14}\,(n-\frac1n)$.
  Note that even if the sample twist operator $\sigma_n$ is the building block of the
  symmetric orbifold theory, and we will deal mostly with it,
  it should be kept in mind that in fact there is one twist operator for each conjugacy class, and
  not for a simple element of the group,
  see (\ref{symhilbert}). CFT operators have the form \cite{lunmat}
 \be
 {\cal O}_n=\frac{c_n}{N!}\sum_{h\in S_N}\sigma_{hnh^-1}~,
 \ee
 where $c_n$ is a normalization constant. Moreover, we should consider twist operators
 with all possible length $\sigma_{n}$ with $n=1, \dots \,N $, and symmetrize in all different ways.
  In the correlation functions
  a simple global combinatorial factor will take into account this
  fact.\\

  The match of the symmetries on both sides of the AdS$_3$$\times$S$^3$/CFT$_2$ correspondence
  tells us that the isometry group $SO(4)$ of the three-sphere should be
  identified with a pair of $SU(2)$ algebras, left and right,
  corresponding to the R-symmetry of the $\mathcal N =(4,4)$ boundary
  theory. Thus, a string moving around S$^3$ has an angular momentum
  that is naturally associated to the R-charge of certain operator on the boundary. Hence it follows
  that high angular momenta
  are dual to operators with large R-charge. In the boundary theory we will only consider chiral primary operators.

  Chiral primary operators (CPO) are operators on
  the super--CFT that are annihilated by half of the supersymmetries
  and
  have same conformal dimension and R-charge\footnote{Same for the right
  contribution, $\bar h=\bar{\jmath}$.}, $h=j$. Since the twist fields we
  defined above
  only permute copies of $\mathcal M$ and do not carry any charge, we see
  that in general the twist $\sigma_n$ is not a CPO, except for the trivial
  $n= 1$. In order to provide the
  twist operator with R-charge, we define the following
  currents  of the $SU(2)_L$ ${\cal N}=(4,4)$ automorphism
  \be
  J_{-m/n}^{+}\equiv \oint \frac{dz}{2\pi i}\sum_{k=1}^n J_z^{k,+}(z)\,e^{-2\pi
  im(k-1)/n}z^{-m/n}~,
  \ee
  where $k$ labels the copy of $\cal M$ and the integral goes around the origin of the $z$-plane.
  Acting with these currents on a twist operator the conformal
  dimension get increased by $m/n$ while the charge is raised by
  one unit. Choosing $m/n<1$ we can repeatedly apply the currents until we reach the point
  where the equality $h=j$ holds, corresponding to the chiral primaries.\\

  It is useful to define the local map $z\to
  at^n$, which states that the $n$ copies of $\mathcal M$ involved in the twist
  operation are translated into a single copy on the covering space $\Sigma$.
  In this space, that we call
  $t$-space in order to distinguish it from the original $z$-plane, $n$
  different currents give rise to a single current on the covering
  space. After mapping we define the currents in the $t$-space as
  \be
  J_{-m}^+\equiv \int \frac{dt}{2\pi i}\,J_t^+(t)\,t^{-m}~.
  \ee
  Denoting the  R-charged twist fields as $\sigma_n^\pm$ and using the
  $t$-space notation above introduced, it is easy to see that we can construct the
  following chiral operators
  \be
  |\sigma_n^-\rangle\equiv J_{-(n-2)}^+\cdots
  J_{-3}^+J_{-1}^+|\,0^-\rangle_{NS}~,\qquad\mathrm{with}\qquad h=j=\frac{n-1}{2}~,
  \ee
  and
  \be
  |\sigma_n^+\rangle\equiv J_{-n}^+J_{-(n-2)}^+\cdots
  J_{-3}^+J_{-1}^+|\,0^-\rangle_{NS}~,\qquad\mathrm{with}
  \qquad h=j=\frac{n+1}{2}~.
  \ee
  These relations are valid only for $n$ odd. In fact, in the covering space $\Sigma$
  the spinor coordinates are identified according to $\psi_t(t)\to
  (-1)^{n-1}\psi_t(t)$, so for $n$ even the spinor is antiperiodic around $t=0$.
  The corresponding Ramond vacua are of two types, depending on their $J^3$ charge.
  We can have $|\,0^{+}\rangle_R$ or $|\,0^{-}\rangle_R$,
  both of them  with $h=\frac14$, but $j=\pm\frac12$ respectively. The Ramond vacua are related between
  them by $|\,0^{+}\rangle_R
  =J_0^+|\,0^{-}\rangle_R$,
  and to the Neveu-Schwarz vacuum by spectral flow transformation.
  From the worldsheet point of view, the field we should insert in $t=0$ in order to
  pass from
  one type of vacuum to the other is the spin field, $|\,0^{\pm}\rangle_R
  =S^{\pm}|\,0\rangle_{NS}$. The chiral primaries are in this case defined
  as
  \be
  |\sigma_n^-\rangle\equiv J_{-(n-2)}^+\cdots
  J_{-2}^+J_{0}^+|\,0^-\rangle_{R}~,\quad
  |\sigma_n^+\rangle\equiv J_{-n}^+J_{-(n-2)}^+\cdots
  J_{-2}^+J_{0}^+|\,0^-\rangle_{R}~,
  \ee
  with the same conformal dimension, and charge, as  $n$ odd.
  We can now put together left and right
  movers in order to write the complete chiral
  primaries of the orbifold theory
  \be
  \sigma_n^{\pm\pm}~,\qquad\mathrm {with}\qquad h=\bar h~,
  \ee
  and
  \be
  \sigma_n^{\pm\mp}~,\qquad\mathrm {with}\qquad h=\bar h \pm1~.
  \ee

 \renewcommand{\theequation}{\arabic{chapter}.\arabic{equation}}
 \chapter{Bosonic String Amplitudes on AdS$_3$$\times$S$^3$ and the PP-Wave Limit}
 \label{str-ads3s3}\setcounter{chapter}{3}
 \noindent

 In this chapter we examine
 the quantum properties of bosonic strings living on AdS$_3$$\times$S$^3$ with
 NS-NS three-form flux background. First we review the main properties of
 the affine algebras\footnote{In this thesis we assume that the reader is
 equipped with
 basic knowledge of conformal field theories
 \cite{ginsparg}.} associated to this background and then we
 generate the spectrum, identifying
 at a quantum level the short and long strings introduced in the previous chapter.
 In the last part we turn to the charge variables formulation of the theory and
 comment further on the holographic screen of
 section \ref{ads3cft2}, an approach that will reveal all its power only in the last chapter.
 After this we pass to the main goal, namely, to prove that we can
 explicitly compute correlation functions in the plane wave limit
 starting from the correlators of primary fields of
 $\widehat{SL}(2,\mathbb{R})$ and $\widehat{SU}(2)$. This chapter
 is in part based on results published in \cite{bdkz}.

 % OOOOOOOOOOOOOOOOOOOOOOOOOOOOOOOOOOOOOOOOO      SPECTRUM     OOOOOOOOOOOOOOOOOOOOOOOOOOOOOOOOOOO

 % OOOOOOOOOOOOOOOOOOOOOOOOOOOOOOOOOOOOOOOOO      SPECTRUM     OOOOOOOOOOOOOOOOOOOOOOOOOOOOOOOOOOO

 \section{Affine algebras and spectra}
 \label{affalg}
 \noindent

 Affine conformal models are two dimensional field theories that, in  addition to
 the conformal invariance we are use to,
 shows a chiral symmetry in some primary fields with unit weight. These
 fields are the so called \emph{affine currents}\index{Affine current|textbf}.
 In other words, the  standard Virasoro
 generators
 are supplemented by a set of chirally conserved currents, $\ie$ they fulfil $\bar\pd J=0$.
 In section \ref{stads3} we showed that this is the chief property
 of WZW models describing strings on group manifolds, see (\ref{chcurr}) and (\ref{currentsym}). In particular
 this is true for strings living on
 AdS$_3$ and S$^3$, so it is fundamental for us to understand in detail how these models work.\\

 Affine Kac-Moody algebras
 by definition meet conformal and chiral invariance, so
 the currents are constraint to have operator product expansion (OPE) given by
 \be\label{kac}
 \mathcal J^i(z)\,\mathcal J^j(w) \sim
 \displaystyle\frac{k\, g^{ij}}{(z-w)^2}+ i \, f^{ijk}\frac{
 \mathcal J^k(w)}{z-w},
 \ee
 where $k$ is a positive integer known as the \emph{level of the algebra}\index{Affine level
 $k$|textbf}. Notice that the general metric $g$ has upper
 indices and it is symmetric on their interchange.

 Expanding in Laurent modes $\mathcal J^i(z)=\sum_{n\in \,\mathbb Z} \mathcal J_n^i \, z^{-n-1}$
 and integrating in the complex plane it is straightforward to get the
 commutation relation of the currents
 \be\label{kaccomm}
 [\mathcal J^i_n,\,\mathcal J^j_m]=\,k\,n\,g^{ij}\,\delta_{n+m,0}+i\,f^{ijk}\mathcal J^k_{n+m}~.
 \ee
 This is the \emph{Affine Kac-Moody algebra}\index{Affine Kac-Moody algebra|textbf}.
 This algebra is an infinite-dimensional extension of ordinary Lie algebras
 (infinite generators since
 $-\infty<n<\infty$\index{Affine Kac-Moody algebras|textbf}). The subalgebra of zero
 modes $\mathcal J_0^i$ is known as the \emph{horizontal
 Lie algebra}\index{Horizontal Lie algebra|textbf} in which the central extension $k$ is absent.
 Moreover, from this general form we can see that the $f^{ijk}$ coefficients are no more than the
 structure constant of the finite-dimensional Lie algebra (the zero mode algebra of the
 currents). As we are dealing with two algebras, the algebra on the target space and the current
 algebra on the worldsheet, we will distinguish them  by putting a hat on the latter. We will
 also restrict our presentation only to $\widehat{SL}(2,{\mathbb R})$ and $\widehat{SU}(2)$
 since these are the current algebras associated with
 strings moving on AdS$_3$$\times$S$^3$ with NS-NS two form field
 $B$.\\

 Due to the fact that $\widehat{SL}(2,{\mathbb R})$\index{Current algebra
 $\widehat{SL}(2,{\mathbb R})$|textbf} and $\widehat{SU}(2)$\index{Current algebra
 $\widehat{SU}(2)$|textbf}
 are analytical continuation of each other, it is not surprising that
 the algebras and other relations
 in the two cases have similar expressions. Thus in order to avoid repetitions we will use
 the letter $\mathcal J$ for the generators of both
 algebras and agree that the upper signs in the following equations
 correspond to $\widehat{SL}(2,{\mathbb R})$ and the lower ones to
 $\widehat{SU}(2)$. For the moment we also suppose that they have
 same level $k$.

 The current algebras have the same form as given in
 (\ref{kac}) with $i,j,k=1,2,3$, in accordance with the number of generators.
 Moreover $g_{ij}=\eta_{ij}=\textrm{diag}\,(++-)$ for $\widehat{SL}(2,{\mathbb R})$ and
 $g_{ij}=\delta_{ij}=\textrm{diag}\,(+++)$ for $\widehat{SU}(2)$.
 If we now define the conventional operators $\mathcal J^{\pm}=\mathcal J^1\pm i \mathcal J^2$, we
 get the OPEs
 {\colsep
 \bea \label{affs2OPE}
 \mathcal J^+(z)\,\mathcal J^-(w) &\sim&
 \frac{k}{(z-w)^2} \mp\frac{2 \mathcal J^3(w)}{z-w} \ , \nonumber\\
 \mathcal J^3(z)\,\mathcal J^\pm(w) &\sim& \pm \frac{\mathcal J^{\pm}(w)}{z-w} \ , \nonumber \\
 \mathcal J^3(z)\,\mathcal J^3(w) &\sim& \displaystyle \mp \frac{k}{2(z-w)^2} \ .
 \eea}
 This set of OPEs are essential for our approach and will be
 used intensely throughout this thesis. From here we can read the
 explicit  $\widehat{SL}(2,{\mathbb R})$ and  $\widehat{SU}(2)$ OPEs or commutators. If needed, the formulas are given in (\ref{sl2}) and
 (\ref{su2}).

 Expanding in modes, the analogue of (\ref{kaccomm}) for $i,j,k=\pm,3$ is
 {\colsep
 \beqa \label{affs2conm}
 \left[\mathcal J_n^+,\,\mathcal J_m^- \right]&=&kn\,\delta_{n+m,0}\mp2\mathcal J_{n+m}^3 ~,\nn \\
 \left[\mathcal J_n^3,\,\mathcal J_m^{\pm} \right]&=& \pm \mathcal J_{n+m}^{\pm}~, \nn \\
 \left[\mathcal J_n^3,\,\mathcal J_m^3 \right]&=&\mp\frac{k}{2}\,n\delta_{n+m,0}~.
 \end{eqnarray}}
 Non-singularity of the fields $\mathcal J^i(z)$ at $z=0$ imposes
 the existence of states $|\,R_i\rangle$
 that are annihilated by the positive modes of the
 currents, $\ie$ ${\mathcal J}_n^i|\,R_i\rangle=0$ for $n\geq
 0\,$.
 This  defines the \emph{primary states}\index{Primary state|textbf} or
 \emph{highest weight states}\index{Highest weight states|textbf}.
 Here we have assumed that all these states $|\,R_i\rangle$
 transform in some representation $R$ of the horizontal algebra, that is $\mathcal J_0^i\,|\,R_i\rangle=
 \,(T^i_R)_{ij}\,|\,R_j\rangle$, where $T^i$ are the generators.
 The OPE counterpart of this relation is
 \be\label{affprim}
 \mathcal J^i(z)R_i(w)\sim
 \frac{{T_R^i}_{ij}}{z-w}\,R_j(w)\,,
 \ee
 where the fields above defined act on the vacuum of the theory as $R_i(0)|\,0\rangle\equiv |\,R_i\rangle$.
 The fields $R_i(z)$ are in consequence called \emph{affine primaries}\index{Affine primary
 fields|textbf}.

 To construct the spectrum of the theory we
 start from the vacuum states
 and apply repeatedly the raising operators with negative modes $\mathcal
 J_{n<0}^{\pm,3}\,\,$. At this point a comment is in order: the
 spectrum of strings on AdS$_3$, or, equivalently, the
 representations of
 $SL(2,\mathbb R)$, contain  ghosts (the generators $\mathcal J^3$
 create states with negative norm);  making the theory ill-defined.
 This longstanding problem was solved in  \cite{mo1}, where
 the authors proved a no-ghost theorem for the model making use of extra unexpected states. These states arise as
 representations of the spectral flowed algebra. We will come back
 to this further down.

 The Sugawara construction establishes that the energy-momentum
 tensor can be written as a bilinear in the currents
 \beq
 T(z)=\frac{1}{k\mp2}\left[ \frac{1}{2}\,(\mathcal J^+\mathcal J^-+\mathcal J^-\mathcal J^+)
 \mp \mathcal J^3\mathcal J^3    \right]~.
 \eeq
 Expanding in modes we get the \emph{Virasoro generators}\index{Virasoro generators|textbf}
 \beq\label{Lnvira}
 L_{n\neq o}=\frac{1}{k\mp2}\sum_{m=1}^{\infty}(\mathcal J_{n-m}^+\mathcal J_m^-
 +\mathcal J_{n-m}^-\mathcal J_m^+ \mp 2 \mathcal J_{n-m}^3\mathcal J_m^3)~.
 \eeq
 As usual, $L_0$ is ambiguous under normal ordering, so we define
 \beq\label{Lovira}
 L_0=\frac{1}{k\mp2}\,\left[\,\frac12\,(\mathcal J_0^+\mathcal J_0^-+\mathcal J_0^-\mathcal J_0^+)
 \mp(\mathcal J_0^3)^2+
 \sum_{m=1}^{\infty}(\mathcal J_{-m}^+\mathcal J_m^-
 +\mathcal J_{-m}^-\mathcal J_m^+ \mp 2 \mathcal J_{-m}^3\mathcal J_m^3)\,\right]~.
 \eeq
 For convenience we remember the \emph{Virasoro
 algebra}\index{Virasoro algebra|textbf}
 \beq \label{vir}
 [L_n,L_m]=(n-m)L_{n-m}+\frac{c}{12}\,(n^3-n)\,\delta_{n+m,0}~.
 \eeq
 Note that the affine current algebra is larger than the conformal Virasoro algebra since it
 comprises conformal and chiral invariance. That is
 why the Virasoro generators can be expressed in terms of the
 modes of the currents. The central extension parameter $k$
 of Kac-Moody algebras and the central extension of the Virasoro algebra are related by
 \be\label{centch}
 c=\frac{3k}{k\mp2}~.
 \ee
 The generators of both algebras have commutator $[L_n,\mathcal J_m^i]=-m\,\mathcal J_{m+n}^i$.

 The Casimir operator of the group is defined by
 \be
 \vec{\mathcal J}^{\,2}=(k\mp2)\,L_0=\frac12\,(\mathcal J_0^+\mathcal J_0^-+\mathcal J_0^-\mathcal J_0^+)
 \mp(\mathcal J_0^3)^2~.
 \ee

 For $\widehat{SL}(2,\mathbb{R})$ we have three types of unitary normalizable
 representations \cite{xpeskin}:\\

 1)~Lowest-weight discrete representations ${\cal D}^+_l$,
 constructed starting from a state $|l \rangle$ which satisfies
 $K^- |l \rangle = 0$, with $l
 >1/2$. The spectrum of $K^3$ is given by $\{l+n\}$, $n \in
 \mathbb{N}$ and the Casimir is ${\cal C}_{SL} =
 -l(l-1)$.\vskip.1cm

 2)~Highest-weight discrete representations ${\cal D}^-_l$,
 constructed starting from a state $|l \rangle$ which satisfies
 $K^+ |l \rangle = 0$, with $l
 >1/2$. The spectrum of $K^3$ is given by $\{-l-n \}$, $n \in
 \mathbb{N}$ and the Casimir is ${\cal C}_{SL} =
 -l(l-1)$.\vskip.1cm

 3)~Continuous representations ${\cal C}_{l,\a}$, constructed
 starting from a state $|l,\a \rangle$ which satisfies $K^\pm |l,\a
 \rangle \ne 0$, with $l = 1/2 + i \s $, $\s \ge 0$. The spectrum
 of $K^3$ is given by $\{\a+n \}$, $n \in \mathbb{Z}$ and $0 \le \a
 <1$. The Casimir is ${\cal C}_{SL} = 1/4 + \s^2$.\\

 Other representations come from the spectral flow 
 \cite{mo1,oogurisb}
 {\colsep
 \bea \label{sflowsl2}
 \mathcal J^3_n \, &\longrightarrow& \, \widetilde{\mathcal J}_n^3\,=\mathcal J_n^3-\frac{k}{2}\,w\,\delta_{n,0}~,\nn \\
 \mathcal J^+_n \, &\longrightarrow& \, \widetilde{\mathcal J}_n^+=\mathcal J_{n+w}^+ ~,\\
 \mathcal J^-_n \, &\longrightarrow& \, \widetilde{\mathcal J}_n^-=\mathcal J_{n-w}^-~.\nn
 \eea}
 For the \emph{Virasoro generators}\index{Virasoro generators|textbf} the automorphism acts as
 \beq
 L_n\, \longrightarrow \,\widetilde
 L_n=L_n+w\,{\mathcal J}^3_n-\frac{k}{4}\,w^2\delta_{n,0}~.
 \eeq
 The full spectrum includes the representations of these
 spectral flowed algebras, $\ie$ representations builded
 from a general highest weight state $|\,h;\omega\rangle$.

 The representations of the spectral flow algebras $\mathcal D_l^{\omega}$ and
 $\mathcal C_{l,\alpha}^{\omega}$ are generated
 applying on the spectral flowed highest weight states the currents of $\widehat{SL}(2,\mathbb R)$. For
 discrete representations these they are defined as
 \be\label{primdisw}
 K_{n+\omega}^+|\,h;\omega\rangle=0, \qquad
 K_{n-\omega-1}^-|\,h;\omega\rangle=0,\qquad
 K_n^3|\,h;\omega\rangle=0, \qquad n=1,2,\dots
 \ee
 and for continuous representations
 \be\label{primconw}
 K_{n\pm\omega}^{\pm}|\,h,\a;\omega\rangle=0, \quad\qquad
 K_n^3|\,h,\a;\omega\rangle=0, \qquad n=1,2,\dots
 \ee

 Since irreps with different $w$ are not equivalent, the full Hilbert space
 including spectral flowed representations is
 \be\label{hilbsl}
 {\mathsf H}_{SL(2,\mathbb R)}=\oplus_{w=-\infty}^{\infty}\left[\, \left(\,
 \int_{\frac12}^{\frac{k-1}{2}}dl\,\,
 \mathcal D_l^{\omega} \otimes   \bar{\mathcal D}_l^{\omega}  \,\right)
 \oplus  \left(\, \int_{\frac12+i{\mathbb R}}dl \int_0^1
 d\alpha\,\,
 \mathcal C_{l,\alpha}^{\omega} \otimes \bar{\mathcal C}_{l,\alpha}^{\omega}   \,\right)
 \,\right]\,\,.
 \ee
 The analysis of the spectrum leads to identify the continuous
 representations, whith discrete energies,
 with \emph{short strings}\index{Short strings AdS$_3$|textbf} moving deeply inside AdS$_3$ (see section \ref{stads3}).
 In \cite{mo1} it was shown that the spectral flow of continuous
 representations lead to physical states with energy given by
 \be
 E=\frac{kw}{2}+\frac{1}{w}\,\left[\, 2\,\frac{s^2+\frac14}{k-2}+\bar N+N+\Delta_{int}+\bar{\Delta}_{int}-2  \,\right]~,
 \ee
 where $N$ is the number of current excitations, left and right contributions
 are considered, before the spectral
 flow is taken. These new states with continuous energies represent
 the long strings\index{Long strings AdS$_3$|textbf}
 also discussed in  section \ref{stads3}.

 For $SU(2)$ the things are much simpler since we have only one
 type of unitary
 representations ${V}_{\tilde{l}}$ with $2\tilde{l} \in \mathbb{N}$,
 $\tilde{m} = -\tilde{l}, -\tilde{l}+1, ..., \tilde{l}$ and
 the spectral flow operation does not give extra states but only maps between conventional representations. Thus, the
 full Hilbert space in this case is
 \be\label{hilbsu}
 {\mathsf H}_{SU(2)}=\oplus\,{}_{\tilde l=0,\frac12\dots\,\frac{k}{2}}
 V_{\tilde l}\, \otimes\, {\bar V_{\tilde l}}~.
 \ee

 % OOOOOOOOOOOOOOOOOOOOOOOOOOOO    P-LIMIT OF AMPLITUDES OOOOOOOOOOOOOOOOOOOOOOOOOOOO

 % OOOOOOOOOOOOOOOOOOOOOOOOOOOO    P-LIMIT OF AMPLITUDES OOOOOOOOOOOOOOOOOOOOOOOOOOOO

 \section{Penrose limit of amplitudes on AdS$_3$$\times$S$^3$}
 \label{pl-ads3s3}
 \noindent

 Starting from three-point correlators for vertex operators of strings on
 AdS$_3$$\times$S$^3$, in this section we proceed to determine the corresponding correlators in the
 plane wave background. The novelty of our approach is that the Penrose limit
 is taken scaling the charge variables already introduced
 for $SL(2,\mathbb R)$ \cite{tesch} and similar for $SU(2)$.  The limit of the $SU(2)$ three-point
 couplings \cite{zf} has been considered in \cite{dak} and we refer
 to that paper for a detailed discussion. Here we provide a similar
 analysis for the $SL(2,\mathbb{R})$ structure constants
 and show that when combined with the $SU(2)$ part
 they reproduce in the limit the $\widehat{\cal H}_6$ structure
 constants we will see in chapter \ref{str-pp}. As a first
 step, we begin by introducing the
 irreps of the Heisenberg algebra $\mathcal
 H_6$ and then show how the quantum numbers of the two models before and
 after the Penrose limit are related to each other. But, first of all let us
 clarify the role of the Euclidean AdS$_3$ (see also section \ref{ads3cft2}).\\

 In general, the AdS$_3$/CFT$_2$ correspondence entails the exact
 equivalence between superstring theory on AdS$_3$$\times$${\cal
 M}$, where ${\cal M}$ is some compact space represented by a
 unitary CFT on the worldsheet, and a CFT defined on the boundary
 of AdS$_3$. Equivalence at the quantum level implies a isomorphism
 of the Hilbert spaces and of the operator algebras of the two
 theories. For various reasons it is often convenient to consider
 the Euclidean version of AdS$_3$ described by an
 $SL(2,\mathbb{C})/SU(2)$ WZW model on the hyperbolic space $H^+_3$
 with $S^2$ boundary, see section \ref{ads3cft2}. Although the Lorentzian $SL(2,\mathbb{R})$
 WZW model and the Euclidean $SL(2,\mathbb{C})/SU(2)$ WZW model are
 formally related by analytic continuation of the string
 coordinates, their spectra are not the same. As observed in
 \cite{mo1, mo3}, except for unflowed ($w=0$) continuous
 representations, physical string states on Lorentzian AdS$_3$
 corresponds to non-normalizable states in the Euclidean
 $SL(2,\mathbb{C})/SU(2)$ model. Yet unitarity of the dual boundary
 CFT$_2$ that follows from positivity of the Hamiltonian and slow
 growth of the density of states should make the analytic
 continuation legitimate. Indeed correlation functions for the
 Lorentzian $SL(2,\mathbb{R})$ WZW model have been obtained by
 analytic continuation of those for the Euclidean
 $SL(2,\mathbb{C})/SU(2)$ WZW model \cite{mo3}. Singularities
 displayed by correlators involving non-normalizable states have
 been given a physical interpretation both at the level of the
 worldsheet, as due to worldsheet instantons, and of the target
 space. Some singularities have been associated to operator mixing
 and other to the non-compactness of the target space of the
 boundary CFT$_2$. The failure of the factorization of some
 four-point string amplitudes has been given an explanation in
 \cite{mo3} and argued not to prevent the validity of the analytic
 continuation from Euclidean to Lorentzian signature. Since we are
 going to take a Penrose limit of $SL(2,\mathbb{R})$ correlation
 functions computed by analytic continuation from
 $SL(2,\mathbb{C})/SU(2)$, we need to assume the validity of this
 procedure. Reversing the argument, the agreement we found between
 correlation functions in the Hpp-wave computed by current algebra
 techniques with those resulting from the Penrose limit (current
 contraction) of the $SL(2,\mathbb{C})/SU(2)$ WZW model should be
 taken as further evidence for the validity of the analytic
 continuation.\\

 To discuss how the three-point couplings in the
 Hpp-wave with ${\mathcal H}_6^L\times \mathcal H_6^R$ symmetry
 are related to the three-point couplings in AdS$_3$$\times$S$^3$,
 the first thing we have to understand is how the ${\mathcal H}_6$
 representations arise in the limit from representations of
 $SL(2,\mathbb{R}) \times SU(2)$. In the next chapter we will examine
 in more details  ${\mathcal H}_6$, here we just want to
 underline that as much as for $SL(2,\mathbb R)\,$,  $\mathcal H_6$ has
 three types of representations: discrete irreps $V^{\pm}_{p,\hj}$ and continuous ones
 $V^0_{s_1,s_2,\hj}$ , where $p,s$ and $\hj$ are the quantum
 numbers labelling the representations.\\

 Let us start with the $V^+_{p,\jh}$ representations. Following
 \cite{dak} we consider states that sit near  the top of an $SU(2)$
 representation
 \be
  \tilde{l} = \frac{k_2}{2} \m_2 p - b  \ ,
 \hspace{1cm} \tilde{m} =  \frac{k_2}{2} \m_2 p - b - n_2  \ .
 \ee
 In order to get in the limit states with a finite conformal
 dimension and well defined quantum numbers with respect to the
 currents in $(\ref{contr})$, we have to choose for
 $SL(2,\mathbb{R})$ a ${\cal D}^-_l$ representation with
 \be
 l =
 \frac{k_1}{2} \m_1 p - a \ , \hspace{1cm} m = -\frac{k_1}{2} \m_1
 p + a -n_1  \ .
 \ee
 In the limit $\jh = - \m_1 a + \m_2 b$.

 Reasoning in a similar way one can see that the $V^-_{p,\jh}$
 representations result from ${\cal D}^+_l \times V_{\tilde l}$
 representations with
 {\colsep
 \ba
 l &=& \frac{k_1}{2} \m_1 p - a \ ,
 \hspace{1cm}
 m = \frac{k_1}{2} \m_1 p - a + n_1  \ , \nb \\
 \tilde{l} &=& \frac{k_2}{2} \m_2 p - b  \ , \hspace{1cm} \tilde{m}
 =  - \frac{k_2}{2} \m_2 p + b + n_2  \ ,
 \ea}
 and  $\jh = \m_1 a - \m_2 b$ in the limit.

 Finally the $V^0_{s_1,s_2,\jh}$
 representations result from ${\cal D}^0_{l,\a} \times
 V_{\tilde l}$ representations with
 \be
 l = \frac{1}{2} + i
 \sqrt{\frac{k_1}{2}}s_1 \ , \hspace{0.4cm} m = \a + n_1  \ ,
 \hspace{0.4cm} \tilde{l} = \sqrt{\frac{k_2}{2}}s_2  \ ,
 \hspace{0.4cm} \tilde{m} = n_2 \ ,
 \ee
 and $\jh = - \m_1 \a$. As we shall see the
 tensor product of these representations reproduces in the limit
 for ${\mathcal H}_6$, see  Eq.
 $(\ref{tensorH6})$.

 We introduce a vertex operator for each unitary representations of
 $SL(2,\mathbb{R})$
 {\colsep
 \ba
 \Psi^+_l(z,x) &=& \sum_{n=0}^\infty c_{l,n} (-x)^{n} R^+_{l,n}(z) \ , \nb \\
 \Psi^-_l(z,x) &=& \sum_{n=0}^\infty c_{l,n} x^{-2l-n} R^-_{l,n}(z) \ , \nb \\
 \Psi^0_{l,\a}(z,x) &=& \sum_{n \in \mathbb{Z}} x^{-l+\a+n}
 R^0_{l,\a,n}(z) \ ,
 \ea}
 where we denote by $R$ the modes of the primary in the $n$ basis
 and $c_{l,n}^2 =\frac{\G(2l+n)}{\G(n+1)\G(2l)}$.\\

 As we anticipated in section \ref{ads3cft2}, we can define
 a two-dimensional holographic screen where the CFT lives.  This space
 is parameterize by the \emph{charge variables}\index{Charge variables|textbf}
 $(x,\bx)$ that are identified as the points where the boundary
 operators are inserted.
 Using this variables we can identify the
 action of the currents of $SL(2,\mathbb R)$ on the affine primaries with some
 differential operators defined on the $x$ space. Specifically we can establish the
 relation
 \be\label{KDx}
 K^A(z)\Psi_{l}(w,\bw;x,\bx)\sim
 \frac{\mathD^A}{z-w}\,\Psi_{l}(w,\bw;x,\bx),\qquad A=\pm, 3
 \ee
 considering the following differential operators
 \be
 {\cal D}_1^- = -
 x^2 \p_x - 2 l x \ , \hspace{0.4cm}  {\cal D}_1^+ = - \p_x \ ,
 \hspace{0.4cm}  {\cal D}_1^3 = l + x \p_x \ .
 \ee
 Similarly for
 $S^3$ we introduce
 \be
 \Omega_{\tilde{l}}(z,y) =
 \sum_{m=-\tilde{l}}^{\tilde{l}}
 \tilde{c}_{\tilde{l},m} y^{\tilde{l}+m} R_{\tilde{l},m}(z) \ ,  \\
 \ee
 where $\tilde{c}_{\tilde{l},m}^2 =
 \frac{\G(2\tilde{l}+1)}{\G(\tilde{l}+m+1)\G(\tilde{l}-m+1)}$ and
 the differential operators that represent the $SU(2)$ action are
 \be
 {\cal D}_2^+ = \p_y \ , \hspace{0.4cm} {\cal D}_2^- = -y^2
 \p_y + 2 \tl y \ , \hspace{0.4cm} {\cal D}_2^3 = y \p_y - \tl \ .
 \ee
 Generalizing the case studied in \cite{dak}, we can now
 implement the Penrose limit on the operators $\Psi^a_l(z,x)
 \Omega_{\tilde{l}}(z,y)$ and determine their precise relation with
 the primaries  $\Phi^a(z,x,y)$ of the Heisenberg algebra $\widehat {\cal H}_6$
 coming from the Penrose limit. In this section
 we introduce two charge variables\index{Charge variables $\widehat {\cal H}_6$|textbf}
 for $\widehat {\cal H}_6$, denoted by $x$ and
 $y$ in order to emphasize that they are related to the charge
 variables of $SL(2,\mathbb{R})$ and $SU(2)$ respectively.

 For the
 discrete representations we have
 {\colsep
 \ba \label{pp+}
 \Phi^+_{p,\hj}(z,x,y) &=& \lim_{k_1, k_2 \to\infty}
 \left(\frac{x}{\sqrt{k_1}}\right)^{-2l}
 \left(\frac{y}{\sqrt{k_2}}\right)^{2\tl}
 \Psi^-_{l}\left(z,\frac{\sqrt{k_1}}{x}\right) \ \Omega_{\tilde l}
 \left(z,\frac{\sqrt{k_2}}{ y}\right) \ ,  \\
 \label{pp-} \Phi^-_{p,\hj}(z,x,y) &=& \lim_{k_1,k_2\to\infty}
 \Psi^+_l\left(z,-\frac{x}{\sqrt{k_1}}\right) \Omega_{\tl}
 \left(z,\frac{y}{\sqrt{k_2}}\right) \ ,
 \ea}
 with
 \be
 l =
 \frac{k_1}{2} \m_1 p - a \ , \hspace{1cm} \tilde{l} =
 \frac{k_2}{2} \m_2 p - b \ ,
 \ee
 where $p$ is the light-cone momentum, $k_1$ is the level of
 the $SL(2,\mathbb R)$ algebra and $k_2$ that of $SU(2)$.

 For the continuous representations we have
 \be \label{pp0}
 \Phi^0_{s_1,s_2,\hj}(z,x,y) = \lim_{k_1,k_2\to\infty} (-i
 x)^{-l+\a} \, y^{\tilde{l}} \ n(k_1, l) \ \Psi^0_{l,\a}\left(z,
 \frac{i}{x} \right) n(k_2, \tilde l) \ \Omega_{\tl}
 \left(z,\frac{1}{y}\right) \ , \ee with
 {\colsep
 \ba l &=& \frac{1}{2} +
 i\sqrt{\frac{k_1}{2}} s_1 \ , \hspace{2cm}
 n(k_1, l) = \sqrt{2\pi} \, (2 k_1)^{\frac{1}{4}} \, 2^{2 l - 1} \ , \nb \\
 \tilde{l} &=& \sqrt{\frac{k_2}{2}} s_2 \ , \hspace{2.9cm} n(k_2,
 \tilde l) = \sqrt{2\pi} \, (2 k_2)^{\frac{1}{4}} \, 2^{-2 \tilde l
 - 1} \ .
 \ea}
 With the help of the previous formulae it is not
 difficult to find the Clebsch-Gordan coefficients of the
 plane-wave three-point correlators. In fact, a similar analysis
 has been performed in \cite{dak} for the three-point correlators
 of the Nappi-Witten gravitational wave considered as a limit of
 $SU(2)_k\times U(1)$. For AdS$_3$ the general form of the three
 point function is fixed, up to normalization, by $\widehat
 {SL}(2,\mathbb{R})_L \times \widehat{SL}(2,\mathbb{R})_R$
 invariance ($x$ dependence) and by $SL(2,\mathbb{C})$ global
 conformal invariance on the world-sheet ($z$ dependence), to be
 \beq \label{CGsl2}
 \left
 \langle\,\prod_{i=1}^3\Psi_{l_i}(z_i,\bz_i,x_i,\bx_i)\, \right
 \rangle =C(l_1,l_2,l_3)\prod_{i<j}^{1,3}
 \frac{1}{|x_{ij}|^{2l_{ij}}|z_{ij}|^{2h_{ij}}}\, ,
 \eeq
 where
 $l_{12}=l_1+l_2-l_3$, $h_{12}=h_1+h_2-h_3$ and cyclic permutation
 of the indexes. Due to the $\widehat{SU}(2)_L \times
 \widehat{SU}(2)_R$ and world-sheet conformal invariance the
 correlation function of three primaries on  $S^3$ is given by
 \beq
 \label{CGsu2} \left
 \langle\,\prod_{i=1}^3\Omega_{\tl_i}(z_i,\bz_i,y_i,\by_i)\, \right
 \rangle
 =\tilde{C}(\tl_1,\tl_2,\tl_3)\prod_{i<j}^{1,3}\frac{|y_{ij}|^{2\tl_{ij}}}{|z_{ij}|^{2h_{ij}}}\, ,
 \eeq
 where $\tl_{ij}$ and $h_{ij}$ are defined as for
 (\ref{CGsl2}).

 Let us consider for instance the limit leading to a
 $\langle ++-\rangle$ correlator. Taking into account that $\sum_i
 \hj_i = -L = - \mu_1 (a_1+a_2-a_3) + \mu_2 (b_1 + b_2 - b_3)$, the
 kinematic coefficient, coming from the Ward identities, receives the following contribution from the
 AdS$_3$ part
 \beq
 K_{++-}(x,\bx)= k_1^{-q_1} \left |e^{-\mu_1
 x_{3}(p_1x_1+p_2x_2)}\right|^2 |x_2-x_1|^{2 q_1} \ ,
 \eeq
 where
 $q_1 = a_1+a_2-a_3$ and a similar contribution from the $S^3$ part
 \be
 K_{++-}(y,\by)= k_2^{-q_2}\left |e^{-\mu_2
 y_{3}(p_1y_1+p_2y_2)}\right|^2 |y_2-y_1|^{2 q_2} \ ,
 \ee
 where
 $q_2 = - b_1-b_2+b_3$.
 Putting the two contributions together
 \beq\label{K++-}
 K_{++-}(x,\bx,y,\by)=  k_1^{-q_1} k_2^{-q_2} \left |e^{-\mu_1
 x_{3}(p_1x_1+p_2x_2)}e^{-\mu_2 y_{3}(p_1y_1+p_2y_2)}\right|^2
 |x_2-x_1|^{2 q_1}|y_2-y_1|^{2 q_2} \ ,
 \eeq
 This result will be reproduce
 in an alternative way in
 $(\ref{cgppm})$. In the $SU(2)$ invariant case $\mu_1=\mu_2$, one
 finds a looser constraint on the $a_i$ and $b_i$ that leads to
 $q_1+q_2= Q = -L/\mu$. Summing over the allowed values of $q_1$
 and $q_2$ one eventually gets an $SU(2)_I$ invariant result, see
 next chapter
 $(\ref{su2ppm})$. Using the above expression for the CG
 coefficients for a coupling of the form $\langle +-0 \rangle $ one
 obtains
 \beq
 K_{+-0} (x,\bx,y,\by) = \left |e^{-\mu_1p_1x_1 x_2-
 \frac{s_{1}}{\sqrt{2}}(x_2 x_3+x_1x_{3})} \right |^2 \left
 |e^{-\mu_2 p_1 y_1 y_2- \frac{s_{2}}{\sqrt{2}}(y_2 y_3+y_1y_{3})}
 \right |^2 |x_3|^{2q_1} |y_3|^{2q_2} \ ,
 \eeq
 where $q_1 =
 a_1-a_2+\a$ and $q_2 = b_2-b_1$.

 For the euclidean AdS$_3$, that is the $H_3^+$ WZW model, the two
 and three-point functions involving vertex operators in unitary
 representations were computed by Teschner \cite{tesch}. The
 two-point functions are given by \be \langle \Psi_{l_1}(x_1,z_1)
 \Psi_{l_2}(x_2,z_2) \rangle = \frac{1}{|z_{12}|^{4h_{l_1}}} \left
 [ \frac{\d^2(x_1-x_2) \d(l_1+l_2-1)}{B(l_1)}
 +\frac{\d(l_1-l_2)}{|x_{12}|^{4 l_1}} \right ] \ , \ee where \be
 B(l) = \frac{\n^{1-2l}}{\pi b^2 \g(b^2(2l-1))}  \ , \hspace{0.4cm}
 \n = \pi \frac{\G(1-b^2)}{\G(1+b^2)} \ , \hspace{0.4cm} b^2 =
 \frac{1}{k_1-2} \ , \ee and $l = \frac{1}{2} + i \s$. The
 three-point functions have the same dependence on the $z_i$ and
 the $x_i$ as displayed in $(\ref{CGsl2})$. The structure constants
 are given by \be \label{sl2qcg} C(l_1,l_2,l_3) = - \frac{b^2
 Y_b(b)G_b(1-l_1-l_2-l_3)}{2 \sqrt{\pi \n} \g(1+b^2)} \prod_{i=1}^3
 \frac{\sqrt{\g(b^2(2l_i-1))}}{G_b(1-2l_i)} \prod_{1=i<j}^3
 G_b(-l_{ij}) \ . \ee In the previous expression we used the entire
 function $Y_b(z)$ introduced in \cite{zz} \be \log
 Y_b(z)=\int_0^{\infty}{dt\over t}\left[\left({Q\over 2}-z\right)^2
 e^{-t}-{\sinh^2\left[\left({Q\over 2}-z\right){t\over
 2}\right]\over \sinh\left[{bt\over 2}\right]\sinh\left[{t\over
 2b}\right]}\right] \ , \ee with \be Q=b+{1\over b} \ . \ee We
 define also  the closely related function $G_b(z)$ given by \be
 G_b(z) \equiv\frac{b^{-b^2z \left ( z+1+\frac{1}{b^2} \right
 )}}{Y_b(-bz)} \ . \label{gtoy} \ee The function $Y_b$ satisfies
 \be Y_b(z+b) = \g(bz)Y_b(z)b^{1-2bz} \ , \hspace{1cm} Y_b(z) =
 Y_b(b+1/b-z) \ . \ee In order to study the Penrose limit of the
 $SL(2,\mathbb{R})$ structure constants we express the function
 $G_b(z)$ in term of the function $P_b(z)$ that appears in the
 $SU(2)$ three-point functions \cite{zf} and whose asymptotic
 behavior was studied in \cite{dak}. For this purpose we write \be
 \ln P_b(z) = f(b^2,b^2|z)-f(1-zb^2,b^2|z) \ , \ee where $f(a,b \,
 |z)$ is the \emph{Dorn-Otto function}\index{Dorn-Otto
 function|textbf} \cite{do} \be f(a,b|s)\equiv
 \int_0^{\infty}{dt\over t}\left[s(a-1)e^{-t}+{bs(s-1)\over
 2}e^{-t}-{se^{-t}\over 1-e^{-t}}+{(1-e^{-tbs})e^{-at}\over
 (1-e^{-bt})(1-e^{-t})}\right]= \ee
 $$=\sum_{j=0}^{s-1}\log \Gamma(a+bj) \ ,
 $$
 where the last relation is valid for integer $s$. From \be
 f(bu,b^2|z) - f(bv,b^2|z) = \ln Y_b(v) - \ln Y_b(u) +zb(u-v) \ln b
 \ , \hspace{1cm} u+v = b + \frac{1}{b} - zb \ , \ee we obtain \be
 G_b(z) = \frac{b \g(-b^2 z)}{Y_b(b)P_b(-z)} \ , \ee and we can
 rewrite the  coupling $(\ref{sl2qcg})$ using the function $P_b$
 \be \label{sl2qcp} C(l_1,l_2,l_3) = - \frac{b^3}{2 \sqrt{\pi \n}
 \g(1+b^2)} \frac{\g(b^2(l_1+l_2+l_3-1))}{P_b(l_1+l_2+l_3-1)}
 \prod_{i=1}^3 \frac{P_b(2l_i-1)}{\sqrt{\g(b^2(2l_i-1))}}
 \prod_{1=i<j}^3 \frac{\g(b^2 l_{ij})}{P_b(l_{ij})} \ . \ee Let us
 consider first the $\langle ++- \rangle$ coupling. As we explained
 before, the AdS$_3$ quantum numbers have to be scaled as follows
 \be
 l_i = \frac{k_1}{2} \m_1 p_i - a_i \ .
 \ee
 The leading
 behavior is
 \be\label{C123}
 C(l_1,l_2,l_3) \sim \ \frac{1}{2 \pi b q_1}
 \frac{1}{P_b(-q_1)} \left [ \frac{\g(\m_1 p_3)}{\g(\m_1
 p_1)\g(\m_1 p_2)} \right ]^{\frac{1}{2}+q_1} \ ,
 \ee where $q_1 =
 a_1+a_2-a_3$.
 Due to the presence of $P_b(-q_1)$ in the
 denominator, the coupling vanishes unless $q_1 \in \mathbb{N}$.
 This result will be reproduced as a classical tensor product, see (\ref{tensorH6}).
 We can then write
 \be
 \lim_{b \to 0} C(l_1,l_2,l_3) = (-1)^{q_1}
 \frac{k_1^{q_1+\frac{1}{2}}}{q_1!} \left [ \frac{\g(\m_1
 p_3)}{\g(\m_1 p_1)\g(\m_1 p_2)} \right ]^{\frac{1}{2}+q_1} \sum_{n
 \in \mathbb{N}} \d(q_1-n) \ .
 \ee
 The sign $(-1)^{q_1}$ does not
 appear in the ${\cal H}_6$ couplings, a discrepancy which might be
 due to some difference between the charge variables used in
 \cite{tesch} and the charge variables used in the present paper.
 The same limit for the $SU(2)$ three-point couplings leads to
 \be\label{Ct123}
 \lim_{\tilde{b} \to 0} \tilde{C}(\tl_1,\tl_2,\tl_3) =
 \frac{k_2^{q_2+\frac{1}{2}}}{q_2!} \left [ \frac{\g(\m_2
 p_3)}{\g(\m_2 p_1)\g(\m_2 p_2)} \right ]^{\frac{1}{2}+q_2} \sum_{n
 \in \mathbb{N}} \d(q_2-n) \ ,
 \ee
 where $\tilde{b}^{-2} = k_2+2$ and
 $q_2 = -b_1-b_2+b_3$.
 Proceeding in a similar way for a $\langle + - 0
 \rangle$ correlator we obtain from AdS$_3$
 \be
 \lim_{b \to 0}
 C(l_1,l_2,l_3) = \frac{2^{- i s_1 \sqrt{2k_1}}}{\sqrt{2 \pi}}
 e^{\frac{s_1^2}{2}(\psi(\m_1 p) + \psi(1-\m_1 p) - 2 \psi(1) )} \
 , \ee and similarly  from $S^3$ \be \lim_{\tilde{b} \to 0}
 \tilde{C}(\tl_1,\tl_2,\tl_3) = \frac{2^{1+s_2
 \sqrt{2k_2}}}{\sqrt{2 \pi}} e^{\frac{s_2^2}{2}(\psi(\m_2 p) +
 \psi(1-\m_2 p) - 2 \psi(1) )} \ .
 \ee

  Let us briefly discuss  how the Penrose limit acts on the
 wave-functions corresponding to the representations considered
 above. We will consider only the limit of the ground states but
 the analysis can be easily extended to the limit of the whole
 $SL(2,\mathbb{R}) \times SU(2)$ representation if we introduce a
 generating function for the matrix elements, which can be
 expressed in terms of the Jacobi functions.

 Using global coordinates for AdS$_3$$\times$S$^3$, the ground
 state of a ${\cal D}^-_l \times V_{\tilde l}$ representation can
 be written as
 \be
 e^{2 i l t - 2 i \tilde{l} \psi} ({\rm
 cosh}\r)^{-2l} ({\rm cos}\theta)^{2 \tilde{l}} \ .
 \ee
 After
 scaling the coordinates and the quantum numbers as required by the
 Penrose limit this function becomes
 \be
 e^{2 i p v + i \jh u
 -\frac{p}{2}(\m_1r^2_1+\m_2r^2_2)} \ , \hspace{1cm} \jh = -\m_1 a
 + \m_2 b \ .
 \ee
 In the same way starting from a ${\cal D}^+_l
 \times V_{\tilde l}$ representation
 \be
 e^{-2 i l t + 2 i
 \tilde{l} \psi} ({\rm cosh}\r)^{-2l} ({\rm cos}\theta)^{2
 \tilde{l}} \ , \ee we obtain \be e^{-2 i p v + i \jh u
 -\frac{p}{2}(\m_1r^2_1+\m_2r^2_2)} \ , \hspace{1cm} \jh = \m_1 a -
 \m_2 b \ .
 \ee
 As anticipated the limit of the generating
 functions lead to semiclassical wave-functions for the
 six-dimensional wave which are a simple generalizations of those
 displayed in \cite{dak}.

 % OOOOOOOOOOOOOOOOOOOOOOOOOOOOOO CONCLUSIONS 3 OOOOOOOOOOOOOOOOOOOOOOOOOOOOOOOOOO

 % OOOOOOOOOOOOOOOOOOOOOOOOOOOOOO CONCLUSIONS 3 OOOOOOOOOOOOOOOOOOOOOOOOOOOOOOOOOO

 \section{Conclusions}
 \label{conc3}
 \noindent

 In this chapter we have seen that it is possible to compute three-point functions for
 strings on the Hpp-wave limit of AdS$_3$$\times$S$^3$. The idea is to use charge
 variables defined on the holographic screen of  AdS$_3$, denoted by
 $(x,\bx)$,
 and also for S$^3$, that we called $(y,\by)$, and then
 reformulate the whole problem in term of them.
 With the $\widehat{SL}(2,\mathbb{R})_{k_1}\times \widehat{SU}(2)_{k_2}$
 correlators in hand we can take the Penrose limit by rescaling the charge variables
 in terms of $k_1$ and $k_2$ and then taking the level of the algebras going to infinite.
 The expressions for the primaries of $\widehat{\mathcal H}_6$ in terms of
 the affine primaries $\Psi_l(z,x)$ and $\Omega_{\tilde l}(z,y)$ are
 given in (\ref{pp+}-\ref{pp0}).
 The Penrose limit as indicated here was taken for both the kinematical contribution as well as for the Clesch-Gordan
 coefficients. In the following chapter we will prove that this correlators are in
 agreement with the results found performing the computation
 directly in the $\widehat{\mathcal H}_6$  affine conformal theory.
 In chapter 6 we will apply a similar analysis to
 the fermionic string and we will examine the correspondence with the boundary theory.
 The main result presented in this section, $\ie$ the plane wave limit from
 charge variables, will finally
 be applied in chapter \ref{holography} to propose an alternative check of the
 AdS$_5$/CFT$_4$ correspondence at the BMN limit.

 % OOOOOOOOOOOOOOOOOOOOOOOOOOOOO     PP-WAVE LIMIT     OOOOOOOOOOOOOOOOOOOOOOOOOOOOOOOOOO

 % OOOOOOOOOOOOOOOOOOOOOOOOOOOOO     PP-WAVE LIMIT     OOOOOOOOOOOOOOOOOOOOOOOOOOOOOOOOOO

 % OOOOOOOOOOOOOOOOOOOOOOOOOOOOO     PP-WAVE LIMIT     OOOOOOOOOOOOOOOOOOOOOOOOOOOOOOOOOO

 % OOOOOOOOOOOOOOOOOOOOOOOOOOOOO     PP-WAVE LIMIT     OOOOOOOOOOOOOOOOOOOOOOOOOOOOOOOOOO

 \renewcommand{\theequation}{\arabic{chapter}.\arabic{equation}}
 \chapter{Bosonic String Amplitudes in
 Plane Waves}
 \label{str-pp}\setcounter{chapter}{4}
 \noindent

 In the present chapter  we extend to $\widehat{\mathcal H}_6$ the
 work done by D'Appollonio and Kiritsis in \cite{dak}.
 Exploiting  current algebra techniques they
 were able to compute string amplitudes for a background with NW model worldsheet
 description. In that case the model under consideration was
 the Penrose limit of the near-horizon geometry of a
 stack of NS5-branes,  realized on the worldsheet by an
 $\widehat{\mathcal H}_4$ current algebra.

 Here we apply the same techniques to the
 pp-wave geometry representing the Penrose limit of
 AdS$_3$$\times$S$^3$$\times$${\mathcal M}$.
 Although we will almost exclusively concentrate ourselves
 on the bosonic string, we will also comment on how to
 extend our results to the superstring (see chapter 6 for more details). We will compute two,
 three and four-point amplitudes with insertions of tachyon vertex
 operators of any of the three types of irreducible representations allowed
 by $\widehat{\mathcal H}_6$ current algebra. The latters defined by
 the value of the light-cone momentum $p^+$: states of
 discrete representations, for $p^+\neq 0 $, and states belonging to continuous
 representations, with $p^+=0$.

 As expected, the amplitudes computed
 here by exploiting the $\widehat{\mathcal H}_6$ current algebra,
 coincide with the ones resulting from the Penrose limit as done in the previous chapter,
 $\ie$ the contraction of the amplitudes on
 AdS$_3$$\times$S$^3$$\times$
 ${\mathcal M}$. This allows us to clarify the crucial role played
 by the charge variables in the fate of holography. They become
 coordinates on a four-dimensional holographic screen for the
 pp-wave \cite{kirpio}.

 This chapter is organized as follows:

 In section \ref{ppw} we briefly describe the general Hpp-waves whose sigma-models
 are WZW models based on the ${\bf H}_{2+2n}$ Heisenberg groups
 and then we concentrate on the six-dimensional plane wave that emerges
 from the Penrose limit of AdS$_3$$\times$S$^3$, discussing the
 corresponding contraction of the
 $[\widehat{SL}(2,\mathbb{R})_{k_1}\times \widehat{SU}(2)_{k_2}]^2$
 currents that leads to the $\widehat{\cal H}_6^L \times \widehat{\cal H}_6^R$
 algebra. In section \ref{specth6} we identify the relevant representations of
 $\widehat{\cal H}_6^L \times \widehat{\cal H}_6^R$ and write down
 the explicit expressions for the tachyon vertex operators. In
 section \ref{3point} we compute two and three-point correlation functions on
 the world-sheet and compare the results with those obtained from
 the limit of AdS$_3$$\times$S$^3$. In section \ref{4point} we compute four-point
 correlation functions on the world-sheet by means of current
 algebra techniques. In section \ref{amplitudes} we
 study string amplitudes in the Hpp-wave and analyze the structure
 of their singularities. Finally, we
 draw our conclusions and indicate lines for future material presented in this thesis.

 % OOOOOOOOOOOOOOOOOOOOOOOOOOOO    PP H_2+2N     OOOOOOOOOOOOOOOOOOOOOOOOOOOOOOOOOOOOOOOO

 % OOOOOOOOOOOOOOOOOOOOOOOOOOOO    PP H_2+2N     OOOOOOOOOOOOOOOOOOOOOOOOOOOOOOOOOOOOOOOO

 \section{The plane wave geometry}
 \label{ppw}
 \noindent

 In section \ref{ppwaves} we introduced plane wave backgrounds\index{Parallel Plane waves|textbf}
 \be
 \label{ppntrans} ds^{2} = - 2 du dv - {1\over 4} du^{2}
 \sum_{\a=1}^{n} \mu_\a^2 y_{\a}\bar{y}_{\a} + \sum_{\a=1}^{n}
 dy_{\a}d\bar{y}_{\a} + \sum_{i=1}^{24-2n}g_{ij} dx^{i} dx^{j} \ ,
 \ee
 where $u$ and $v$ are light-cone coordinates, $y_\a = r_\a e^{i
 \f_\a}$ are complex coordinates parameterizing the $n$ transverse
 planes and  $x^{i}$ are the remaining $24-2n$ dimensions of the
 critical bosonic string that we assume compactified on some
 internal manifold ${\cal M}$ with metric $g_{ij}$. In the
 following we will concentrate on the $2+2n$ dimensional part of
 the metric in Eq. $(\ref{ppntrans})$. The wave is supported by a
 non-trivial NS-NS antisymmetric tensor field strength
 \be
 H =\sum_{\a=1}^{n}\mu_\a du\wedge dy_\a\wedge d \bar y_\a \ ,
 \label{atfs}
 \ee
 while the dilaton is constant and all the other
 fields are set to zero.

 The background defined in $(\ref{ppntrans})$ and $(\ref{atfs})$
 with generic $\m_\a$ has a $(5n+2)$-dimensional isometry group
 generated by translations in $u$ and $v$, independent rotations in
 each of the $n$ transverse planes and $4n$ ``magnetic
 translations''. When $2 \le k \le n$ of the $\m_\a$ coincide the
 isometry group is enhanced: the generic $U(1)^n$ rotational
 symmetry of the metric is enlarged to $SO(2k) \times U(1)^{n-k}$,
 broken to $U(k) \times U(1)^{n-k}$ by the field strength of the
 antisymmetric tensor. The dimension of the resulting isometry
 group is therefore $5n+2+k(k-1)$.

 As first realized in \cite{nw} for the case $n=1$ and then in \cite{kehagias} for
 generic $n$, the $\s$-models corresponding to Hpp-waves are WZW
 models based on the ${\bf H}_{2+2n}$ Heisenberg group. The
 $\widehat{\mathcal H}_{2+2n}$ current algebra is defined by the
 following OPEs
 {\colsep
 \ba
 P^+_{\alpha}(z) {P}^{-\beta}(w) &\sim&
 {2\delta_{\alpha}^{\beta} \over (z-w)^{2}} - {2 i \m_\a
 \delta_{\alpha}^{\beta}\over (z-w)}
 K(w)\nonumber \ , \\
 J(z) P^+_{\alpha}(w) &\sim& -{i \m_\a \over (z-w)}P^+_{\alpha}(w)
 \ , \nb
 \\
 J(z) {P}^{-\alpha}(w) &\sim& +{i \m_\a \over (z-w)}
 {P}^{-\alpha}(w) \ , \nb \\
 J(z)K(w) &\sim& \frac{1}{(z-w)^2} \ , \label{opes}
 \ea}
 where $\a,\b = 1, ..., n$. From here we can easily deduce the
 corresponding algebra already displayed in (\ref{h2+2n}).
 The anti-holomorphic currents satisfy a similar
 set of OPEs \footnote{As usual we will distinguish the right
 objets by putting a bar on them.} and the total affine symmetry of
 the model is $\widehat{\mathcal H}^L_{2+2n} \times
 \widehat{\mathcal H}^R_{2+2n}$.

 A few clarifications are in order. First of all the zero modes of
 the left and right currents only realize a $(4n+3)$-dimensional
 subgroup of the whole isometry group. The left and right central
 elements  \footnote{Notice that we use the same letter for both a
 (spin $s$) current $W(z)$ and the corresponding charge $W\equiv
 W_0=\oint \frac{dz}{2\pi i}\,z^{s-1} W(z) $. In order to avoid any
 confusion  we try  always to emphasize the two-dimensional nature
 of the world-sheet fields by showing their explicit $z$
 dependence.} $K$ and $\bar K$ are identified and generate
 translations in $v$; $P^+_\a$ and $P^{-\a}$ together with their
 right counterparts generate the $4n$ magnetic translations;
 $J+\bar J$ generates translations in $u$ and $J - \bar J$ a
 simultaneous rotation in all the $n$ transverse planes. In the
 following we will refer to the subgroup of the isometry group that
 is not generated by the zero modes of the currents as $G_I$. For
 the supersymmetric ${\bf H}_{2+2n}$ WZW models the existence of
 enhanced symmetry points for particular choices of the parameters
 $\m_i$ should be related to the existence of supernumerary Killing
 spinors, as discussed in \cite{ppwaves}.

 The position of the index $\a = 1, ..., n$ carried by the $P^\pm$
 generators is meant to emphasize that at the point where the
 generic $U(1)^n$ part of the isometry group is enhanced to $U(n) =
 SU(n)_I \times U(1)_{J-\bar J}$ they transform respectively in the
 fundamental and in the anti-fundamental representation of
 $SU(n)_I$. The left and right current modes satisfy the same
 commutation relations with the generators of the $SU(n)_I$
 symmetry of the background.

 Let us discuss some particular cases. When $n=1$ we have the
 original NW model and all the background isometries are realized
 by the zero-modes of the currents. When $n=2$ there is an
 additional $U(1)_I$ symmetry which extends to  $SU(2)_I$ for
 $\m_1=\m_2$. In this paper we will describe in detail only the
 six-dimensional Hpp-wave, because of its relation to the BMN limit
 of the AdS$_3$/CFT$_2$ correspondence. Higher-dimensional
 Hpp-waves also arise as Penrose limits of interesting backgrounds:
 the ${\bf H}_8$ WZW model is for instance the Penrose limit of a
 non-standard brane intersection  whose near horizon geometry is
 AdS$_3$$\times$S$^3$$\times$S$^3$$\times$S$^1$.
 However these models do not display any new features as far as the
 spectrum and the computation of the correlation functions are
 concerned and they can be analyzed in precisely the same way as
 the ${\bf H}_6$ WZW model. When we discuss the Wakimoto
 representation for the ${\bf H}_6$ WZW model, the following change
 of variables
 \be
 y^{\alpha} = e^{i \m_\a u/2} w^{\alpha} \ ,
 \hspace{1cm} \bar{y}_{\alpha}= e^{-i \m_\a u/2} \bar{w}_{\alpha} \
 ,
 \ee
 which yields a metric with a $U(2)$ invariant form
 \be
 \label{pp2trans} ds^{2} = - 2 du dv + {i\over 4} du
 \sum_{\a=1}^{2} \m_\a( w^{\alpha}d\bar{w}_{\alpha} -
 \bar{w}_{\alpha}dw^{\alpha}) + \sum_{i=1}^{2}
 dw^{\alpha}d\bar{w}_{\alpha} ~,
 \ee
 will prove useful.

 As it is well known, the background $(\ref{ppntrans})$,
 $(\ref{atfs})$ with $n=2$ and $\m_1=\m_2$ arises from the Penrose
 limit of AdS$_3$$\times$S$^3$, the near horizon geometry of an
 F1-NS5 bound state. The general metric with $\m_1 \ne \m_2$ can
 also be obtained as a Penrose limit but starting with different
 curvatures for AdS$_3$ and $S^3$. In global coordinates the metric
 can be written as in (\ref{ads3s3})
 \be
 ds_6^{2} = R_1^{2} \ [ -(\cosh\rho)^{2}
 dt^{2} + d\rho^{2} + (\sinh\rho)^{2} d\varphi_{1}^{2}] + R_2^2 \
 [(\cos\theta)^{2} d\psi^{2} + d\theta^{2} + (\sin\theta)^{2}
 d\varphi_{2}^{2}] \ ,
 \ee
 where $(t,\rho,\varphi_1)$ parameterize
 the three dimensional anti-de Sitter space with curvature radius
 $R_1$ and $(\theta,\psi,\varphi_2)$ parameterize $S^3$ with
 curvature radius $R_2$. In the Penrose limit\index{Penrose
 limit|textbf} we focus on a null geodesic of a particle moving
 along the axis of AdS$_3$ ($\rho \to 0$) and spinning around the
 equator of the three sphere ($\theta\to 0$). We then change
 variables according to
 \be
 t = \frac{\m_1 u}{2} + \frac{v}{\m_1
 R_1^{2}} \ , \qquad \psi = \frac{\m_2 u}{2} - \frac{v}{\m_2
 R_2^{2}} \ , \qquad \rho = {r_{1}\over R_1} \ , \qquad \theta =
 {r_{2}\over R_2} \ ,
 \ee
 and take the limit sending $R_1, R_2 \to
 \infty$ while keeping $\m_1^2 R_1^2 = \m_2^2 R_2^2$.\\

 From the world-sheet point of view, the Penrose limit of AdS$_3$
 $\times$S$^3$ amounts to a contraction of the current algebra of
 the underlying $\widehat{SL}(2,\mathbb{R})\times \widehat{SU}(2)$
 WZW model. The $\widehat{SL}(2,\mathbb{R})$ current
 algebra\index{Current algebra $\widehat{SL}(2,{\mathbb R})$|textbf}
 at level $k_1$ is given by
 {\colsep
 \ba
 \label{sl2}
  K^+(z)K^-(w) &\sim&
 \frac{k_1}{(z-w)^2} -\frac{2 K^3(w)}{z-w} \ , \nonumber\\
 K^3(z)K^\pm(w) &\sim& \pm \frac{K^{\pm}(w)}{z-w} \ , \nonumber \\
 K^3(z)K^3(w) &\sim& \displaystyle - \frac{k_1}{2(z-w)^2} \ .
 \ea}
 Similarly the $\widehat{SU}(2)$ current algebra\index{Current algebra $\widehat{SU}(2)$|textbf} at level $k_2$ is
 {\colsep
 \ba
 J^+(z)J^-(w) &\sim&
 \frac{k_2}{(z-w)^2} + \frac{2 J^3(w)}{z-w} \ , \nonumber\\
 J^3(z)J^\pm(w) &\sim& \pm \frac{J^{\pm}(w)}{z-w} \ , \nonumber \\
 J^3(z)J^3(w) &\sim& \frac{k_2}{2 (z-w)^2} \ . \label{su2}
 \ea}

 The contraction\index{Current contraction|textbf} \cite{sfetsos} to the $\widehat{\mathcal H}_6$ algebra defined in
 $(\ref{opes})$ is performed by first introducing the new currents
 {\colsep
 \ba
 P_1^\pm &=& \sqrt{\frac{2}{k_1}} K^\pm \ , \hspace{1.6cm}
 P_2^\pm = \sqrt{\frac{2}{k_2}} J^\pm \ , \nb \\
 J &=& - i (\m_1 K^3 + \m_2 J^3) \ , \hspace{0.6cm} K = - i \left
 (\frac{K^3}{\m_1 k_1} - \frac{J^3}{\m_2 k_2}  \right ) \ ,
 \label{contr}
 \ea}
 and then by taking the limit $k_1, k_2
 \rightarrow \infty$ with $\m_1^2 k_1 = \m_2^2 k_2$.

 In view of possible applications of our analysis to the
 superstring, and in order to be able to consider flat space or a
 torus with $c_{int}=20$ as a consistent choice for the internal
 manifold ${\cal M}$ of the bosonic string before the Penrose limit
 is taken, one should choose $k_{1}-2=k_{2}+2=k$ so that the
 central charge is $c=6$.

 % OOOOOOOOOOOOOOOOOOOOOOOOOOOO      SPECTRUM    H6      OOOOOOOOOOOOOOOOOOOOOOOOOOOOOOOOOOOOOOOO

 % OOOOOOOOOOOOOOOOOOOOOOOOOOOO      SPECTRUM    H6      OOOOOOOOOOOOOOOOOOOOOOOOOOOOOOOOOOOOOOOO

 \section{Spectrum of the model}
 \label{specth6} \noindent

 Our aim in this section is to determine the spectrum of the string
 in the Hpp-wave with ${\bf H}_6$ Heisenberg symmetry. As in the
 ${\bf H}_4$ case,  in addition to `standard' highest-weight
 representations, new modified highest-weight (MHW) representations
 should be included. In the ${\bf H}_4$ case as well as in ${\bf
 H}_6$ with $SU(2)_I$ symmetry, such MHW representations are
 actually spectral flowed representations. However, in the general
 ${\bf H}_6$ $\mu_1\neq \mu_2$ case, we have the novel phenomenon
 that spectral flow cannot generate the MHW representations.

 The MHW representations are difficult to handle in the current
 algebra formalism. Fortunately they are  easy to analyze in the
 quasi-free field representation \cite{kk,dak} where their
 unitarity and their interactions are straightforward.

 The representation theory of the extended Heisenberg algebras,
 such as ${\mathcal H}_6$, is very similar to the ${\cal H}_4$ case
 \cite{kk,dak}. The ${\mathcal H}_6$\index{Heisenberg algebra
 ${\mathcal H}_6$|textbf} commutation relations are \be
 [P^+_\a,P^{-\b}] = -2 i \m_\a \d_\a^\b K \ , \hspace{0.4cm}
 [J,P^{+}_\a] = - i \m_\a P^+_\a \ , \hspace{0.4cm} [J,P^{-\a}] = i
 \m_a P^{-\a} \ . \ee

 As explained in the previous paragraph this algebra generically
 admits an additional $U(1)_I$ generator $I^3$ that satisfies \be
 [I^3,P^+_\a] = -i  (\s^3)_\a{}^\b P^+_\b \ , \hspace{1cm}
 [I^3,P^{-\a}] = i  (\s^{3,t})^\a{}_\b P^{-\b} \ . \ee When $\m_1 =
 \m_2 \equiv \m$ the $U(1)_I$ symmetry is enhanced to $SU(2)_I$
 \be[I^{a}, P^+_{\alpha}] = - i
 (\sigma^{a})_{\alpha}^{\phantom{\alpha}\beta}P^+_{\beta} \ ,
 \qquad [I^{a}, {P}^{-\alpha}] = i
 (\sigma^{a,t})^{\alpha}{}_{\beta}{P}^{-\beta}  \ , \hspace{0.4cm}
 a = 1,2,3 \ .\ee

 For ${\mathcal H}_6$ there are two \emph{Casimir
 operators}\index{Casimir operators ${\mathcal H}_6$|textbf}: the
 central element $K$ and the combination \be {\cal
 C}=2JK+\frac{1}{2} \sum_{\a=1}^2
 (P^+_{\alpha}P^{-\alpha}+P^{-\alpha}P^+_{\alpha}) \ . \ee

 There are three types of unitary representations:\\

 1) \emph{Lowest-weight representations LWR}\index{LWR $\mathcal H_6$|textbf} $ V_{p,\jh}^+$, where $p
 > 0$. They are constructed starting from a state $|p,\jh\rangle$
 which satisfies $P^+_{\alpha}|p,\jh \rangle=0$, $K|p,\jh\rangle =
 i p |p,\jh\rangle$ and $J|p,\jh\rangle = i \jh |p,\jh\rangle$. The
 spectrum of $J$ is given by $\{ \jh +\m_1 n_1+\m_2 n_2 \}$,
 $n_1,n_2 \in \mathbb{N}$ and the value of the Casimir is
 $\mathcal{C}=-2p\jh+(\m_1+\m_2)p$ .\vskip.1cm

 2) \emph{Highest-weight
 representations HWR}\index{HWR $\mathcal
 H_6$|textbf} $ V_{p,\jh}^-$,
 where $p >
 0$. They are constructed starting from a state $|p,\jh\rangle$
 which satisfies $P^{-\alpha}|p,\jh \rangle=0$, $K|p,\jh\rangle = -
 i p |p,\jh\rangle$ and $J|p,\jh\rangle = i \jh |p,\jh\rangle$. The
 spectrum of $J$ is given by $\{ \jh -\m_1 n_1-\m_2 n_2 \}$,
 $n_1,n_2 \in \mathbb{N}$ and the value of the Casimir is
 $\mathcal{C}= 2p\jh+(\m_1+\m_2)p$. The representation
 $V^-_{p,-\jh}$ is conjugate to $V^+_{p,\jh}$.\vskip.1cm

 3) \emph{Continuous representations}\index{Continuous
 irreps $\mathcal H_6$|textbf} $V _{s_1,s_2,\jh}^0$
 with $p=0$. These
 representations are characterized by $K |s_1,s_2,\hj \rangle = 0$,
 $J |s_1,s_2,\hj \rangle = i \jh|s_1,s_2,\hj \rangle $ and
 $P^\pm_{\alpha}|s_1,s_2,\hj \rangle\neq 0$. The spectrum of $J$ is
 then given by $\{ \jh +\m_1 n_1+\m_2 n_2 \}$, with $n_1,n_2 \in
 \mathbb{Z}$ and $| \jh | \le \frac{\m}{2}$ where $\m = {\rm
 min}(\m_1,\m_2)$. In this case we have two other Casimirs besides
 $K$: ${\cal C}_1 = P^+_1 P^{-1}$ and ${\cal C}_2 = P^+_2 P^{-2}$.
 Their values are ${\cal C}_\a = s^2_\a$, with $s_\a \ge 0$ and
 $\a=1, 2$. The one dimensional representation can be considered as
 a particular continuous representation, where the charges $s_\a$
 and $\jh$ are zero.\\

 The ground states of all these representations are assumed to be
 invariant under the $U(1)_I$ ($SU(2)_I$) symmetry. This follows
 from comparison with the spectrum of the scalar Laplacian in the
 gravitational wave background, described  below.

 Since we are dealing with infinite dimensional representations, it
 is very convenient to introduce charge variables in order to keep
 track of the various components of a given representation in a
 compact form. We introduce two doublets of charge variables $x_\a$
 and $x^\a$, $\a = 1, 2$. The action of the ${\mathcal H}_6$
 generators and of the additional generator $I^3$ on the
 $V_{p,\jh}^+$ representations is given by
 {\colsep
 \ba
 P_\a^+ &=& \sqrt{2}
 \m_\a p x_\a \ , \hspace{1cm}
 P^{-\a} = \sqrt{2} \p^\a \ , \hspace{1cm} K = i p \ , \nb \\
 J &=& i \left ( \jh + \m_\a x_\a \p^\a \right ) \ , \hspace{1cm}
 I^3 =i x_\a (\s^{3,t})^\a{}_\b \p^\b \ . \label{difp}
 \ea}
 Similarly for the $V_{p,\jh}^-$ representations we have
 {\colsep
 \ba
 P_\a^+
 &=&  \sqrt{2} \p_\a \ , \hspace{1cm}
 P^{-\a} = \sqrt{2} \m_\a p x^\a\ , \hspace{1cm} K = - i p \ , \nb \\
 J &=& i \left ( \jh - \m_\a x^\a \p_\a \right ) \ , \hspace{1cm}
 I^3 = - i x^\a (\s^{3})_\a{}^\b \p_\b \ . \label{difm}
 \ea}
 Finally
 for the $V_{s_1,s_2,\jh}^0$ representations we have
 \be
 P_\a^+ =
 s_\a x_\a \ , \hspace{0.4cm} P^{-\a} = s_\a x^\a \ ,
 \hspace{0.4cm} J = i \left ( \jh + \m_\a x_\a \p^\a \right ) \ ,
 \hspace{0.4cm} I^3 =i x_\a (\s^{3,t})^\a{}_\b \p^\b \ ,
 \label{dif0}
 \ee
 with the constraints $x^1 x_1 = x^2 x_2 = 1$, \ie
 \ $x_\a = e^{i\phi_\a}$. Alternative representations of the
 generators are possible. In particular, acting on $V_{s,\jh}^0$,
 it may prove convenient to introduce charge variables $\xi_\alpha$
 such that $\sum_{\alpha} \xi_{\alpha}\xi^{\alpha}=1$. The
 $\xi_\alpha$ are related to the $x_\alpha$ in (\ref{dif0}) by
 $\xi_\a = \frac{s_\a}{s} x_\a$ where $s^2 = s_1^2+s_2^2$.

 We can easily organize the spectrum of the D'Alembertian in the
 plane wave background in representations of ${\mathcal H}^L_6
 \times {\mathcal H}_6^R$. Using radial coordinates in the two
 transverse planes the covariant scalar D'Alembertian reads \be
 \nabla^2 = - 2\p_u \p_v + \sum_{\a=1}^2 \left ( \p_{r_\a}^2
 +\frac{1}{r_\a^2} \p^2_{\f_\a}+\frac{1}{r_\a} \p_{r_\a}
 +\frac{\m^2_\a}{4} r_\a^2 \p^2_v \right )  \ , \label{dalam} \ee
 and its scalar eigenfunctions may be taken to be of the form \be
 f_{p^+,p^-}(u,v,r_\a,\f_\a) = e^{i p^+ v + i p^- u} g(r_\a,\f_\a)
 \ . \ee For $p^+ \neq 0$, $g(r_\a,\f_\a)$ is given by the product
 of wave-functions for two harmonic oscillators in two dimensions
 with frequencies $\omega_\a = \left|p^+\right| \mu_\a/2$ \be
 g_{l_\a,m_\a}(r_\a,\f_\a) = \left ( \frac{l_\a!}{2 \pi
 (l_\a+|m_\a|)!} \right )^{\frac{1}{2}} e^{i m_\a \f_\a}
 e^{-\frac{\xi_\a}{2}} \xi_\a^{\frac{|m_\a|}{2}}
 L_{l_\a}^{|m_\a|}(\xi_\a) \ , \label{2dho} \ee with $\xi_\a =
 \frac{\m_\a p^+ r_\a^2}{2}$ and $l_\a \in \mathbb{N}$, $m_\a \in
 \mathbb{Z}$. The resulting eigenvalue is \be \Lambda_{p^+\ne0} =
 2p^+p^- - \sum_{\a=1}^2 \m_\a \left | p^+ \right |
 (2l_\a+|m_\a|+1) \ . \label{daleig} \ee and by comparison with the
 value of the Casimir on the  ${\mathcal H}^L_6 \times {\mathcal
 H}_6^R$ representations we can identify \be p = \left | p^+ \right
 | \ , \hspace{0.68cm} \jh =  p^- - \sum_{\a=1}^2
 \m_\a(2l_\a+|m_\a|)  \ , \hspace{0.68cm} m_\a = n_\a - \bar{n}_\a
 \ , \hspace{0.68cm} l_\a = {\rm Max}(n_\a,\bar{n}_\a)   \ . \ee
 For $p^+=0$ the $g(r_\a,\f_\a)$ can be taken to be Bessel
 functions and they give the decomposition of a plane wave whose
 radial momentum in the two transverse planes is $s_\a^2$,
 $\a=1,2$.

 The representations of the affine Heisenberg algebra
 $\widehat{\mathcal H}_{6}$ that will be relevant for the study of
 string theory in the six-dimensional Hpp-wave are the
 highest-weight representations with a unitary base and some new
 representations with a modified highest-weight condition that we
 will introduce below and that in the case $\m_1=\m_2$ coincide
 with the spectral flowed representations.

 The OPEs in $(\ref{opes})$ correspond to the following commutation
 relations for the $\widehat{\mathcal H}^L_6$ left-moving current
 modes \be [P^+_{\alpha\,n},P^{-\beta}_m] = 2n
 \delta_{\alpha}^{\beta}\, \delta_{n+m}-2i \m_\a
 \delta_{\alpha}^{\beta}K_{n+m} \ , \hspace{0.8cm} [J_n,K_m]=n
 \delta_{n+m,0} \ , \nb \ee \be [J_n,P^+_{\alpha\,m}] = -i \m_\a
 P^+_{\alpha\,n+m}  \ , \hspace{2.5cm} [J_n,P^{-\alpha}_m]=  i
 \m_\a P^{-\alpha}_{n+m} \ . \ee

 There are  three types of highest-weight representations. Affine
 representations based on $V^\pm_{p,\jh}$ representations of the
 horizontal algebra, with conformal dimension \be h = \mp p \jh +
 \frac{\m_1 p}{2}(1-\m_1 p)+ \frac{\m_2 p}{2}(1-\m_2 p) \ , \ee and
 affine representations based on $V^0_{s_1,s_2,\jh}$
 representations, with conformal dimension \be h = \frac{s_1^2}{2}
 + \frac{s_2^2}{2} = {s^2\over 2} \ . \ee

 In the current algebra formalism we can introduce a doublet of
 charge variables and regroup the infinite number of fields that
 appear in a given representation of $\widehat{\mathcal H}^L_6$ in
 a single field \be \Phi_{p,\hj}^+(z;x_{\alpha}) = \sum_{n_1, n_2 =
 0}^{\infty} \prod_{\a=1}^2 \frac{(x_\a \sqrt{\m_\a
 p})^{n_\a}}{\sqrt{n_\a!}} R_{p,\hj; n_1, n_2}^+(z) \ , \quad p>0
 \ , \ee \be \Phi_{p,\hj}^-(z;x^{\alpha}) = \sum_{n_1, n_2 =
 0}^{\infty} \prod_{\a=1}^2 \frac{(x^\a \sqrt{\m_\a
 p})^{n_\a}}{\sqrt{n_\a!}} R_{p,\hj; n_1, n_2}^-(z) \ , \quad p>0 \
 , \ee \be \Phi_{s_1,s_2,\hj}^0(z;x_\a) =
 \sum_{n_1,n_2=-\infty}^{\infty} \prod_{\a=1}^2 (x_\a)^{n_\a}
 R_{s_1,s_2,\jh; n_1,n_2}^0(z) \ , \quad s_1,s_2\geq 0 \ . \ee

 Highest-weight representations of the current algebra lead to a
 string spectrum free from negative norm states only if they
 satisfy the constraint \be {\rm Max}(\m_1 p,\m_2 p)  < 1 \ .
 \label{bound} \ee

 When $\m_1 = \m_2 = \m$ new representations should be considered
 that result from spectral flow of the original representations
 \cite{mo1}. Spectral flowed\index{Spectral flowed irrep. $\mathcal
 H_6$|textbf} representations are highest-weight representations of an
 isomorphic algebra whose modes are related to the original ones by
 {\colsep
 \ba
 \tilde{P}^+_{\a, n} &=& P^+_{\a, n - w} \ , \hspace{1cm}
 \tilde{P}^{- \a}_{n} = P^{- \a}_{n + w} \ , \hspace{1cm}
 \tilde{J}_n = J_n \ , \nb \\
 \tilde{K}_n &=& K_n - i w \d_{n,0} \ ,  \hspace{1cm} \tilde{L}_n =
 L_n - i w J_n \ .
 \ea}
 The long strings\index{Long strings
 $\mathcal H_6$|textbf} in this case can move freely in the two transverse
 planes and correspond to the spectral flowed type 0
 representations, exactly as for the ${\bf H}_4$ NW model
 \cite{dak}.

 In the general case $\m_1 \ne \m_2$ a similar interpretation is
 not possible. However instead of introducing new representations
 as spectral flowed representations we can still define them
 through a modified highest-weight condition. Such \emph{Modified
 Highest-Weight (MHW)}\index{MHW $\mathcal H_6$|textbf}
 representations are a more general concept compared to spectral
 flowed representations, as the analysis for $\mu_1\not=\mu_2$
 indicates.

 In order to understand which kind of representations are needed
 for the description of states with $p$ outside the range
 $(\ref{bound})$, it is useful to resort to a free field
 realization of the $\widehat{\mathcal H}_{2+2n}$ algebras, first
 introduced for the original NW model in \cite{kk}. This
 representation provides an interesting relation between primary
 vertex operators and twist fields in orbifold models. For
 $\widehat{\mathcal H}_6$ we introduce a pair of free bosons
 $u(z),v(z)$ with $\langle v(z)u(w) \rangle  = \log{(z-w)}$ and two
 complex bosons $y_\a(z) = \xi_\alpha(z) + i \eta_\alpha(z) $ and
 $\tilde{y}^\a(z) = \xi_\alpha(z) - i \eta_\alpha(z) $ with
 $\langle y_\a(z)\tilde{y}^\b(w) \rangle = -2 \d^\b_\a
 \log{(z-w)}$. The currents
 {\colsep
 \ba
 J(z) &=& \p v(z) \ , \hspace{2cm} K(z) =  \p u(z) \ , \nb \\
 P^+_\a(z) &=& i e^{-i \m_\a u(z)} \p y_\a(z) \ , \hspace{1cm}
 P^{-\a}(z) = i e^{i \m_\a u(z)} \p \tilde{y}^\a(z) \ , \label{ff1}
 \ea}
 satisfy the $\widehat{\mathcal H}_6$ OPEs $(\ref{opes})$. The
 ground state of a $V^{\pm}_{p,\jh}$ representation is given by the
 primary field
 \be
 R^{\pm}_{p,\jh;0}(z) = e^{i [\jh u(z) \pm p
 v(z)]} {\s}^{\mp}_{\m_1 p}(z) {\s}^{\mp}_{\m_2 p}(z) \ .
 \label{ff2}
 \ee
 The $ {\s}^{\mp}_{\m p}(z)$ are \emph{twist
 fields}\index{Twist fields $\mathcal H_6$|textbf}, characterized
 by the following OPEs
 {\colsep
 \ba
 \p y(z) {\s}^-_{\m p}(w) &\sim&
 (z-w)^{-\m p} \ {\tau}_{\m p}^-(w) \ , \hspace{1cm} \p
 \tilde{y}(z) {\s}^-_{\m p}(w) \sim (z-w)^{-1+\m p} \ {\s}_{\m
 p}^{-
 (1)}(w)\ , \nb \\
 \p y(z) {\s}^+_{\m p}(w) &\sim& (z-w)^{-1+\m p} \ {\s}_{\m p}^{+
 (1)}(w)\ , \hspace{0.4cm} \p \tilde{y}(z) {\s}^+_{\m p}(w) \sim
 (z-w)^{-\m p} \ {\tau}_{\m p}^+(w) \ , \label{ff3}
 \ea}
 where
 ${\tau}_{\m p}^\pm(z)$ and ${\s}_{\m p}^{\pm (1)}(z)$ are excited
 twist fields.  The ground state of a $V_{s_1,s_2,\jh}^0$
 representation is determined by the primary field
 \be
 R^0_{s_1,s_2,\jh;0}(z) = e^{i \jh u(z)} R^0_{s_1}(z)R^0_{s_2}(z) \
 , \label{ff4} \ee where \be R^0_{s_\a}(z) = \frac{1}{2\pi}
 \int_0^{2 \pi} d \t_\a e^{ \frac{i s_\a }{2} \left ( y_\a(z) e^{-i
 \t_\a} + \tilde{y}^\a(z) e^{i \t_\a}  \right )} \ . \label{ff5}
 \ee
 are essentially free vertex operators.

 In analogy with $\widehat{\mathcal H}_4$ we define for arbitrary
 $\m p > 0$
 {\colsep
 \ba
 R^{\pm}_{p,\jh;0}(z) &=& e^{i [\jh u(z) \pm p
 v(z)]} {\s}^{\mp}_{\{\m_1 p\}}(z) {\s}^{\mp}_{\{\m_2 p\}}(z) \ ,
 \hspace{1cm}
 \{ \m_1 p \} \ne 0 \ , \{ \m_2 p \} \ne 0 \ , \nb \\
 R^{\pm}_{p,\jh,s_1;0}(z) &=& e^{i [\jh u(z) \pm p v(z)]}
 R^0_{s_1}(z) {\s}^{\mp}_{\{\m_2 p\}}(z) \ ,
 \hspace{1.4cm}
 \{ \m_1 p \} = 0 \ , \{ \m_2 p \} \ne 0 \ , \\
 R^{\pm}_{p,\jh,s_2;0}(z) &=& e^{i [\jh u(z) \pm p v(z)]}
 {\s}^{\mp}_{\{\m_1 p\}}(z) R^0_{s_2}(z) \ ,
 \hspace{1.4cm} \{ \m_1
 p \} \ne 0 \ , \{ \m_2 p \} = 0 \ , \nb \label{ff6}
 \ea}
 where $[\m p]$
 and $\{ \m p \}$ are the integer and fractional part of $\m p$
 respectively. Quantization of the model in the light-cone gauge
 shows that the resulting string spectrum is unitary. {}From the
 current algebra point of view the states that do not satisfy the
 bound $(\ref{bound})$ belong to new representations which satisfy
 a modified highest-weight condition and are defined as follows.
 When $K_0 |p,\jh \rangle = i \m p |p,\jh \rangle$ with  $\{ \m_\a
 p \} \ne 0$, $\a =1, 2$, the affine representations we are
 interested in are defined by
 {\colsep
 \ba P^+_{\a, \ n} |p,\jh\rangle &=& 0
 \ , \hspace{0.4cm} n \ge - [\m_\a p] \ , \hspace{1cm} P^{-\a}_{n}
 |p,\jh\rangle = 0 \ , \hspace{0.4cm} n \ge 1 + [\m_\a p] \ ,
 \nb \\
 J_{n} |p,\jh\rangle &=& 0 \ , \hspace{0.4cm} n \ge 1 \ ,
 \hspace{1cm} K_{n} |p,\jh\rangle = 0 \ , \hspace{0.4cm} n \ge 1  \
 . \label{ff7}
 \ea}
 Similarly when $K_0 |p,\jh \rangle = - i \m p
 |p,\jh \rangle$ with  $\{ \m_\a p \} \ne 0$, $\a =1, 2$, the
 affine representations we are interested in are defined by
 {\colsep
 \ba
 P^+_{\a, \ n} |p,\jh\rangle &=& 0 \ , \hspace{0.4cm} n \ge  1 +
 [\m_\a p] \ , \hspace{1cm} P^{-\a}_{n} |p,\jh\rangle = 0 \ ,
 \hspace{0.4cm} n \ge - [\m_\a p] \ ,  \nb
 \\
 J_{n} |p,\jh\rangle &=& 0 \ , \hspace{0.4cm} n \ge 1 \ ,
 \hspace{1cm} K_{n} |p,\jh\rangle = 0 \ , \hspace{0.4cm} n \ge 1  \
 . \label{ff8}
 \ea}
 Finally whenever either  $\{ \m_1 p \} = 0$ or
 $\{ \m_2 p \} = 0$ we introduce new ground states $|p,s_1,\jh
 \rangle$ and  $|p,s_2,\jh \rangle$ which satisfy the same
 conditions as in $(\ref{ff7})$, $(\ref{ff8})$ except that
 \be
 P^+_{\a, \ n} |p,\jh,s_\a \rangle = 0 \ , \hspace{0.4cm} n \ge -
 [\m_\a p] \ , \hspace{1cm} P^{- \a}_{n} |p,\jh,s_\a \rangle = 0 \
 , \hspace{0.4cm} n \ge  [\m_\a p] \ , \label{ff9}
 \ee
 for either $\a =1$ or $\a=2$.

 These states correspond to strings that do not feel any more the
 confining potential in one of the two transverse planes. The
 presence of these states in the spectrum can be justified along
 similar lines as for AdS$_3$ \cite{mo1} or the ${\bf H}_4$
 \cite{kirpio,dak} WZW models.

 % OOOOOOOOOOOOOOOOOOOOOOOOOOOOOOOO     3 POINT FUNCTIONS  OOOOOOOOOOOOOOOOOOOOOOOOOOO

 % OOOOOOOOOOOOOOOOOOOOOOOOOOOOOOOO     3 POINT FUNCTIONS  OOOOOOOOOOOOOOOOOOOOOOOOOOO

 \section{Three-point functions}
 \label{3point}
 \noindent

 We now turn to compute the simplest interactions in the Hpp-wave,
 encoded in the three-point functions of the scalar (tachyon)
 vertex operators identified in the previous section. We will
 initially discuss the non symmetric $\mu_1\neq \mu_2$ case, where
 global Ward identities can be used to completely fix the form of
 the correlators. We will then address the $SU(2)_I$ symmetric case
 and argue that the requirement of non-chiral $SU(2)_I$ invariance
 is crucial in getting a unique result. We will finally describe
 the derivation of the two and three-point functions starting from
 the corresponding quantities in AdS$_3$$\times$S$^3$.\\

 In the last section we have seen that the primary fields of the
 $\widehat {\mathcal H}^L_6\times \widehat {\mathcal H}^R_6$ affine
 algebra are of the form \beq \Phi_{\n}^{a}(z,\bz;x,\bx) \ , \eeq
 where $a={\pm,0}$ labels the type of representation and $\n$
 stands for the charges that are necessary in order to completely
 specify the representation, {\it i.e}. $\n=(p,\hj)$ for $V^{\pm}$
 and $\n=(s_1,s_2,\hj)$ for $V^0$. Finally $x$ stands for the
 charge variables we introduced to keep track of the states that
 form a given representation: $x = x_\a$ for ${V}^{+}$, $x = x^\a$
 for ${V}^{-}$ and  $x = x_\a$ with $x_\a = 1/x^\a$ ({\it i.e.}
 $x_\a = e^{i\phi_\a}$) for ${V}^{0}$ . In the following we will
 leave the dependence of the vertex operators on the
 anti-holomorphic variables $\bar z$ and $\bar x$ understood. The
 OPE between the currents and the primary vertex operators can be
 written in a compact form \be {\cal J}^A(z) \F^a_\n(w;x) = {\cal
 D}^A_a \frac{\F^a_\n(w;x)}{z-w} \ , \ee where $A$ labels the six
 $\widehat{\mathcal H}_6$ currents and the $ {\cal D}^A_a$ are the
 differential operators that realize the action of ${\cal J}^A_0$
 on a given representation $(a,\n)$, according to $(\ref{difp})$,
 $(\ref{difm})$ and $(\ref{dif0})$.

 We fix the normalization of the operators in the
 ${V}^{\pm}_{p_1,\hj_1}$ representations by choosing the overall
 constants in their two-point functions, which are not determined
 by the world-sheet or target space symmetries, to be such that \be
 \langle\Phi^+_{p_1,\hj_1}(z_1,x_{1\alpha})\Phi^-_{p_2,\hj_2}(z_2,x_2^{\alpha})
 \rangle = \frac{|\prod_{\a=1}^2 e^{-p_1 \m_\a
 x_{1\alpha}x_2^{\alpha}}|^2}
 {|z_{12}|^{4h}}\delta(p_1-p_2)\delta(\hj_1+\hj_2) \ , \label{pm}
 \ee where we introduced the shorthand notation
 $f(z,x)f(\bz,\bx)=|f(z,x)|^2$. Similarly, the other non-trivial
 two-point functions are chosen to be \be
 \langle\Phi^0_{s_{1\alpha},\hj_1}(z_1,x_{1\alpha})
 \Phi^0_{s_{2\alpha},\hj_2}(z_2,x_{2\alpha})\rangle = \small
 (2\pi)^4 \prod_{\alpha=1,2} {\delta(s_{1\alpha}- s_{2\alpha})
 \over s_{1\alpha}} \delta(\phi_{1\alpha} - \phi_{2\alpha}-\pi)
 \delta(\bar\phi_{1\alpha} - \bar\phi_{2\alpha}-\pi)
 \d(\jh_1+\jh_2) \ , \ee where we set $x_{i\alpha} =
 e^{i\phi_{i\alpha}}$.

 Three-point functions, denoted by $G_{abc}(z_i,x_i)$ or more
 simply by $\langle a b c\rangle$ in the following, are determined
 by conformal invariance on the world-sheet to be of the form \be
 \label{three-point}
 \langle\Phi_{\n_1}^a(z_1,x_1)\Phi_{\n_2}^b(z_2,x_2)
 \Phi_{\n_3}^c(z_3,x_3)\rangle=
 \frac{C_{abc}(\n_1,\n_2,\n_3)K_{abc}(x_1,x_2,x_3)}
 {|z_{12}|^{2(h_1+h_2-h_3)}|z_{13}|^{2(h_2+h_3-h_2)}|z_{23}|^{2(h_2+h_3-h_1)}}
 \ , \ee where $C_{abc}$ are the quantum structure constants of the
 CFT and  the `kinematical' coefficients $K_{abc}$ contain all the
 dependence on the ${\mathcal H}^L_6\times {\mathcal H}^R_6$ charge
 variables $x$ and $\bx$. For generic values of $\m_1$ and $\m_2$
 ($\frac{\m_1}{\m_2} \notin \mathbb{Q}$), the functions $K_{abc}$
 are completely fixed by the global Ward identities, as it was the
 case for the ${\bf H}_4$ WZW model \cite{dak}. When $\m_1 = \m_2$
 we will have to impose the additional requirement of $SU(2)_I$
 invariance. An important piece of information for understanding
 the structure of the three-point couplings is provided by the
 decomposition of the tensor products between representations of
 the ${\mathcal H}_6$ horizontal algebra
 {\colsep
 \ba
 V^+_{p_1,\jh_1}
 \otimes V^+_{p_2,\jh_2} &=& \sum_{n_1,n_2=0}^\infty
 V^+_{p_1+p_2,\jh_1+\jh_2+\m_1 n_1+\m_2n_2}
 \ , \nb \\
 V^+_{p_1,\jh_1} \otimes V^-_{p_2,\jh_2} &=&
 \sum_{n_1,n_2=0}^\infty V^+_{p_1+p_2,\jh_1+\jh_2-\m_1 n_1-\m_2n_2}
 \ , \hspace{0.5cm}  p_1 > p_2  \ , \nb \\
 V^+_{p_1,\jh_1} \otimes V^-_{p_2,\jh_2} &=&
 \sum_{n_1,n_2=0}^\infty V^-_{p_1+p_2,\jh_1+\jh_2+\m_1 n_1+\m_2n_2}
 \ , \hspace{0.5cm} p_1 < p_2 \ .  \label{tensorH6}
 \ea}
 Note that
 when $\m_1=\m_2$ there are $n+1$ terms with the same $\jh =
 \jh_1+\jh_2 \pm \m n$ in $(\ref{tensorH6})$. The existence of this
 multiplicity is precisely what is necessary in order to obtain
 $SU(2)_I$ invariant couplings, as we will explain in the
 following. We will also need
 {\colsep
 \ba
 V^+_{p \, ,\jh_1} \otimes V^-_{p
 \, ,\jh_2} &=& \int_0^\infty s_1 ds_1
 \int_0^\infty s_2 ds_2 V^0_{s_1,s_2,\jh_1+\jh_2} \ , \nb \\
 V^+_{p_1,\jh_1} \otimes V^0_{s_1,s_2,\jh_2} &=&
 \sum_{n_1,n_2=-\infty}^\infty V^+_{p_1+p_2,\jh_1+\jh_2+\m_1
 n_1+\m_2n_2} \ . \label{tensorH6b}
 \ea}

 Let us first discuss the generic case $\m_1 \ne \m_2$, starting
 from $\langle++-\rangle$. According to $(\ref{tensorH6})$ this
 coupling is non-vanishing only when $p_1+p_2 = p_3$ and $L =
 -(\jh_1+\jh_2+\jh_3) = \m_1 q_1 + \m_2 q_2$, with $q_1,q_2\in
 \mathbb{N}$. The global Ward identities can be unambiguously
 solved and the result is\footnote{The standard $\delta$-function
 for the Cartan conservation rules are always implied. We do not
 write them explicitly. } \be K_{++-}(q_1,q_2) = \left |
 \prod_{\a=1}^2 e^{-\m_\a
 x^\a_3(p_1x_{1\a}+p_2x_{2\a})}(x_{2\a}-x_{1\a})^{q_\a} \right |^2
 \label{cgppm}\ . \ee

 The corresponding three-point couplings are \be {\it
 C}_{++-}(q_1,q_2) = \prod_{\a=1}^2\frac{1}{q_\a!} \left [
 \frac{\g(\m_\a p_3)}{\g(\m_\a p_1)\g(\m_\a p_2)} \right
 ]^{\frac{1}{2}+q_\a} \ , \label{qpppm}  \ee where $\gamma(x) =
 \Gamma(x)/ \Gamma(1-x)$. All other couplings that only involve
  $\F^\pm$ vertex operators follow from $(\ref{cgppm})$,
 $(\ref{qpppm})$ by permutation of the indices and by using the
 fact that $K_{++-}{C}_{++-} \rightarrow K_{--+}{C}_{--+}$ up to
 the exchange $x_i^\a \leftrightarrow x_{i\a}$ and the inversion of
 the signs of all the $\jh_i$.

 Similarly the $\langle +-0 \rangle$ coupling can be non-zero only
 when $p_1=p_2$ and $L = -(\jh_1+\jh_2+\jh_3) = \sum_\alpha \m_\a
 q_\a$, with $q_\a \in \mathbb{Z}$. Global Ward identities yield

\be K_{+-0} = \left | \prod_{\a=1}^2 e^{- \m_\a p_1
x_{1\a}x_{2}^\a-\frac{s_\a}{\sqrt{2}} \left (x_2^\a
x_{3\a}+x_{1\a} x_3^\a \right)} x_{3\a}^{q_\a} \right |^2 \ .
\label{cpmo1}  \ee  Moreover \be
{C}_{+-0}(p,\jh_1;p,\jh_2;s_1,s_2,\jh_3) = \prod_{\a=1}^2
e^{\frac{s_\a^2}{2}[\psi(\m_\a p)+\psi(1-\m_\a p)-2\psi(1)]} \ ,
\label{cpmo2} \ee where $\psi(x) = \frac{d \ln{\G(x)}}{dx}$ is the
\emph{digamma function}\index{Digamma function|textbf}.

Finally the coupling between three $\F^0$ vertex operators simply
reflects momentum conservation in the two transverse planes.
Therefore it is non-zero only when  \be s_{3\a}^2 =
s_{1\a}^2+s_{2\a}^2+2s_{1\a}s_{2\a} \cos{\xi_{\a}} \ ,
\hspace{1cm} s_{3\a}e^{i\h_{\a}} = -s_{1\a}-s_{2\a}e^{i \xi_\a} \
, \hspace{0.6cm} \a = 1, 2 \ , \label{conserv} \ee where
$\xi_\a=\phi_{2\a}-\phi_{1\a}$ and $\h_\a=\phi_{3\a}-\phi_{1\a}$.
It can be written as \be K_{000}(\phi_{1\a},\phi_{2\a},\phi_{3\a})
= \prod_{\a=1}^2 \frac{ 8 \pi \d(\xi_\a+\bar{\xi}_\a)\d(\h_\a +
\bar{\h}_\a)}{\sqrt{4s_{1\a}^2s_{2\a}^2-(s_{3\a}^2-s_{1\a}^2-s_{2\a}^2)^2}}
e^{-iq_\a(\phi_{1\a}+\bar \phi_{1\a})} \ , \label{cooo} \ee where
the angles $\xi_\a$ and  $\h_\a$ are fixed by the Eqs.
(\ref{conserv}) and again $L= \sum_\a \m_\a q_\a$ with $q_\a \in
\mathbb{Z}$.

 As discussed in section \ref{ppw}, when $\m_1=\m_2=\m$ the plane
 wave background displays an additional $SU(2)_I$ symmetry. At the
 same time we see from $(\ref{tensorH6})$ that there are also new
 possible couplings and they precisely combine to give an $SU(2)_I$
 invariant result. Let us start again from three-point couplings
 containing only $\F^\pm$ vertex operators. In this case the
 $SU(2)_I$ invariant result is obtained after summing over all the
 couplings $C_{++-}(q_1,q_2)$ with $(q_1+q_2)=L/\mu = Q$
 {\colsep
 \ba \label{su2ppm} K_{++-}(Q){\it C}_{++-}(Q) &=& \sum_{q_1=0}^{Q}
 K_{++-}(q_1,Q-q_1) C_{++-}(q_1,Q-q_1)  \\
 &=& \frac{1}{Q!}\left [ \frac{\g(\m p_3)}{\g(\m p_1)\g(\m p_2)}
 \right ]^{\frac{1}{2}+Q}\left |e^{-\m \sum_{\a=1}^2
 x_3^\a(p_1x_{1\a}+p_2x_{2\a})} \right |^2 ||x_{2}-x_{1}||^{2Q} \ ,
 \nb
 \ea}
 where $||x||^2 \equiv \sum_{\a} |x_\a|^2$ is indeed
 $SU(2)_I$ invariant.

 Similarly the $\langle +-0 \rangle$ correlator becomes, after
 summing over $q_1 \in \mathbb{Z}$ ,
 {\colsep
 \ba & &
 K_{+-0}(Q){C}_{+-0}(p,\jh_1;p,\jh_2;s_1,s_2,\jh_3) =
 \prod_{\a=1}^2  \left | e^{- \m p_1
 x_{1\a}x_{2}^\a-\frac{s_\a}{\sqrt{2}} \left (x_2^\a
 x_{3\a}+x_{1\a} x_3^\a \right)} \right |^2
 \left ( \frac{||x_{3}||^2}{2} \right )^{Q} \nb \\
 & &
 %\d\left(\sum_{\a}x_{3\a}\bar{x}_{3}^{\a}\right )
 e^{\frac{s_1^2+s_2^2}{2}[\psi(\m p)+\psi(1-\m p)-2\psi(1)]} \ ,
 \label{cpmo2}
 \ea}
 with the constraint $x_{31}\bar{x}_3^1 =
 x_{32}\bar{x}_3^2$. The $\langle 000 \rangle$ coupling gets
 similarly modified.

 % OOOOOOOOOOOOOOOOOOOOOOOOOOOOOOOO     4 POINT FUNCTIONS  OOOOOOOOOOOOOOOOOOOOOOOOOOO

 % OOOOOOOOOOOOOOOOOOOOOOOOOOOOOOOO     4 POINT FUNCTIONS  OOOOOOOOOOOOOOOOOOOOOOOOOOO

 \section{Four-point functions}
 \label{4point}
 \noindent

Four-point correlation functions of worldsheet primary operators
are computed in this section by solving the relevant
\emph{Knizhnik - Zamolodchikov (KZ) equations}\index{KZ
equations|textbf} \cite{kz}. As we will explain the resulting
amplitudes are a simple generalization of the amplitudes of the
${\bf H}_4$ WZW model. In appendix \ref{app A} the same results
will be reproduced by resorting to the Wakimoto free-field
representation. As in the previous section we find it convenient
to first discuss the non-symmetric ($\mu_1\neq\mu_2$) case and
then pass to the symmetric ($\mu_1=\mu_2$) case where $SU(2)_I$
invariance is needed in order to completely fix the correlators.

In general, world-sheet conformal invariance and global Ward
identities allow us to write \be G(z_i, \bar{z}_i, x_{i},
\bar{x}_{i}) = \prod_{i<j}^4 |z_{ij}|^{2 \left
(\frac{h}{3}-h_i-h_j \right) } K(x_{i},\bar{x}_{i}){\cal
G}(z,\bar{z}, x, \bar{x}) \ , \label{f1} \ee where
$h=\sum_{i=1}^4h_i$ and the $SL(2,\mathbb{C})$ invariant
cross-ratios $z$, $\bar{z}$ are defined according to \be z =
\frac{z_{12}z_{34}}{z_{13}z_{24}} \ , \hspace{1cm} \bar z =
\frac{\bar{z}_{12}\bar{z}_{34}} {\bar{z}_{13}\bar{z}_{24}} \ .
\label{f2} \ee The form of the function $K$ and the expression of
the $\widehat{\mathcal H}_6$ invariants $x$ in terms of the $x_i$
are fixed by the global symmetries but are different for different
types of correlators and therefore their explicit form will be
given in the next sub-sections.

The four-point amplitudes are non trivial only when \be L = -
\sum_{i=1}^4 \jh_i = \m_1q_1 + \m_2 q_2 \ , \label{f3} \ee for
some integers $q_\a$. In the generic case for a given $L$ these
integers are uniquely fixed and the Ward identities fix the form
of the functions $K$ up to a function of two ${\mathcal H}_6$
invariants\footnote{Sometimes we will collectively denote the
${\mathcal H}_6$ invariants $x_1$ and $x_2$ by $x_\a$ with
$\a=1,2$. They should not be confused with the components of the
charge variables $x_{i\a}$ that carry an additional label
associated to the insertion point $i=1, ..., 4$.} $x_1$ and $x_2$.
The KZ equations can be schematically written in the following
form \be \p_z {\cal G}(z,x_1,x_2) = \sum_{\a=1}^2 D_{{\mathcal
H}_4,q_\a}(z,x_\a){\cal G}(z,x_1,x_2) \ , \label{f4} \ee where the
$D_{{\mathcal H}_4,q_\a}$ are differential operators closely
related to those that appear in the KZ equations for the NW model
based on the $\widehat{\mathcal H}_4$ affine algebra \cite{dak}.
The equations are therefore easily solved by setting \be {\cal
G}_{q_1,q_2}(z,x_1,x_2) = {\cal G}_{{\mathcal
H}_4,q_1}(z,x_1){\cal G}_{{\mathcal H}_4,q_2}(z,x_2) \ .
\label{f5} \ee

When $\m_1=\m_2$, there are several integers that satisfy
$(\ref{f3})$ and the $SU(2)_I$ invariant correlators can be
obtained by summing over all possible pairs $(q_1,q_2)$ such that
$(q_1+q_2)=L/\mu = Q$ \be {\cal G}_{Q}(z,x_1,x_2) = \sum_{q_1=0}^Q
{\cal G}_{{\mathcal H}_4,q_1}(z,x_1){\cal G}_{{\mathcal
H}_4,Q-q_1}(z,x_2) \ . \label{f6} \ee This is the same procedure
we used for the three-point functions and reflects the existence
of new couplings between states in $\widehat{\cal H}_6$
representations at the enhanced symmetry point.  In the following
we will describe the various types of four-point correlation
functions.

\subsection{$\langle +++- \rangle$ correlators}

Consider a correlator of the form \be G_{+++-} = \langle
\F^+_{p_1,\jh_1} \F^+_{p_2,\jh_2} \F^+_{p_3,\jh_3}
\F^-_{p_4,\jh_4} \rangle \ , \hspace{1cm} p_1+p_2+p_3 = p_4 \ .
\label{ex1} \ee This is the simplest `extremal' case. {}From the
decomposition of the tensor products of ${\mathcal H}_6$
representations displayed in Eq. (\ref{tensorH6}) it follows that
the correlator vanishes for $L < 0$ while for $L \ge 0$, $L = \m_1
q_1 + \m_2 q_2$  it decomposes into the sum of a finite number $N=
(q_1+1)(q_2+1)$ of conformal blocks which correspond to the
propagation in the $s$-channel of the representations
$\F^+_{p_1+p_2,\jh_1+\jh_2+\m_1 n_1+\m_2 n_2}$ with $n_1 = 0, ...,
q_1$ and $n_2 = 0,..., q_2$. Global ${\mathcal H}_6$ symmetry
yields \be K(q_1,q_2) =\prod_{\a=1}^2\left|e^{-\m_\a
x_{4}^\a(p_1x_{1\a}+p_2x_{2\a}+p_3x_{3\a})} \right |^2
|x_{3\a}-x_{1\a}|^{2q_\a}  \label{ex2} \ , \ee up to a function of
the two invariants ($\alpha =1,2$) \be x_\a =
\frac{x_{2\a}-x_{1\a}}{x_{3\a}-x_{1\a}} \ . \ee We decompose the
amplitude in a sum over the conformal blocks and write
 \be {\cal
 G}_{q_1,q_2}(z, \bar z, x_\a,\bar{x}^\a) \sim  \sum_{n_1=0}^{q_1}
 \sum_{n_2=0}^{q_2} {\cal F}_{n_1,n_2}(z,x_\a) \bar{{\cal
 F}}_{n_1,n_2} (\bar{z},\bar{x}^\a) \ . \label{ex3}
 \ee
 We set
 ${\cal F}_{n_1,n_2} = z^{\ka_{12}}(1-z)^{\ka_{14}}F_{n_1,n_2}$
 where
 {\colsep
 \ba \ka_{12} &=&
 h_1+h_2-\frac{h}{3}-\jh_2p_1-\jh_1p_2-(\m_1^2+\m_2^2)p_1p_2 \ ,
   \\
 \ka_{14} &=&
 h_1+h_4-\frac{h}{3}-\jh_4p_1+\jh_1p_4+(\m_1^2+\m_2^2)p_1p_4
 -(\m_1+\m_2)p_1+L(p_2+p_3) \ , \nb \label{ex4}
 \ea}
 and where the
 $F_{n_1,n_2}$ satisfy the following KZ equation
 {\colsep
 \ba && \p_z
 F_{n_1,n_2}(z,x_1,x_2) = \frac{1}{z} \sum_{\a=1}^2 \m_\a \left [
 -(p_1x_\a+p_2x_\a(1-x_\a))\p_{x_\a} - q_\a p_2x_\a \right ]
 F_{n_1,n_2}(z,x_1,x_2) \nb \\
 &-& \frac{1}{1-z} \sum_{\a=1}^2 \m_\a \left [
 (1-x_\a)(p_2x_\a+p_3)\p_{x_\a} - q_\a p_2(1-x_\a) \right ]
 F_{n_1,n_2}(z,x_1,x_2) \ . \label{ex5}
 \ea}
 The explicit form of
 the conformal blocks is
 \be
 F_{n_1,n_2}(z,x_1,x_2) =
 \prod_{\a=1}^2 f(\m_\a,z,x_\a)^{n_\a} g(\m_\a,z,x_\a)^{q_\a-n_\a}
 \ , \hspace{0.6cm} n_\a = 0, ..., q_\a  \ . \label{ex6}
 \ee
 Here
 {\colsep
 \ba
 f(\m_\a,z,x_\a) &=& \frac{\m_\a
 p_3}{1-\m_\a(p_1+p_2)}z^{1-\m_\a(p_1+p_2)}\f_0(\m_\a) - x_\a
 z^{-\m_\a(p_1+p_2)}\f_1(\m_\a) \ , \nb \\
 g(\m_\a,z,x_\a) &=& \g_0(\m_\a) -\frac{x_\a p_2}{p_1+p_2}
 \g_1(\m_\a) \ , \label{ex7}
 \ea}
 and
 {\colsep
 \ba \label{ex8} \f_0(\m) &=&
 F(1\!\!-\!\!\m p_1,1+\m p_3,2\!\!-\!\!\m (p_1 + p_2),z) \ ,
 \hspace{0.3cm}
 \g_0(\m) = F(\m p_2,\m p_4,\m (p_1+p_2),z) \   \\
 \f_1(\m) &=& F(1\!\!-\!\!\m p_1,\m p_3,1\!\!-\!\!\m(p_1+p_2),z) \
 , \hspace{0.3cm} \g_1(\m) = F(1+\m p_2,\m p_4,1+\m (p_1+p_2),z) \
 , \nb
 \ea}
 where $F(a,b,c,z)$ is the standard $_1F_2$
 hypergeometric function.

 We can now reconstruct the four-point function as a monodromy
 invariant combination of the conformal blocks and the result is
 \colsep\beqa{\cal G}_{q_1,q_2}(z,\bar{z},x_\a, \bar{x}^\a) &=& |z|^{2
 \ka_{12}}|1-z|^{2 \ka_{14}} \prod_{\a=1}^2
 \frac{\sqrt{\tau(\m_\a)}}{q_\a!}  [
 C_{12}(\m_\a)|f(\m_\a,z,x_\a)|^2 + \nn\\
 &+&C_{34}(\m_\a)|g(\m_\a,z,x_\a)|^2 ]^{q_\a} \label{cor-pppm}
 \eeqa}
\noindent
 where $\tau(\m)
 =C_{12}(\m)C_{34}(\m) $ and
 \be
 C_{12}(\m) =
 \frac{\g(\m(p_1+p_2))}{\g(\m p_1)\g(\m p_2)} \ , \hspace{1cm}
 C_{34}(\m) = \frac{\g(\m p_4)}{\g(\m p_3)\g(\m (p_4-p_3))} \ .
 \ee
 When $\m_1=\m_2=\mu$ we set $Q=L/\mu=\sum_{\a}q_{\a}$ and find the
 $SU(2)_I$ invariant combination
 {\colsep
 \ba  & & {\it
 K}_Q(x_\a,\bar{x}^\a) {\cal G}_Q(z,\bar{z},x_\a,\bar{x}^\a) =
 \sum_{q_1=0}^{Q}{\it
 K}(q_1,Q-q_1){\cal G}_{q_1,Q-q_1}(z,\bar{z},x_\a, \bar{x}^\a)  \nb \\
 &=& |z|^{2 \ka_{12}}|1-z|^{2 \ka_{14}} \frac{\tau(\m)}{Q!}
 \prod_{\a=1}^2
 \left|e^{-\m x_{4}^\a(p_1x_{1\a}+p_2x_{2\a}+p_3x_{3\a})} \right |^2  \times\nn \\
 &\times & \left [ \sum_{\a=1}^2 \left (
 C_{12}(\m)|x_{13\a}f(\m,z,x_\a)|^2+C_{34}(\m)|x_{13\a}g(\m,z,x_\a)|^2
 \right ) \right ]^Q \ . 
 \ea}

 \subsection{$\langle +-+- \rangle$ correlators}

 The next class of correlators we want to discuss is of the
 following form
 \be
 G_{+-+-} = \langle \F^+_{p_1,\jh_1}
 \F^-_{p_2,\jh_2} \F^+_{p_3,\jh_3} \F^-_{p_4,\jh_4} \rangle \ ,
 \hspace{1cm} p_1+p_3=p_2+p_4 \ . \label{pm1}
 \ee
 Also in this case
 we write $L=-\sum_i \jh_i=\sum_\alpha \m_\alpha q_\alpha $. The
 Ward identities give
 \be
 K(q_1,q_2) = \prod_{\a=1}^2 \left |
 e^{-\m_\a p_2 x_{1\a}x^\a_2-\m_\a p_3x_{3\a}x^\a_4
 -\m_\a(p_1-p_2)x_{1\a}x^\a_4} (x_{1\a}-x_{3\a})^{q_\a} \right |^2
 \ , \label{pm3}
 \ee
 and the two invariants (no sum over $\alpha
 =1,2$)
 \be
 x_\a = (x_{1\a}-x_{3\a})(x^\a_2-x^\a_4) \ . \label{pm4}
 \ee
 We pass to the conformal blocks and set  ${\cal F}_{n_1,n_2} =
 z^{\ka_{12}}(1-z)^{\ka_{14}}F_{n_1,n_2}$ where
 {\colsep
 \ba \ka_{12} &=&
 h_1+h_2-\frac{h}{3}+(\m_1^2+\m_2^2)p_1p_2-\jh_2p_1+\jh_1p_2
 -(\m_1+\m_2)p_2 \ ,  \nb \\
 \ka_{14} &=&
 h_1+h_4-\frac{h}{3}+(\m_1^2+\m_2^2)p_1p_4-\jh_4p_1+\jh_1p_4-(\m_1+\m_2)p_4
 \ . \label{pm5}
 \ea}
 The functions $F_{n_1,n_2}$ solve the
 following KZ equation
 {\colsep
 \ba
 && z(1-z) \p_z F_{n_1,n_2}(z,x_1,x_2) =
 \sum_{\a=1}^2 \left [ x_\a\p^2_{x_\a} +\left ( \m_\a(p_1-p_2)
 x_\a+1+q_\a \right )\p_{x_\a}
 \right ] F_{n_1,n_2}(z,x_1,x_2) \nb \\
 &+&  z \sum_{\a=1}^2 \left [ -\m_\a (p_1+p_3) x_\a \p_{x_\a} +
 x_\a \m_\a^2 p_2p_3 - (1+q_\a) \m_\a p_3 \right ]
 F_{n_1,n_2}(z,x_1,x_2) \ . \label{kz-pmpm-nf}
 \ea}
 The conformal
 blocks are very similar to the conformal blocks for the ${\bf
 H}_4$ WZW model \cite{dak} \be F_{n_1,n_2}(z,x_1,x_2) =
 \prod_{\a=1}^2 \n_{n_\a} \frac{e^{\m_\a x_\a z p_3 -z(1-z) \m_\a
 \p \ln{f_1(\m_\a,z)}}} {(f_1(\m_\a, z))^{1+q_\a}}
 L^{q_\a}_{n_\a}[x_\a  g(\m_\a,z)] \left (
 \frac{f_2(\m_\a,z)}{f_1(\m_\a,z)} \right )^{n_\a} \ , \label{pm98}
 \ee where $n_\a \in \mathbb{N}$ and $L^{q}_{n}$ is the n-th
 generalized Laguerre polynomial. We also introduced the functions
 {\colsep
 \ba
 f_1(\m,z) &=& F(\m p_3,1-\m p_1,1-\m p_1+\m p_2,z) \ , \nb \\
 f_2(\m,z) &=& z^{\m (p_1-p_2)}F(\m p_4,1-\m p_2,1-\m p_2+\m p_1,z)
 \ , \label{pm8}
 \ea}
 and \be g = -z(1-z) \p \ln{\left ( f_2/f_1
 \right )} \ , \hspace{1cm} \n_{n_\a} =
 \frac{n_\a!}{[\m_\a(p_1-p_2)]^{n_\a}} \ . \ee The four-point
 correlator can be written in a compact form using the combination
 \be S(\m_\a,z,\bar{z}) = |f_1(\m_\a,z)|^2 - \rho(\m_\a)
 |f_2(\m_\a,z)|^2 \ , \hspace{1cm} \rho(\m) = \frac{
 \tilde{C}_{12}(\m) \tilde{C}_{34}(\m)}{\m^2(p_1-p_2)^2} \ ,
 \label{pm2} \ee where we defined \be  \tilde{C}_{12}(\m) =
 \frac{\g(\m p_1)}{\g(\m p_2)\g(\m(p_1-p_2))} \ , \hspace{1cm}
 \tilde{C}_{34}(\m) = \frac{\g(\m p_4)}{\g(\m p_3)\g(\m(p_4-p_3))}
 \ . \label{pm9} \ee The four-point function reads
 {\colsep
 \ba {\cal
 G}_{q_1,q_2}(z,\bar{z},x_\a, \bar{x}^\a) &=& |z|^{2\ka_{12}}
 |1-z|^{2\ka_{14}}\prod_{\a=1}^2
 \frac{\tau(\m_\a,q_\a)}{S(\m_\a,z)} \left |e^{ \m_\a p_3 x_\a
 z-x_\a z(1-z)\p_z \ln{S(\m_\a,z)}} \right |^2\times \nb \\
 &\times& |x_\a z^{b_\a}(1-z)^{c_\a}|^{-q_\a}I_{q_\a}(\zeta_\a) \ ,
 \label{pm12}
 \ea}
 where $I_{q}(\zeta)$ is a modified Bessel
 function and \be \zeta_\a = \frac{2\sqrt{\rho(\m_a)}|\m_\a
 (p_1-p_2) x_\a z^{b_\a}(1-z)^{c_\a}|}{S(\m_\a, z)} \ ,
 \hspace{1cm} \tau(\m,q) =
  \tilde{C}_{12}(\m)^{\frac{1-q}{2}}  \tilde{C}_{34}(\m)^{\frac{1+q}{2}} \ .
 \label{pm11} \ee When $\m_1=\m_2=\mu$ the $SU(2)_I$ invariant
 correlator is given by the sum over $q_1 \in \mathbb{Z}$ with $q_2
 = Q - q_1$ and $Q=L/\mu$. The addition formula for Bessel
 functions leads to
 {\colsep
 \ba && K_{Q}(x_\a,\bar x^\a){\cal
 G}_Q(z,\bar{z},x_\a, \bar{x}^\a) = \frac{\tau(\m,Q)
 |z|^{2\ka_{12}- b Q }|1-z|^{2\ka_{14}- c Q }}{S(\mu,z)^2} \times \\
 &\times& \prod_{\a=1}^2 \left | e^{-\m_\a p_2 x_{1\a}x^\a_2-\m_\a
 p_3x_{3\a}x^\a_4 -\m_\a(p_1-p_2)x_{1\a}x^\a_4} \right |^2
 \frac{||x_{13}||^{Q}}{||x_{24}||^{Q}} \left |e^{x z [\mu p_3 -
 (1-z) \p_z \ln{S}(\mu,z)]} \right |^2 I_{Q}(\zeta) \ ,
 \label{pmsu2inv} \nb
 \ea}
 where
 \be \zeta =
 \frac{2\sqrt{C_{12}C_{34}}|z^b(1-z)^c|}{S(\mu,z)} ||x_{13}||
 ||x_{24}|| \ , \ee and $x= x_{13}\cdot x_{24} = \sum_{\a}
 (x_{1\a}-x_{3\a})(x^\a_2-x^\a_4)$ as well as $||x_{ij}||^2 =
 \sum_{\a} |x_{i\a} - x_{j\a}|^2$ are $SU(2)_I$ invariant.

 The factorization properties of these correlators can be analyzed
 following \cite{dak}. In this way one can check that the modified
 highest-weight representations introduced  actually appear in the intermediate channels.

 \subsection{$ \langle ++- \ 0 \rangle$ correlators}

 Let us describe now a correlator of the form \be G_{++-0} =
 \langle \F^+_{p_1,\jh_1} \F^+_{p_2,\jh_2}
 \F^-_{p_3,\jh_3}\F^0_{s_1,s_2,\jh_4} \rangle \ , \hspace{1cm}
 p_1+p_2 = p_3 \ . \label{po1} \ee {}From the global symmetry
 constraints we derive \be K(q_1,q_2) = \prod_{\a=1}^2 \left |
 e^{-\mu_\a x^\a_3(p_1x_{1\a}+p_2x_{2\a})- \frac{s_\a}{\sqrt{2}}
 x^\a_3 x_{4\a} - \frac{s_\a}{2\sqrt{2}}(x_{1\a}+x_{2\a})x^\a_4}
 x_{4\a}^{q_\a}\right |^2  \ , \label{po2} \ee up to a function of
 the two invariants (no sum over $\alpha=1,2$) \be x_\a =
 (x_{1\a}-x_{2\a})x^\a_4 \ .\ee
 We rewrite the conformal blocks as
 {\colsep
 \be
 {\cal F}_{n_1,n_2} = z^{\ka_{12}}(1-z)^{\ka_{14}} F_{n_1,n_2}
 \ , \ee where \ba \ka_{12} &=& h_1+h_2-\frac{h}{3}
 -p_1\jh_2-p_2\jh_1-(\m_1^2+\m_2^2)p_1p_2 \
 , \nb \\
 \ka_{14} &=& h_1+h_4-\frac{h}{3}  -p_1\jh_4 - L p_1 -
 \frac{s_1^2+s_2^2}{4} \ . \label{po4}
 \ea}
 The KZ equation then
 reads
 {\colsep
 \ba && z(1-z) \p_z F_{n_1,n_2}(z,x_1,x_2) = -\sum_{\a=1}^2
 \left[ \m_\a p_3 x_\a\p_{x_\a} + \frac{s_\a}{2\sqrt{2}}\m_\a
 (p_1-p_2)x_\a \right ]
 F_{n_1,n_2}(z,x_1,x_2) \nb \\
 &+& z \sum_{\a=1}^2 \left [ \left ( \m_\a p_2 x_\a -
 \frac{s_\a}{\sqrt{2}} \right ) \p_{x_\a} - \frac{s_\a \m_\a
 p_2}{2\sqrt{2}}x_\a \right ] F_{n_1,n_2}(z,x_1,x_2) \ ,
 \label{kz-ppmo}
 \ea}
 and the solutions are \be F_{n_1,n_2}(z,x_\a)
 = \prod_{\a=1}^2[s_\a \f(\m_\a,z)+x_\a \omega(\m_\a,z)]^{n_\a}
 e^{s_\a^2 \h(\m_\a, z) +s_\a x_\a \chi(\m_\a,z)} \ , \label{po5}
 \ee with $n_1, n_2 \ge 0$. We have introduced the following
 functions
 {\colsep
 \ba \f(\m,z) &=& \frac{z^{1-\m p_3}}{\sqrt{2}(1-\m
 p_3)}F(1-\m p_1,1-\m p_3,2-\m
 p_3,z) \ , \nb \\
 \omega(\m,z) &=& -z^{-\m p_3}(1-z)^{\m p_1} \ , \nb \\
 \chi(\m,z) &=& -\frac{1}{2\sqrt{2}}
 +\frac{p_2}{\sqrt{2}p_3}(1-z)F(1+\m
 p_2,1,1+\m p_3,z) \ , \nb \\
 \h(\m,z) &=& -\frac{z p_2}{2p_3} \ {}_3F_2(1+\m p_2,1,1;1+\m
 p_3,2;z)  - \frac{1}{4} \ln{(1-z)} \ . \label{po6}
 \ea}
 The
 four-point function is then given by
 {\colsep
 \ba {\cal
 G}_{q_1,q_2}(z,\bar{z},x_\a, \bar{x}^\a) &=&
 |z|^{2\ka_{12}}|1-z|^{2\ka_{14}} \prod_{\a=1}^2
 C^{1/2}_{12}(\m_\a) C_{+-0}(\m_\a,p_3,s_\a) \times
 \nb \\
 & & e^{C_{12}(\m_\a)|s_\a \f(\m_\a,z)+x_\a\omega(\m_\a,z)|^2}
 \left |e^{s_\a^2 \h(\m_\a,z) + s_\a x_\a\chi(\m_\a,z)} \right |^2
 \ , \label{po7}
 \ea}
 where \be C_{12}(\m) =
 \frac{\g(\m(p_1+p_2))}{\g(\m p_1)\g(\m p_2)} \ , \hspace{0.6cm}
 C_{+-0}(\m,p_3,s) = e^{\frac{s^2}{2}[\psi(\m p_3)+\psi(1-\m
 p_3)-2\psi(1)]}
   \ .
 \label{po8} \ee The $SU(2)_I$ invariant correlator at the point
 $\m_1=\m_2=\mu$ is obtained after summing over $q_1 \in
 \mathbb{Z}$ with $q_1 + q_2 = Q = L/\mu$.

 \subsection{$\langle +- \ 0 \ 0 \rangle$ correlators}

 The last correlator we have to consider is of the form \be \langle
 \F^+_{p,\jh_1} \F^-_{p,\jh_2}\F^0_{s_{3\a},\jh_3}
 \F^0_{s_{4\a},\jh_4} \rangle \ , \quad p_1 = p_2 . \label{pmo1}
 \ee The Ward identities give \be K(q_1,q_2) = \prod_{\a=1}^2 \left
 | e^{-\m_\a p x_{1\a}x^\a_2 - \frac{x_{1\a}}{\sqrt{2}} \left (
 s_{3 \a} x_3^\a + s_{4 \a} x_4^\a \right )-
 \frac{x^\a_2}{\sqrt{2}} ( s_{3 \a} x_{3 \a} +s_{4 \a} x_{4 \a})}
 x_{3 \a}^{q_\a} \right |^2   \ , \label{pmo2} \ee up to a function
 of the two invariants (no sum over $\alpha=1,2$) $x_\a = x^\a_3
 x_{4\a}$.

 We decompose this correlator around $z=1$ setting $u = 1-z$, since
 the conformal blocks turn out to be simpler and rewrite them as
 \be {\cal F}_{n_1,n_2} = z^{\ka_{12}}(1-z)^{\ka_{14}} F_{n_1,n_2}
 \ , \ee where \be \ka_{14} = h_1+h_4-\frac{h}{3} -p \jh_4 -
 \sum_{\a=1}^2\frac{s_{4\a}^2}{2} \ , \hspace{1cm} \ka_{12} =
 \sum_{\a=1}^2 \frac{s_{3\a}^2+s_{4\a}^2}{2} - \frac{h}{3} \ .
 \label{pmo4} \ee The KZ equation
 {\colsep
 \ba
 \p_u F_{n_1,n_2}(u,x_1,x_2)
 &=& - \frac{1}{u} \prod_{\a=1}^2 \left [ \m_\a p x_\a \p_{x_\a}  +
 \frac{s_{3\a}s_{4\a} x_\a}{2} \right ]
 F_{n_1,n_2}(u,x_1,x_2) \nb \\
 &-& \frac{1}{1-u} \prod_{\a=1}^2\frac{s_{3\a}s_{4\a}}{2} \left (
 x_\a + \frac{1}{x_\a} \right ) F_{n_1,n_2}(u,x_1,x_2) \ ,
 \label{pmo99}
 \ea}
 has the solutions \be F_{n_1,n_2}(u,x_\a) =
 \prod_{\a=1}^2(x_\a u^{-\m_\a p})^{n_\a} e^{x_\a \omega(\m_\a, u)
 + x^\a \chi(\m_\a,u)} \ , \label{pmo5} \ee with $n_1, n_2 \in
 \mathbb{Z}$, $x^\a = x_{3 \a} x_{4}^{\a}=1/x_\a$ and \be
 \omega(\m,u) = - \frac{s_3s_4}{2 \m p} \ F(\m p,1,1+\m p,u) \ ,
 \hspace{0.4cm} \chi(\m,u) =  - \frac{s_3s_4}{2(1-\m p)} \ u \
 F(1-\m p,1,2 - \m p, u) \ . \label{pmo6} \ee

 The four-point function is then given by \be {\cal
 G}(u,\bar{u},x_\a,\bar{x}^\a) =  |u|^{2\ka_{14}}|1-u|^{2\ka_{12}}
 \prod_{\a=1}^2 \tau(\m_\a) \left | e^{x_\a \omega(\m_\a,u) + x^\a
 \chi(\m_\a, u)} \right |^2 \sum_{n_\a \in \mathbb{Z}} \left | x_\a
 u^{-\m_\a p} \right |^{2n_\a} \ , \label{pmo7} \ee where $\tau(\m)
 =  {C}_{+-0}(\m,p,s_{3}) {C}_{+-0}(\m,p,s_{4})$

 The $SU(2)_I$ invariant correlator at the point $\m_1=\m_2=\mu$ is
 obtained after summing over $q_1 \in \mathbb{Z}$ with $q_1 + q_2 =
 Q = L/\mu$.

 % OOOOOOOOOOOOOOOOOOOOOOOOOOOOOOOO     STRING AMPLITUDES  OOOOOOOOOOOOOOOOOOOOOOOOOOO

 % OOOOOOOOOOOOOOOOOOOOOOOOOOOOOOOO     STRING AMPLITUDES  OOOOOOOOOOOOOOOOOOOOOOOOOOO

 \section{String amplitudes}
 \label{amplitudes}
 \noindent

 In this section we study the string amplitudes in the Hpp-wave.
 After combining the results of the previous sections with the ones
 for the internal CFT and for the world-sheet ghosts, one can
 easily extract irreducible vertices and decay rates in closed
 form. The world-sheet integrals needed for the computation of
 four-point scattering amplitudes of scalar (tachyon) vertex
 operators are not elementary and we only study the appropriate
 singularities and interpret them in terms of OPE. As mentioned in
 section \ref{ppw} the Hpp-wave with $\widehat{\mathcal
 H}_6$ affine Heisenberg symmetry that emerges in the Penrose limit
 of AdS$_3$$\times$S$^3$ should be combined with extra degrees of
 freedom in order to represent a consistent background for the
 bosonic string. Quite independently of the initial values of
 $k_{SL(2,\mathbb{R})}=k_1$ and $ k_{SU(2)}=k_2$, one needs to
 combine the resulting CFT that has $c=6$ with some internal CFT
 with $c=20$. For definiteness let us suppose this internal CFT to
 correspond to flat space $\mathbb{R}^{20}$ or to a torus $T^{20}$,
 but this choice is by no means crucial in the following.

 In a covariant approach, such as the one followed throughout the
 paper, string states correspond to BRS invariant vertex operators.
 As usual, negative norm states correspond to unphysical
 `polarizations'. These are absent for the scalar (tachyon) vertex
 operators we have constructed in section \ref{specth6}. Let us
 focus on the left-movers. Starting from a `standard' HW ($
 \mu_\alpha p <1$ for $\alpha=1,2$) primary state $|\Psi\rangle$ of
 $\widehat{\mathcal H}_6$, the Virasoro constraints \be
 L_n|\Psi\rangle = 0 \ , \hspace{2cm} {\rm for} \quad n>0 \ , \ee
 together with \be L_0|\Psi\rangle =|\Psi\rangle \ , \ee project
 the Hilbert space on positive norm states. The mass-shell
 condition becomes
 \be
 h^a_{p,\hj} + h_{int} + N =1 \ ,
 \label{dimcon} \ee where $N$ is the total level, $h_{int}$ is the
 contribution of the internal CFT, \ie \ $h_{int} = |\vec{p}|^2/2$
 and for $p\neq 0$ \be h^{\pm}_{p,\hj} = \mp p \hj + {1\over 2}
 \sum_{\alpha=1}^2 \mu_\alpha p (1 - \mu_\alpha p) \ , \ee while
 for $p=0$,  \be h^{0}_{s,\hj} = {1\over 2} s^2 = {1\over 2}
 \sum_{\alpha=1}^2 s_\alpha^2 \ .
 \ee

 Outside the range $\mu_\alpha p <1$ one has to consider spectral
 flowed representations when $\mu_1=\mu_2=\mu$ or MHW
 representations when $\mu_1\neq \mu_2$, as discussed in section
 \ref{specth6}. Let us concentrate for simplicity on
 $\mu_1=\mu_2=\mu$ with enhanced (non-chiral) $SU(2)_I$ invariance.
 In this particular case, spectral flow yields states with \be
 h^{\pm, w}_{p,\hj} = \mp \left ( p + {w \over \mu} \right ) \hj +
 \mu p (1 - \mu p) \mp w \lambda \ , \ee where $\lambda = n_- -
 n_+$ is the total `helicity' and, for $p=0$, \be h^{0, w}_{s,\hj}
 ={w \over \mu} \hj - {1\over 2} s^2 - w \lambda \ . \ee The
 physics is similar to the case of the NW background \cite{dak}:
 whenever $\mu p$ reaches an integer value in string units, stringy
 effects become important and one has to resort to spectral flow in
 order to make sense of the resulting state \cite{hik}. The string
 feels no confining potential and is free to move along the
 `magnetized planes'. The analysis of AdS$_3$ leads qualitatively
 to the same conclusions \cite{mo1}. Spectral flowed states can
 appear both in intermediate channels and as external legs. Even
 though in this paper we have only considered  correlation
 functions with states in highest-weight representations with
 $\mu|p|<1$ as external legs, it is not difficult to generalize our
 results to include spectral flowed external states along the lines
 of \cite{dak}.

 In order to compute covariant string amplitudes in the Hpp-wave
 one has to combine the correlators computed in sections
 \ref{3point}, \ref{4point} with the
 contributions of the internal CFT and of the bosonic $b,c$ ghosts.
 Contrary to the AdS case discussed in \cite{mo1,mo3}, we do not
 expect any non-trivial reflection coefficient in the Hpp-wave
 limit, so, given the well known normalization problems in the
 definition of two-point amplitudes, let us start considering
 three-point amplitudes. The irreducible three-point couplings can
 be directly extracted from the tree-point correlation functions
 computed in section \ref{3point}, where we also showed that they
 agree with those resulting from the Penrose limit of
 AdS$_3$$\times$
 S$^3$. Trading the integrations over the insertion points for the
 volume of the $SL(2,\mathbb{C})$ global isometry group of the
 sphere and combining with the trilinear coupling $T_{IJK}(h_i)$ in
 the internal CFT one simply gets \be {\cal A}_{abc}^{IJK}(\nu_i,
 x_i; h_i) = K_{abc}(\nu_i, x_i) C_{abc}(\nu_i) T_{IJK}(h_i) \ ,
 \ee where $a_i = \pm,0$, $\nu_i$ denote the relevant quantum
 numbers and the $\delta$-functions associated to the conservation
 laws are understood. Except for $T_{IJK}(h_i)$ all the relevant
 pieces of information can be found in section 4. For ${\mathcal
 M}= \mathbb{R}^{20}$ or $T^{20}$, $T_{IJK}(h_i)$ is essentially
 purely kinematical, \ie \ $\delta(\sum_i \vec{p}_i)$. Other
 consistent choices require a case by case analysis. Depending on
 the kinematics, amputated three-point amplitudes can be
 interpreted as decay or absorption rates. In particular
 kinematical regimes (for the charge variables) they allow one to
 compute mixings, to determine the $1/k \approx g_s$ corrections to
 the string spectrum in the Hpp-wave and to address the problem of
 identifying `renormalized' BMN operators
 \cite{bmn,ppreviews}.

 Additional insights can be gained from the study of four-point
 amplitudes. In particular the structure of their singularities
 provides interesting information on the spectrum and couplings of
 states that are kinematically allowed to flow in the intermediate
 channels. Needless to say, one would have been forced to discover
 spectral flowed states or non highest-weight states even if one
 had not introduced them in the external legs.

 As usual, $SL(2,\mathbb{C})$ invariance allows one to fix three of
 the insertion points and integrate over the remaining one or
 rather their $SL(2,\mathbb{C})$ invariant cross ratio denoted by
 $z$ in previous sections. Schematically
 \be
 {\cal A}_4 = \int d^2z
 |z|^{\sigma_{12} - 4/3} |1- z|^{\sigma_{14} - 4/3} K(x_i, \nu_i)
 {\cal G}_{Hpp}(\nu_i,x_i, z) {\cal G}_{\mathcal M}(h_i,z) \ ,
 \ee
 where, for a flat ${\mathcal M}$, $\sigma_{ij} = \kappa_{ij} +
 \vec{p}_i\cdot \vec{p}_j$ with $\kappa_{ij}$ defined in section 5.

 At present, closed form expressions for ${\cal A}_4$ are not
 available. Still the OPE allows one to extract interesting
 physical information. Let us consider, for a flat ${\mathcal M}$,
 the two cases ${\cal A}_{+++-}$ and ${\cal A}_{+-+-}$. The
 relevant $\widehat {\cal H}_6$ four-point functions have been computed
 both solving the KZ equation (in section \ref{4point}) and by
 means of the Wakimoto representation (in section \ref{app A}).
 Expanding ${\mathcal A}_{+++-}$ in the s-channel yields
 {\colsep
 \beqa
 {\mathcal A}_{+++-} &=& \int d^2z |z|^{2(h_{12}-2)} \sum_{q=0}^Q
 C_{++}{}^+(\nu_1,\nu_2; q) C_{+-}{}^-(\nu_3,\nu_4; Q-q) \nonumber \\
 &\times& |z|^{-2q(p_1+p_2)} ||x_{12}||^{2q} ||x_{13}||^{2(Q-q)} + \dots
 \eeqa}
 where $h_{12}= h^+(p_1 + p_2, \hj_1 +\hj_2) + {1\over 2}
 (\vec{p}_1 + \vec{p}_2)^2$. Studying the $z$ integration near the
 origin determines the presence of singularities whenever $h_{12} -
 q(p_1+p_2) = 1 - N$ that coincides with the mass-shell condition
 for the intermediate state in the $V^+$ representation. The
 amplitudes ${\cal A}_{+-+-}$ are more interesting in that they
 feature the presence of logarithmic singularities in the s-channel
 when $p_1 = p_2$ and $p_3 = p_4$, that is when the amplitudes
 factorize in the continuum of type 0 representations parameterized
 by $s$. Using the explicit form of the OPE coefficients already determined
 and integrating $z$ in a small disk around the
 origin yields
 \be
 {\cal A}_{+-+-} \approx \small \int d^2z
 |z|^{2h_{12}-4-2Q} \Psi(p_1,p_3)^{Q+1}\left |e^{p_3 x z + x
 \Psi(p_1,p_3)}\right |^2 ||x_{13}||^{2Q} \sum_{q=0}^{\infty}
 {(||x_{13}||||x_{24}||)^{2q}| \Psi(p_1,p_3)|^{2q} \over {q! (Q +
 q)!}} \ ,
 \ee
 where as usual $Q = L/\mu =-\sum_i \hj_i/\mu$ and
 $\Psi(p_1,p_3) = [- \log|z|^2 - 4\psi(1) - \psi(p_1) -
 \psi(1-p_1)- \psi(p_3) - \psi(1-p_3)]^{-1}$. For $q=Q=0$ one has
 \be
 {\cal A}_{+-+-} \approx \int_{|z|<\epsilon} {d|z| \over
 |z|^\delta \log|z|} \ ,
 \ee
 where $\delta = 3 - 2h_{12}$ that
 converges for $\delta<1$ but diverges logarithmically as ${\cal
 A}_{+-+-} \approx \log(h_{12} -1)$ for $\delta\approx 1$. The
 logarithmic branch cut departing from $h_{12}=1$ signals the
 presence of a continuum mass spectrum of intermediate states with
 $s=0$. Expanding in the u-channel for $p_1 +p_3=w$ one can proceed
 roughly in the same way and identify the continuum of intermediate
 states in spectral flowed type 0 representations. They signal the
 presence of branch cuts for each string level.

 % OOOOOOOOOOOOOOOOOOOOOOOOOOOOOOOOO   CONCLUSIONS 4        OOOOOOOOOOOOOOOOOOOOOOOOOOOOOOOO

 % OOOOOOOOOOOOOOOOOOOOOOOOOOOOOOOOO   CONCLUSIONS 4        OOOOOOOOOOOOOOOOOOOOOOOOOOOOOOOO

  \section{Conclusions}
  \label{conc4}
  \noindent

 In this chapter we have computed two, three and four-point amplitudes for
 tachyon vertex operators of bosonic strings in AdS$_3$$\times$S$^3$. We used conformal techniques
 developed in \cite{dak}. The expressions for such correlators were found to agree
 with previous results gotten in chapter \ref{str-ads3s3}.
 Instead of taking the Penrose limit in the $SU(2,\mathbb R)_{k_1}\times SU(2)_{k_2}$ solutions by
 rescaling the charge variables, here we have 
 calculated the correlators using conformal skills  for the $\widehat{\mathcal
 H}_6$ model. In this sense, we saw that
 (\ref{K++-}), (\ref{C123}) and (\ref{Ct123})
 correspond respectively to (\ref{cgppm}), (\ref{tensorH6}) and (\ref{qpppm}).
 In appendix \ref{app A} we give a further prove of the
 correctness of these results by computing the same quantities from
 the free-field realization.

 The main novelties we have found here with respect to the $\widehat{\mathcal
 H}_4$ case are the existence of non-chiral symmetries that
 correspond to background isometries not realized by the zero-modes
 of the currents and the presence in the spectrum of new
 representations of the current algebra that satisfy a modified
 highest-weight condition. The results are compactly encoded in
 terms of auxiliary charge variables, which form doublets of the
 external $SU(2)$ symmetry. On the other hand, global Ward identities represent powerful
 constraints on the form of the correlation functions and in the
 next chapter we argue that higher dimensional generalizations, even in the
 presence of R-R fluxes where no chiral splitting is expected to
 take place, should follow the same pattern. We thus believe that
 some of the pathologies of the BMN limit pointed out in the
 literature should rather be ascribed to an incomplete knowledge of
 the scaling limit in the computation of the relevant amplitudes.
 Taking fully into account the rearrangement, technically speaking
 a Saletan contraction,  of the superconformal
 generators in a $\widehat{\mathcal H}_{2+2n}$ Heisenberg algebra
 is imperative in this sense.

 % OOOOOOOOOOOOOOOOOOOOOOOOOOOOOOOO      HOLOGRAPHY   OOOOOOOOOOOOOOOOOOOOOOOOOOOOOOOOOOOOO

 % OOOOOOOOOOOOOOOOOOOOOOOOOOOOOOOO      HOLOGRAPHY   OOOOOOOOOOOOOOOOOOOOOOOOOOOOOOOOOOOOO

 % OOOOOOOOOOOOOOOOOOOOOOOOOOOOOOOO      HOLOGRAPHY   OOOOOOOOOOOOOOOOOOOOOOOOOOOOOOOOOOOOO

 % OOOOOOOOOOOOOOOOOOOOOOOOOOOOOOOO      HOLOGRAPHY   OOOOOOOOOOOOOOOOOOOOOOOOOOOOOOOOOOOOO

 \renewcommand{\theequation}{\arabic{chapter}.\arabic{equation}}
 \chapter{Holography and the BMN Limit}
 \label{holography}
 \noindent

 Having explicit control on the detailed action of the Penrose
 limit on string theory in AdS$_3$$\times$S$^3$, we can employ the
 original AdS$_3$/CFT$_2$ recipe to provide a concrete formula for
 the holographic correspondence in the Hpp-wave background. On the
 string side we end up with S-matrix elements as anticipated
 earlier \cite{kirpio} and defined unambiguously in \cite{dak},
 alternative approaches can be found in \cite{bn,dgr,lor}. On
 the CFT$_2$ side we can produce an explicit formula for the
 Penrose limit of CFT correlators, to be compared with the string
 theory S-matrix elements\footnote{An interesting study of D-branes in pp-wave backgrounds 
 is given in \cite{dak2}. In a couple of examples, D-branes preserving half of the background isometries, 
 the authors have found explicit espressions for the bulk to boundary and three point coupling, 
 as well as for boundary four point functions.}

 The key ingredients of such a holographic formula are:

 $\bullet$ The original AdS$_3$/CFT$_2$ equality between ``S-matrix"
 elements\footnote{These are not the standard S-matrix elements,
 but their closest analogue in AdS. They can be defined as the
 on-shell action evaluated on a solution of the (quantum) equations
 of motion with specified sources on the boundary. For AdS$_3$ such
 elements were conjectured by Maldacena and Ooguri \cite{mo3}.} for
 vertex operators in Minkowskian signature AdS$_3$ and CFT
 correlation functions. Introducing two charge variables $\vec x$
 for $SL(2,\mathbb{R})$ and as many $\vec y$ for $SU(2)$, the
 ``S-matrix elements" depend on both $\vec x$  and $\vec y$. On the
 CFT side, $\vec x$ represent the positions of CFT operators ${\cal
 O}_{l,\tl}(\vec x,\vec y)$, while $\vec y$ are charge variables
 for the $SU(2)_L\times SU(2)_R$ R-symmetry. The conformal weight
 of the operators ${\cal O}_{l,\tl}$ is given by $\Delta=l$.

 $\bullet$ The limiting formulae (\ref{pp+}), (\ref{pp-}) and
 (\ref{pp0}) that describe the precise way operators of the
 original theory map to the operators of the pp-wave theory under
 the Penrose contraction.

 In the expressions below, $\vec z_i$ are the coordinates of the
 vertex operators on the string world-sheet, $\vec x_i$ are the
 $SL(2,\mathbb{R})$ charge variables, that represent the insertion
 points on the boundary, and $\vec y_i$ are the $SU(2)$ R-charge
 variables. $\Psi^{\pm}_{l}\left(\vec z,\vec x\right)$ are
 $SL(2,\mathbb{R})$ primary fields of string theory on AdS$_3$
 corresponding to the ${\cal D}^{\pm}_l$ representations,
 $\Omega_{\tilde l}\left(\vec z,\vec x\right)$ are $SU(2)$ primary
 fields of string theory on $S^3$ corresponding to the $SU(2)$
 representation of spin $\tilde l$, and $\Psi^{0}_{l,\a}\left(\vec
 z,\vec x\right)$ are the $SL(2,\mathbb{R})$ primary fields of
 string theory corresponding to the continuous representations of
 spin $l$. We neglect the internal CFT part of the operators as it
 is not relevant for the structure of our formulae.

 The left and right charge variables $x,\bar x$ are related to the
 Cartesian ones used here by $x=x^1+ix^2, \bar x=x^1-ix^2$. Thus,
 the transformation that inverts the chiral charge variables, $x\to
 1/x, \bar x\to 1/\bar x$ corresponds in the cartesian basis to
 $\vec x \to \vec x^c/|\vec x|^2$ where the superscript stands for
 a parity transformation, $(x^1,x^2)^c=(x^1,-x^2)$. Since we
 consider lorentzian AdS$_3$ also a Minkowski continuation of the
 charge variables is necessary, and this can readily be implemented
 in the CFT correlators by $x\to x^+,\bar x\to x^-$.

 We will denote by ${\cal O}_{l,\tilde l}(\vec x,\vec y)$ operators
 in the CFT that correspond to the appropriate ones in  AdS$_3$
 \be
 \Psi_{l}\left(\vec z,\vec x\right)\Omega_{\tilde l}\left(\vec
 z,\vec y\right) \Leftrightarrow {\cal O}_{l,\tilde l}\left({\vec
 x,\vec y}\right) \ .
 \ee
 The AdS$_3$ ``S-matrix elements" are
 functions of the spins $(l,\tl)$ as well as of the charge
 variables $\vec x_i,\vec y_i$.
 They can be obtained by standard
 techniques by integrating the CFT correlators appropriately over
 the positions of the vertex operators \cite{mo3}. We will split
 the AdS$_3$ states into three families, distinguished by the type
 of ${\cal H}_6$ representation they will asymptote to in the
 Penrose limit, namely $\Phi^+$, $\Phi^-$ and $\Phi^0$. Thus the
 starting string ``S-matrix elements" are of the form \be
 S^{AdS_3}_{N_{\pm,0}}(l_i,\tilde l_i,\vec x_i,\vec y_i|l_j,\tilde
 l_j,\vec x_j,\vec y_j|l_k,\a_k,\tilde l_k,\vec x_k,\vec y_k) \ ,
 \ee where the index $i = 1, ..., N_+$ labels the operators that
 asymptote to the $\Phi^+_{p_i,\jh_i}$ operators,  the index $j =
 1, ..., N_-$ labels the operators that asymptote to the
 $\Phi^-_{p_j,\jh_j}$ operators and the index $k = 1, ..., N_0$
 labels the operators that asymptote to the
 $\Phi^0_{s^1_k,s^2_k,\jh_k}$ operators. As shown in section
 \ref{3point}, by taking the Penrose limit the AdS$_3$$\times$S$^3$
 S-matrix elements asymptote to the pp-wave S-matrix elements we
 computed, we have
 $$
 \lim_{k_1\to\infty\atop k_2\to \infty} \prod_{i=1}^{N_+}
 \left( {k_1\over |\vec x_i|^2}\right)^{2l_i}~ \left( {k_2\over
 |\vec y_i|^2}\right)^{-2\tilde l_i} \prod_{k=1}^{N_0} |\vec
 x_k|^{-2l_k+2\a_k}|\vec y_k|^{2\tl_k} \ n(k_1,l_k) \ n(k_2,\tilde
 l_k)~ \times
 $$
 \be
 \times ~S^{AdS_3}_{N_{\pm,0}} \left( l_i,\tilde
 l_i,{\sqrt{k_1}\vec x_i^c\over  | \vec x_i|^2},{\sqrt{k_2}\vec
 y_i^c\over  |\vec y_i|^2} \left | l_j,\tilde l_j,{\vec x_j\over
 \sqrt{k_1}},{\vec y_j\over \sqrt{k_2}} \right | l_k,\a_k,\tilde
 l_k,\vec x_k,\vec y_k \right )= \label{sads3}
 \ee
 $$
 =C_{N_+,N_-,N_0}(k_1,k_2)~ S^{Hpp}_{N_{\pm,0}}(p_i,\jh_i,\vec
 x_i,\vec y_i| p_j,\jh_j,\vec x_j,\vec y_j|s_{k}^{1,2},\jh_k, \vec
 x_k,\vec y_k) \ .
 $$
 In the previous formula the limit on the
 spins is taken as explained in section \ref{3point}. For the first
 two classes of operators (labeled by $i$ and $j$) we have
 \be
 l = \frac{k_1}{2} \m_1 p - a \ , \hspace{1cm} \tl = \frac{k_2}{2} \m_2
 p - b \ ,
 \ee
 with the subleading terms $a$ and $b$ related to
 $\jh$ in the limit as follows
 \be
 \jh_i = -\m_1 a_i + \m_2 b_i \ ,
 \hspace{1cm} \jh_j = \m_1 a_j - \m_2 b_j \ .
 \ee
 For the third
 class of operators we set
 \be l = \frac{1}{2} + i
 \sqrt{\frac{k_1}{2}} s_1 \ , \hspace{1cm} \tl =
 \sqrt{\frac{k_2}{2}} s_2 \ ,
 \ee
 and in the limit  $\jh_k$ is
 given by the fractional part of the $SL(2,\mathbb{R})$ spin $\jh_k
 = - \m_1 \a_k$. The coefficients $C_{N_+,N_-,N_0}(k_1,k_2)$ are
 divergent in the limit $k_{1,2}\to \infty$ and can be computed in
 principle directly. Using the results obtained in section
 \ref{3point} we have for instance \be C_{2,1,0}(k_1,k_2) =
 \sqrt{k_1 k_2} \ . \ee By employing the holographic recipe of
 AdS/CFT we can now write the relation between pp-wave S-matrix
 elements and limits of CFT correlators\footnote{We ignore type
 $V^0$ operators since, although their definition and dynamics are
 clear on the string theory side, they are less clear in the CFT
 side. They are  related to the continuous spectrum and the
 associated instabilities of the NS5/F1 system in analogy with
 the discussion in \cite{sw}.}
 \be
 S^{Hpp}_{N_{\pm}}(p_i,\jh_i,\vec
 x_i,\vec y_i| p_j,\jh_j, \vec x_j,\vec
 y_j)=\lim_{k_1\to\infty\atop k_2\to \infty} {\prod_{i=1}^{N_+}
 \left( {k_1\over |\vec
 x_i|^2}\right)^{2l_i}\left( {k_2\over |\vec
 y_i|^2}\right)^{-2\tilde l_i}\over C_{N_+,N_-}(k_1,k_2)}~\times
 \ee
 $$
 \times ~\left \langle \prod_{i=1}^{N_+} {\cal O}_{l_i,\tilde
 l_i}\left(\sqrt{k_1}{\vec x_i^c\over |\vec x_i|^2},\sqrt{k_2}{\vec
 y_i^c\over |\vec y_i|^2}\right)\prod_{j=1}^{N_-} {\cal
 O}_{l_j,\tilde l_j}\left({\vec x_j\over \sqrt{k_1}},{\vec y_j\over
 \sqrt{k_2}}\right) \right \rangle \ .
 $$
 The $SL(2,\mathbb{R})$ spin is the conformal dimension of the CFT
 operator while the $SU(2)$ spin determines its transformation
 properties under the $SU(2)$ R-symmetry. The level $k$ in the
 space-time CFT is interpreted as the number of $NS5$ branes used
 to build the background \cite{bort}.

 The interpretation of the limit in the CFT is as follows. CFT
 operators that asymptote to $V^-$ representations (with negative
 values of $p^+$) have their position and charge variables scaled
 to zero. Operators that asymptote to $V^+$ representations (with
 positive values of $p^+$) are instead placed at antipodal points
 and then their positions are scaled to infinity. Finally all the
 spins are scaled as indicated and there is an  overall
 renormalization. The limit of the two-point functions of the CFT
 is particularly simple. In this case $C_{1,1,0}=1$ and we obtain
 in the Penrose limit \be S(p_1,\jh_1,\vec x_1,\vec
 y_1|p_2,\jh_2,\vec x_2,\vec y_2)= \exp\left[-\mu_2p(y_1y_2+\bar
 y_1\bar y_2)- \mu_1p(x^+_1x^+_2+x^-_1x^-_2)\right] \ , \ee where
 $\vec y_i$ are in Euclidean space and $\vec x_i$ are in Minkowski
 space.\\

 The same procedure can be
 applied to correlation functions of nearly BPS operators with large R-charge
 in ${\cal N}=4$ SYM theory. While for
 AdS$_3$$\times$S$^3$ one has two charge variables
 $x$ and $\bar{x}$
 and two, $y$ and $\bar{y}$, for S$^3$,
 in the case of AdS$_5$$\times$S$^5$ one introduces four charge variables $x^\mu$
 (coordinates on the boundary) for $SO(4,2)\approx SU(2,2)$ and as many $y^a$
 for $SO(6)\approx SU(4)$. The latter may be regarded as harmonic variables
 in the so-called harmonic superspace approach and
 in our approach it plays a crucial role.

 These charge variables help to make sense of
 correlation functions in the BMN limit.
 Indeed, if one sends $N$ and $J$ to infinity with
 $J\approx \sqrt{N}$,
 keeping the insertion points fixed,
 even protected two-point functions of CPO's $~{\cal O}_J(x) = {\rm Tr}(Z^J)(x)$,
 become meaningless
 $$
 \lim_{J\to \infty} \langle \lambda_J {\cal O}_J(x_1)
 \bar{\lambda}_J {\cal O}^\dagger_J(x_2) \rangle
 \equiv \lim_{J\to \infty} {\lambda_J \bar{\lambda}_J
 \over (x_1 - x_2)^{2J}}~,
 $$
 this no matter how one rescales the local operators.

 For a properly nearly BPS
 operator ${\cal O}_{\Delta,J}(x_1,y_1)$, with
 $\Delta - J \neq 0$ and $K =\sqrt{N}$,
 one can rescale $x\to \tilde{x}/\sqrt{K}$
 and $y\to \tilde{y}/\sqrt{K}$, invert and rescale the coordinate of the
 conjugate operator ${\cal O}^\dagger_{\Delta,J}(x,y)$
 in the opposite way and get
 \bea
 &&\lim_{J\to \infty} \langle {\cal O}_{\Delta,J}(x_1,y_1)
 {\cal O}^\dagger_{\Delta,J}(x_2,y_2)
 \rangle = \nn \\
 &&=\lim_{K\to \infty} \left({ K\over \tilde{x}_2^2}\right)^\Delta
 \left({  \tilde{y}_2^2\over K}\right)^J
 \langle {\cal O}_{\Delta,J}\left({\tilde{x}_1\over \sqrt{K}},
 {\tilde{y}_1\over \sqrt{K}}\right)
 {\cal O}^\dagger_{\Delta,J}\left({\sqrt{K}\tilde{x}_2\over \tilde{x}_2^2},
 {\sqrt{K} \tilde{y}_2\over\tilde{y}_2^2}\right) \rangle = \nn\\
 &&= \exp\{\mu p^+( \tilde{x}_1\cdot \tilde{x}_2 \pm \tilde{y}_1 \cdot
 \tilde{y}_2\} + \cdots\nn
 \eea
 where $\Delta = \half \mu K p^+ + h$ and  $J = \half \mu K p^+ + j$.

 The essence of our proposal is that one has to `smear' the original local
 operators in order to get a sensible result. Moreover, the extra charge variables
 $y,\,\by$ should rescale in roughly the same
 way as the spacetime coordinates $x,\,\bx$.
 We expect a correct kinematical structure, the one
 dictated by the Ward identity of the super-Heisenberg group, to result from
 the Saletan contraction of the superconformal group $PSU(2,2|4)$.
 We would like also to show that the procedure applies equally
 well to all correlation functions that are expected to survive the BMN limit,
 including those whose structure is not fixed by symmetry such as 4-point
 functions \cite{bdkz2}.

 Despite the presence of tachyons and other
 limitations of the bosonic string,
 tree-level amplitudes of states with large R-charge were shown to
 display the following pattern:
 conformal invariance
 $\rightarrow$ Saletan contraction $\rightarrow$ Heisenberg symmetry.
 We expect this pattern to be reproduced by the superstring amplitudes.

 % OOOOOOOOOOOOOOOOOOOOOOOOOOOOOOOOO     SUPERSTRINGS       OOOOOOOOOOOOOOOOOOOOOOOOOOOOOOOOOOOO

 % OOOOOOOOOOOOOOOOOOOOOOOOOOOOOOOOO     SUPERSTRINGS       OOOOOOOOOOOOOOOOOOOOOOOOOOOOOOOOOOOO

 % OOOOOOOOOOOOOOOOOOOOOOOOOOOOOOOOO     SUPERSTRINGS       OOOOOOOOOOOOOOOOOOOOOOOOOOOOOOOOOOOO

 % OOOOOOOOOOOOOOOOOOOOOOOOOOOOOOOOO     SUPERSTRINGS       OOOOOOOOOOOOOOOOOOOOOOOOOOOOOOOOOOOO

 \renewcommand{\theequation}{\arabic{chapter}.\arabic{equation}}
 \chapter{Outlook}
 \label{outlook}
 \noindent

 Exploring superstring theory on AdS$_3\times$S$^3$ gives an ideal platform to
 test the AdS/CFT correspondence beyond the supergravity approximation and
 provides useful ideas and insights into the very issue of
 holography \cite{bort,gks,kutseib,kll,ags}. Here we
 propose a method to compute superstring amplitudes on AdS$_3\times$S$^3$
 supported by NS-NS three-form flux.
 We take advantage of the formulation
 of the theory in terms of current algebras and their representations in terms of
 charge variables. The latter play the role of coordinates on a holographic screen.
 We conclude that correlation functions are simply written in terms of
 differential operators on the already known bosonic
 amplitudes. This chapter is based on \cite{bpz}.

 The purpose is four-fold. First, we want to examine the pathologies
 displayed in the bosonic case and see if they are cured or ameliorated by considering the
 superstring.
 Second, we would
 like to set the discrepancy between bulk supergravity and
 boundary CFT results within the full string framework. Third, it would
 be interesting to see how this procedure can help to compute similar amplitudes in the case where a
 RR three-form flux is also present. Fourth we
 would like to eventually take the Penrose limit of the amplitudes in order to
 clarify the role of holography in the resulting plane-wave background.

 %%%%%%%%%%%%%%%% Superstrings on $AdS_3 \times S^3$  %%%%%%%%%%%%%%%%%%%%%%%%%%

 %%%%%%%%%%%%%%%% Superstrings on $AdS_3 \times S^3$  %%%%%%%%%%%%%%%%%%%%%%%%%%

 %%%%%%%%%%%%%%%% Superstrings on $AdS_3 \times S^3$  %%%%%%%%%%%%%%%%%%%%%%%%%%

 \section{Superstring amplitudes on AdS$_3\times$S$^3$}
 \label{superads3s3}
 \subsection{Superstrings on AdS$_3$}
 \noindent

 The bosonic string on AdS$_3$ has been extensively studied in the not so recent past,
 see chapter \ref{str-ads3s3} and references therein.\\

 Unitary irreducible representations of the horizontal algebra of
 $\widehat{SL}(2,{\mathbb R})$ are typically infinite dimensional and come in
 three different kinds:
 \begin{quote}
 \item --  Discrete representations: spin
  $h>0$ and third component $m= h +p$ ($p=0,1,2 \dots $) and their conjugate with
 $m=-h-p$.
 \item -- Continuous representations: spin $h=-\frac{1}{2}+is$ and $m = \alpha \pm p$
 with $0\le \alpha <1 $.
 \item -- Complementary representations: spin $1/2<h<1/2 + |\alpha -1/2|$ and
 $m = \alpha \pm p$ with $0\leq \alpha <1 $.
 \end{quote}

 The trivial one-dimensional representation with $j=0$ is the only unitary
 finite dimensional representation. All other finite dimensional
 representations are non-unitary.\\

  Denoting $\widehat{SL}(2,{\mathbb R})$ primaries by $\Phi_{h,n,\bn}(z)$,
  the action of the currents is defined according to
  {\colsep\bea\label{kprim}
  K^{\pm}(z)\,\Phi_{h,n,\bn}(w)&\sim& \frac{n\mp
  (h-1)}{z-w}\,\,\Phi_{h,n\pm1,\bn}(w)~,\\
  K^{3}(z)\,\Phi_{h,n,\bn}(w)&\sim&
  \frac{n}{z-w}\,\,\Phi_{h,n,\bn}(w)~.
  \eea}
  The action of the Casimir operator $K^2=\frac12(K^+K^-+K^-K^+)-(K^3)^2\,$ fixes
  {\colsep \bea\label{k2}
  K^2(z)\,\Phi_{h,n,\bn}(w)\sim
  \frac{-h(h-1)}{z-w}\,\,\Phi_{h,n,\bn}(w)~.
  \eea}
  Following \cite{mo1,kutseib} we consider only scalar primary operators
  with $h=\bar h$. The consistency of the classical theory
  imposes this condition on the dimension of the primaries.

  In order to compactly encode the infinite components of an
  irreducible representation it is convenient to introduce a complex
  variable $x$ and its conjugate $\bx$, that may viewed as complex
  coordinates on the two-dimensional boundary of AdS$_3$. In the
  $x$-basis, the bosonic primaries of $\widehat{SL}(2, {\mathbb R})$
  read
  \bea \label{Phix}
  \Phi_{h}(x,\bx)=\sum_{n,\bn=0}^{\infty}\Phi_{h-1,n,\bn}x^{-n-h}\bx^{-\bn-h}~.
  \eea
  Hence, in terms of the complex variables $x$ and $\bx$ the
  primary operators written in the $n$ basis are no more than
  standard Laurent expansions coefficients of the field $\Phi_{h}(x,\bx)$.

  Inverting (\ref{Phix}) one gets the following integral transform
  \bea\label{Phin}
  \Phi_{h-1,n,\bn}=\oint\, {d^2 x}\,
  x^{h-1+n} \bx^{h-1+\bn}\Phi_{h} (x,\bx)~,
  \eea
  where for simplicity we have dropped all the $2\pi i\,$ factors. \\

  We now identify the
  action of the currents on the primaries with some operators
  defined on the $x$ space. Specifically, we can establish the
  relation
  \be\label{KDx}
  K^A(z)\Phi_{h}(w,\bw;x,\bx)\sim
  \frac{\mathD^A}{z-w}\,\Phi_{h}(w,\bw;x,\bx)~,
  \ee
  satisfied by the differential operators
  \bea\label{Dx}
  \mathD^-(x)=-\pd_x~, \quad \mathD^3(x)=-(x\pd_x +h)~, \quad
  \mathD^+(x)=-(x^2\pd_x+2hx)~.
  \eea
  Similar formulas would appear for the right-moving part.

  From the correspondence we expect the $\widehat{SL}(2,{\mathbb R})$ current algebra of
  the worldsheet to have its counterpart in the
  boundary CFT, this means, in  the ($x,\bx$) space.
  The authors of \cite{kutseib} proposed that
  any observable, and in particular the currents, can be
  expressed in terms of the charge variables $x$ according to
  \be \label{ThetaK}
  K^A(z,x)=e^{-xK_0^-}K^A(z,0)e^{xK_0^-}~.
  \ee
  With this definition, the current $K(z;x)$ involving both the worldsheet and boundary
  variables is
  \bea\label{Kzx}
   K(z;x)=2xK^3(z)-K^+(z)-x^2K^-(z)~.
  \eea

  In terms of this bi-field, the $\widehat{SL}(2, {\mathbb R})$ algebra
 can compactly be written as
  \bea\label{KKx}
  K(z;x_1)K(w;x_2)\sim
  k\frac{(x_2-x_1)^2}{(z-w)^2}+\frac{1}{z-w}\,[(x_2-x_1)^2\partial_{x_2}
  -2(x_2-x_1)]K(w;x_2)~,
  \eea
   while primary operators satisfy
  \bea
  \label{KPhix} K(z;x_1)\Phi_{h}(w;x_2)\sim
 \frac{1}{z-w}\,[(x_2-x_1)^2\partial_{x_2}+2h(x_2-x_1)]\Phi_{h}(w;x_2)~.
  \eea

  The Sugawara stress tensor can also be expressed in the $x$-basis
  {\colsep
  \bea\label{STx}
  T(z)&=& \frac{1}{(k-2)}\,\eta_{AB}:K^A K^B:~=\frac{1}{(k-2)}\,[K^+K^--K^3K^3]\nonumber\\
  &=&\frac{1}{2(k-2)}\left[ K(x)\partial_x^2K(x)-\frac{1}{2}\left(
  \partial_xK(x) \right)^2 \right]~.
  \eea}
  using $\eta_{+-}=1/2$ and $\eta_{33}=-1$. Notice that in this formula any $x$ dependence drops since $T(z)$
  is a singlet.\\

  Superstring theory on AdS$_3$ introduces three fermionic fields
  $\psi^A$ which transform in the adjoint presentation of $SL(2,
  {\mathbb R})$ and have OPEs given by
  \bea\label{psipsi}
  \psi^{A}(z)\psi^{B}(w)\sim \frac{k\,
  \eta^{AB}}{2\,(z-w)},\qquad\qquad A,B=\pm,3 \eea
  where the metric elements are $\eta^{+-}=2$, $\eta^{33}=-1$~.

  The fermions also modify the currents of the model,
  adding to the the bosonic currents already introduced a fermionic
  contribution $K^A_F=-{i\over k}\,{\epsilon^A}_{BC}\psi^B\psi^C$.
  This generates an affine algebra for the total current
  \bea \label{sl2ferm}
  K^A_{T}(z)K^B_{T}(w)\sim \frac{(k+2)\,\eta^{AB}}{2\,(z-w^2)}+i\,
  {\epsilon^A}_{BC}\,\frac{K^C_T(w)}{z-w}~,
  \eea
  where we can see that the introduction of the fermions have shifted
  the level of the algebra from $k$ to $k+2$. The unitary bound on
  $h$ is also modified to $1/2<h<(k+1)/2$.\\

  In complete analogy with the
  bosonic currents, we introduce a fermionic field that depends
  also on the charge variables
  \be\label{psix}
  \psi(z;x)=2x\psi^3(z)-\psi^+(z)-x^2\psi^-(z)~,
  \ee with OPE
  \be
  \psi(z;x_1)\psi(w;x_2)\sim\frac{k(x_2-x_1)^2}{z-w}~.
  \ee
  In addition to the fact that using the $x$ dependent field we can keep track of information of the
  boundary theory, this also simplifies the computations notably since we
  are working with scalars fields.\\

  The total stress tensor in terms of the charge variables is given
  by
  {\colsep
  \bea\label{STSx}
  T(z) &=&
  \frac{1}{k}\,\eta_{AB}:K^A K^B:-
  \frac{1}{2k}\,\eta_{AB}:\psi^A\pd_z\psi^B: \cr & \cr
  &=&\frac{1}{4k}\left[ 2\,K\pd^2_x K-(\pd_x K)^2-
  \pd_x\psi\pd_z\pd_x\psi+\pd_x^2\psi\pd_z\psi+\psi\pd_z
  \pd_x^2\psi\right]~,
  \eea}
  while for the worldsheet supercurrent we get
  {\colsep
  \bea \label{Gsl2}
  G(z)&=&\frac{2}{k}\left[ \eta_{AB}\psi^A
  K^B-\frac{i}{3k}f_{ABC}\psi^A\psi^B \psi^C  \right] \cr & \cr
  &=&\frac{1}{2k}\left[-\pd_x\psi\pd_x K+\pd_x^2\psi K+
  \psi\pd_x^2K+\frac{1}{2}\psi\pd_x\psi\pd_x^2\psi\right]~.
  \eea}

  %   OOOOOOOOOOOOOOOOOOOOOOOOOOOOOOOOOOOO      SU(2)    OOOOOOOOOOOOOOOOOOOOOOOOOOOOOOOOOO

  \subsection{The S$^3$ contribution}
  \noindent

  As we have done for $\widehat{SL}(2,{\mathbb R})$, in this subsection we give
  for the $\widehat{SU}(2)$ WZW model an interpretation in terms of charge variables
  that in some sense allows us to
  keep the information of the two-dimensional holographic theory. But,
  since S$^3$ does not have a boundary, the picture does not look so
  appealing as for the AdS$_3$ case. Nevertheless, in the next sections we will be able to display some
  correlation functions and
  extract valuable information from that.\\

  The action of the
  currents of $\widehat{SU}(2)$ on the chiral primary operators is
  {\colsep\bea
  J^{\pm}(z)\,\Omega_{j,m,\bm}(w)&\sim& \frac{j\mp 
  m}{z-w}\,\,\Omega_{j,m\pm1,\bm}(w),\\
  J^{3}(z)\,\Omega_{j,m,\bm}(w)&\sim&
  \frac{m}{z-w}\,\,\Omega_{j,m,\bm}(w)~.
  \eea}
  With this definition, the action of
  the Casimir $J^2={1\over 2} (J^+ J^- + J^- J^+) + {(J^3)}^2$ gives
  {\colsep\bea J^2(z)\,\Omega_{j,m,\bm}(w)\sim
  \frac{j(j+1)}{z-w}\,\,\Omega_{j,m,\bm}(w)~,
  \eea}
  where $j$ is the $SU(2)$
  spin and $m$ the component along an arbitrary direction.\\

  Once again, it is convenient to introduce a complex variable $y$
  and its conjugate $\bar{y}$ (\,different from the previous $x$ and
  $\bx$\,).
  In this space the bosonic
  primaries can be expressed as
  \be \label{su2y}
  \Omega_{j}(y,\by)=\sum_{m,\bm=-j}^{j}\Omega_{j-1,m, \bar m}
  y^{-j-m}\by^{-j-\bm}~.
  \ee
  Inverting this relation, one gets
  back \be \Omega_{j-1,m,\bar m}=\oint\, d^2 y\,
  y^{j-1+m} {\bar y}^{j-1+\bar m} \Omega_{j}(y,\bar y)~.
   \ee
  In contrast to $\widehat{SL}(2, {\mathbb R})$,
  the variables
  $y$ and $\bar{y}$ cannot be viewed as boundary coordinates. They
  are rather `charge' variables, only a very convenient book-keeping
  device for the components of $\Omega_{j}(z)$, which thanks to the
  compactness of $SU(2)$ are finite in number\footnote{Alternatively, $y$
  and $\bar{y}$ can be regarded as harmonic coordinates on the coset
  $SU(2)/U(1)$. This is geometrically an $S^2$ and represents the
  basis of a Hopf fibration of $S^3 = SU(2)$. From this
  point of view, the components are nothing but spherical
  harmonics.}.

  The  action of the generators of the algebra on the primary fields
  can be realized in terms of the following differential operators
  \bea\label{Ty}
  \mathT^-(y)=-\pd_y~, \quad
  \mathT^3(y)=-(y\pd_y +\eta)~, \quad \mathT^+(y)=(y^2\pd_y+2\eta y)~,
  \eea
  with $\eta=j+1$. It can be easily checked that these operators
  satisfy the analogous of (\ref{KDx}) for $SU(2)$.\\

   As for AdS$_3$, it is
  rewarding to define a single operator for the current that involves both the
  worldsheet and boundary variables. The differential operators defined in
  (\ref{Ty}) suggest that the currents are given by
   \be \label{ThetaJ}
  J^a(z,y)=e^{-yJ_0^-}J^a(z)e^{yJ_0^-}~,
  \ee
   \be J(z;y)=2y J^3 (z) - J^+ (z)
  + y^2 J^- (z).
   \ee
  Notice that in spite of the similarity of the equations (\ref{ThetaK}) and
  (\ref{ThetaJ}), the two are different.

   In terms of these `bi-current', the
  $\widehat{SU}(2)$ affine algebra can be written in a more suggestive
  manner
  \be J(z;y_1)J(w;y_2)\sim
  -k\frac{(y_2-y_1)^2}{(z-w)^2}+\frac{1}{z-w}[(y_2-y_1)^2\partial_{y_2}
  - 2(y_2-y_1)]J(w;y_2)~.
  \ee
  The primary operators can also be shown to satisfy
  \be
  J(z;y_1)\Omega_{j}(w;y_2)\sim
  -\frac{1}{(z-w)}[(y_2-y_1)^2\partial_{y_2}+2\eta(y_2-y_1)]\Omega_{j}(w;y_2)~.
  \ee

 For the superstring on $S^3$, one introduces three fermions
 $\chi$'s `tangent` to S$^3$. They satisfythe OPE's
 \be
 \chi^{a}(z)\chi^{b}(w)\sim \frac{k\,
 \delta^{ab}}{2\,(z-w)},\qquad\qquad a,b=\pm,3~.
 \ee
where $\delta^{+-}=2$ and $\delta^{33}=1$.
 The total current has a bosonic and a
 fermionic contribution
 $J^a_{T}=J^a+J^a_{F}=J^a-\frac{i}{k}\,{\epsilon^a}_{bc}\chi^b\chi^c$,
 with affine algebra
 \be \label{su2ferm}
 J^a_{T}(z)J^b_{T}(w)\sim \frac{(k-2)\,\delta^{ab}}{2\,(z-w^2)}+i\,
 {\epsilon^a}_{bc}\,\frac{J^c_T(w)}{z-w}~.
 \ee
 The total current $J^a_{T}$ has two contributions:
 a level $k$ bosonic current and a level $-2$ fermionic current.

 It is straightforward to define
 fermionic fields that also depend on the `charge' variable~$y$
 \be
 \chi(z;y)=2y\chi^3(z)-\chi^+(z)+y^2\chi^-(z)~,
 \ee
 with OPE
 \be
 \chi(z;y_1)\chi(w;y_2)\sim-\frac{k(y_2-y_1)^2}{z-w}~.
 \ee
 In the $y$-basis, the stress tensor is given by (see \cite{kutseib})
 {\colsep\bea
 T(z)&=&\frac{1}{k}:J^a J^b:-\frac{1}{2}\chi^a\pd_z\chi^b \nonumber\\
 &=&-\frac{1}{4k}\left[ 2\,J\pd^2_y J-(\pd_y J)^2+
 \pd_y\chi\pd_z\pd_y\chi-\pd_y^2\chi\pd_z\chi-\chi\pd_z
 \pd_y^2\chi\right]~,
 \eea}
 as expected it is $y$ independent, being a group singlet.

 Similarly, the worldsheet supercurrent is given by
 {\colsep \bea
 G(z) &=&\frac{2}{k}\left[ \chi^a J^b-\frac{i}{3k}f_{abc}\chi^a\chi^b\chi^c
 \right] \nonumber\\
 &=&-\frac{1}{2k}\left[-\pd_y\chi\pd_y J+\pd_y^2\chi J+
 \chi\pd_y^2J-\frac{1}{2}\,\chi\pd_y\chi\pd_y^2\chi\right]~.
 \eea}

 % OOOOOOOOOOOOOOOOOOOOOOOOOOOOOOOOOO    Vertex Operators on AdS$_3$$\times$S\,$^3$  OOOOOOOOOOOOOOOOOOOOOOOOOOOOOOOOOOOOOOOO

 % OOOOOOOOOOOOOOOOOOOOOOOOOOOOOOOOOO    Vertex Operators on AdS$_3$$\times$S\,$^3$  OOOOOOOOOOOOOOOOOOOOOOOOOOOOOOOOOOOOOOOO

 % OOOOOOOOOOOOOOOOOOOOOOOOOOOOOOOOOO    Vertex Operators on AdS$_3$$\times$S\,$^3$  OOOOOOOOOOOOOOOOOOOOOOOOOOOOOOOOOOOOOOOO

  \subsection{Vertex operators}
  \noindent

  In the canonical (-1) picture for the superghost
  $\varphi$, NS physical vertex operators have the following general
  form
  \be\label{vertgen}
  V_{phys}(h,n,\bn; j,m,\bm; q;
  N)=e^{-\varphi}e^{-\bar{\varphi}}\, V_{h,n,\bn}\, V'_{j,m,\bm}W_q~.
  \ee
  Here $V_{h,n,\bn}$ is a superconformal primary of AdS$_3$,
  \ie, a primary of the total algebra (\ref{sl2}), and
  $V'_{j,m,\bm}$ is a  superconformal primary of S$^3$
  (\ref{su2}). On the other hand,
  $W_q$ is a primary of the internal $N=1$ worldsheet
  superconformal algebra (with internal manifold T$^4$ or $K3$),
  labelled by some set of quantum numbers collectively denoted by
  $q$.

  The worldsheet scaling dimension is given by
  \be
  \Delta(h,j,q,N)=
  \frac{1}{2}+N-\frac{h(h-1)}{k}+\frac{j(j+1)}{k}+ \Delta_{int}(q)~,
  \ee
  with similar relations for the right moving part.

  BRST invariance requires $\Delta = \bar\Delta = 1$. Additional
  conditions arise from the OPE of the vertex operator
  $V_{phys}(h,n,\bn; j,m,\bm; q; N)$ with the total worldsheet
  supercurrent $G=G_{SL(2)} + G_{SU(2)} + G_{int} $. In
  particular,
  the `dressing' of the bosonic primary operators $\Phi$ and
  $\Omega$ with the fermions $\psi$ and $\chi$, $\widehat{SL}(2,{\mathbb R})$ and
  $\widehat{SU}(2)$ respectively, is rather constrained
  by the requirement of primarity w.r.t. the full combined affine
  current algebras (see below).

  It can be shown that BRST physical $(h=j+1)$ chiral primary operators have
  $N=1/2$ and $h_{int}(q)=0$. With these conditions, the NS physical operators we can
  construct are of two types. The first involves a dressing
  including AdS$_3$ fermions\footnote{From now on we drop the ghost contribution.}
  {\colsep
  \bea\label{Wmn}
  \mathcal{W}_{h,n,\bn,m,\bm}(z,\bz)&=&\left[\,\psi(z)\,\bar{\psi}(\bz)
  \,\Phi_{h,n,\bn}(z,\bz)\,\right]_{h-1,h-1}\,\Omega_{j,m,\bm}(z,\bz)\nonumber\\
  &=&P_{h-1}\bar
  P_{h-1}\,\left[\,\psi(z)\,\bar{\psi}(\bz)\,\Phi_{h,n,\bn}(z,\bz)\,\right]\,
  \Omega_{j,m,\bm}(z,\bz)~,
  \eea}
  with
  {\colsep
  \bea\label{psiPhi}
  (\psi \Phi_{h})_{h-1,n}=
  \psi^3 \Phi_{h,n}-\frac{1}{2}\,\psi^+
  \Phi_{h,n-1}-\frac{1}{2}\,\psi^- \Phi_{h,n+1}~.
  \eea}
  Notice that we have included the right fermion in order to have a
  non-chiral vertex operator. A formula  similar to (\ref{psiPhi}) is valid for the right
  part.\\

  The other primary giving rise to BRST physical states is
  {\colsep
  \bea\label{Xmn}
  \mathcal{X}_{h,n,\bn,m,\bm}(z,\bz)&=&\Phi_{h,n,\bn}(z,\bz)
  \,\left[\,\chi(z)\,\bar{\chi}(\bz)\,\Omega_{j,m,\bm}(z,\bz)\,\right]_{j+1,j+1}\nonumber\\
  &=&\Phi_{h,n,\bn}(z,\bz)\,P_{j+1}\bar P_{j+1}\,\left[\,\chi(z)\bar{\chi}(\bz)\,
  \Omega_{j,m,\bm}(z,\bz)\,\right]~,
  \eea}
  that admits the following decomposition
  {\colsep
  \bea
  (\chi \Omega_{j})_{j+1,m}&=&(j+1-m)(j+1+m)\, \chi^3\,
  \Omega_{j,m}-\frac{1}{2}\,(j+m)(j+1+m)\,\chi^+\,
  \Omega_{j,m-1}\nonumber\\&+&\frac{1}{2}\,(j-m)(j+1-m)\,\chi^-\,
  \Omega_{j,m+1}~.
  \eea}

  Other possible vertex operators come combining left and right movers spinors with
  the primaries as follows
  \be\label{Vmn}
  \mathcal V_{h,n,\bn,m,\bm}(z,\bz)=
  \left[\,\psi(z)\,\bar\chi(\bz)\Phi_{h,n,\bn}(z,\bz)\,\right]_{h-1,h}
  \left[\,\psi(z)\bar\chi(\bz)\Omega_{j,m,\bm}(z,\bz)\,\right]_{j,j+1}~,
  \ee
  \be\label{tildeVmn}
  \widetilde{\mathcal V}_{h,n,\bn,m,\bm}(z,\bz)=
  \left[\,\bar\psi(\bz)\,\chi(z)\Phi_{h,n,\bn}(z,\bz)\,\right]_{h,h-1}
  \left[\,\bar\psi(\bz)\chi(z)\Omega_{j,m,\bm}(z,\bz)\,\right]_{j+1,j}~,
  \ee
  where the first index indicates the left spin component, for $SL(2, {\mathbb
  R})$ or $SU(2)$, and the
  second one the right part. It can be seen that these satisfy the
  physical conditions given above.\\

  The spectrum of the Ramond sector can be analyze in a similar
  way. The chiral primaries were given in \cite{kll}
  {\colsep
   \bea\label{Ymn}
  \mathcal{Y}_{h,n,\bn,m,\bm}(z,\bz)=
  \left[\,S(z)\,\bar{S}(z)\,\Phi_{h,n,\bn}(z,\bz)\,\Omega_{j,m,\bm}(z,\bz)\,\right]_{h-\frac12,j+\frac12}~,
  \eea}
  where now the first and second indices indicate the spins of $SL(2, {\mathbb
  R})$ and $SU(2)$ respectively, both of them for the left part.
  We have omitted an easy deal contribution containing $H_4$ and $H_5$, the two (free) scalar fields that
  bosonize the four (free) internal fermions $\lambda^I$. It is of
  the form $\Sigma^{\dot a}=\exp[\,\dot a {1\over 2}(H_4-H_5)\,](z)$, with $\dot a=\pm$. Same for the right
  part.\\

  The spin field $S$ form an $SO(5,1)$ spinor that can be
  decomposed as
  \be S \rightarrow S_{\alpha}\cdot \widetilde{S}_{\alpha}~,
  \ee
    where $S_{\alpha}$,
  $\alpha=\pm\frac12$, transforms in the spin ($\frac{1}{2},0$) of
  $SL(2,{\mathbb R})\times SU(2)$ and $\widetilde{S}_{\alpha}$,
  $\alpha=\pm\frac12$, transforms in the spin ($0,\frac{1}{2}$).\\

  Using the projector notation, (\ref{Ymn}) can be written
  {\colsep
  \bea
  \mathcal{Y}_{h,n,\bn,m,\bm}(z,\bz)&=&
  P_{h-\frac12} \,\bar P_{h-\frac12} \left[\,S_{\alpha}(z)\bar S_{\alpha}(\bz)\,\Phi_{h,n,\bn}(z,\bz)\,\right]\times \nonumber \\
  &\times& P_{j+\frac12} \,\bar P_{j+\frac12}
  \left[\,\widetilde{S}_{\alpha}(z)\bar
  {\widetilde{S}}_{\alpha}(\bz)\,\Omega_{j,m,\bm}(z,\bz)\,\right]~,
  \eea}
  plus an internal contribution $\Sigma^{\dot a}(z)\Sigma^{\dot b}(\bz)$.\\

  The low energy limit of the spectrum of spacetime chiral primaries
  here displayed was shown to be in agreement with the
  supergravity spectrum \cite{sezgin,kll}.
  \\

  In order to compute correlation functions in the bulk that could
  be related to the dual correlators on the boundary CFT, we now write
  the vertex operators in terms of the charge variables.

  Neveu-Schwarz operators given in (\ref{Wmn}) and (\ref{Xmn}) look like
  {\colsep
  \bea\label{W-X}
  \mathcal{W}_h(x,\bx,y,\by)&=&
  \lim_{\scriptsize
  \begin{array}{c}
    x'\rightarrow x \\
    \bx'\rightarrow \bx \\
  \end{array}}
  \,P_{h-1}(x,x')\bar P_{h-1}(\bx,\bx')
  \left[\,\psi(x')\,\bar\psi(\bx')\,
  \Phi_h(x,\bx)\,\right]\Omega_j(y,\by)~,
  \nonumber\\
  \mathcal{X}_h(x,\bx,y,\by)&=&
  \lim_{\scriptsize
  \begin{array}{c}
    y'\rightarrow y \\
    \by'\rightarrow \by \\
  \end{array}}
  \,\Phi_{h}(x,\bx)\,P_{j+1}(y,y')\bar P_{j+1}(\by,\by')\left[\,
  \chi(y')\,\bar\chi(\by')\,\Omega_j(y,\by)\,\right]~.
  \eea}
  The NS mixed operator we proposed in (\ref{Vmn}) in terms of
  the charge variables looks like
  {\colsep
  \bea\label{V}
  \mathcal{V}_h(x,\bx,y,\by)&=&\lim_{x'\rightarrow
  x}P_{h-1}(x,x')\,\left[\,\psi(x')\,\bar\chi(\by)\,\Phi_h(x,\bx)\,\right]\times
  \nonumber\\&\times& \lim_{\by'\rightarrow \by}
  \bar
  P_{j+1}(\by,\by')\,\left[\,\psi(x)\,\bar\chi(\by')\,\Omega_j(y,\by)\,\right]~,
  \eea}
  {\colsep
  \bea\label{tildeV}
  \mathcal{\widetilde V}_h(x,\bx,y,\by)&=&\lim_{\bx'\rightarrow
  \bx}\bar
  P_{h-1}(\bx,\bx')\,\left[\,\bar\psi(\bx')\,\chi(y)\,\Phi_h(x,\bx)\,\right]\times
  \nonumber\\&\times& \lim_{y'\rightarrow y}
   P_{j+1}(y,y')\,\left[\,\bar\psi(\bx)\,\chi(y')\,\Omega_j(y,\by)\,\right]~.
  \eea}
  On the other hand, for the Ramond sector we have
  {\colsep
   \bea\label{Y}
    \mathcal{Y}_h(x,\bx,y,\bx)&=&
   \lim_{\scriptsize
  \begin{array}{c}
    x'\rightarrow x \\
    \bx'\rightarrow \bx \\
  \end{array}}
  P_{h-{1\over2}}(x,x')\,\bar P_{h-{1\over2}}(\bx,\bx')\left[\,S_{\alpha}(x')\,\bar S_{\alpha}(\bx')\,
  \Phi_{h}(x,\bx)\,\right]\times\nonumber\\
  &\times&
   \lim_{\scriptsize
  \begin{array}{c}
    y'\rightarrow y \\
    \by'\rightarrow \by \\
  \end{array}}
  P_{j+{1\over2}}(y,y')\,\bar P_{j+{1\over2}}(\by,\by')\,[\,\widetilde{S}_{\alpha}(y')
  \,\bar{\widetilde{S}}_{\alpha}(\by') \,\Omega_{j}(y,\by)\,]~.
  \eea}

 \subsection{Some useful OPEs}
 \noindent

 \emph{ \underbar {Spin Field OPEs for AdS$_3$}}\\

 Spinors of AdS$_3$ has the following OPEs with the fermionic field:
 \bea
 \psi^A (z) S^{\al} (w) \sim \sqrt{k\over 2} {(\tau^A)}^{\al}_{\bt}
 {{S^{\bt} (w)}\over{(z-w)^{1/2}}}
 \eea
 On the other hand, two spin fields have the OPE, which looks like:
 \bea
 S^{\al}(z) S^{\bt} (w) \sim {{\epsilon^{\al \bt}} \over{(z-w)^{3/8}}}
 + C^{(\psi)}_{SS}(z-w)^{1/8} \eta_{AB} \psi^A {(\tau^B)}^{\al \bt}.
 \eea
 where,
 \bea
 \tau^1 = -i\sigma^2,\>\>\>\> \tau^2 = i\sigma^1,\>\>\>\> \tau^3 = \sigma^3.
 \eea

 In our notation, $A =\pm,3$ and $\alpha = \pm\frac12$.

 The spin fields in terms of charge variable can be written as:
 \bea
 S(x, z) = e^{xK^+_0} S(z)^{-} e^{-xK^+_0} = S^- (z) - xS^+ (z).
 \eea
 We have then the following OPEs
 \bea
 \psi(z;x_1) S(w,x_2) = {\sqrt{k\over 2} \over{(z-w)^{1/2}}}
 \left[ ((x_1 + x_2)^2 + 2x_1 x_2)\pd_{2} + (x_2 -2x_1)\right] S(w; x_2)~,
 \nonumber
 \eea
 \bea
 S(z;x_1) S(w; x_2) &=& {(x_1 - x_2)\over{(z -w)^{3/8}}}
 + {1\over2} \Big[(1+ x_1 x_2)\pd_2 + 2x_1)\psi (w; x_2)\Big] (z-w)^{1/8}.
 \nonumber
 \eea

 \emph{\underbar {Spin Field OPEs for S$^3$}}\\

 Spin fields on S$^3$ has the following OPE with the fermionic fields:
 \bea
 \chi^A (z) {\widetilde S}^{\al} (w) = {\sqrt{k\over 2}} {(\sigma^A)}^{\al}_{\bt}
 \frac{{\widetilde S}^{\bt}}{(z-w)^{\frac12}}~.
 \eea
 On the other hand, the two spin fields have
 \bea
 {\widetilde S}^{\al}(z) {\widetilde S}^{\bt} (w) = {{\epsilon^{\al \bt}}
 \over{(z-w)^{3/8}}} + C^{(\chi)}_{\tilde S \tilde S}(z-w)^{1/8}
 \delta_{AB} \chi^A {(\sigma^B)}^{\al \bt}.
 \eea
 where the $\sigma^i$'s are the Pauli matrices.
 One can write the spin fields in terms of charge variables, as we did
 for AdS$_3$, and we get
 \bea
 {\widetilde S}(y;z) = e^{y J^-_0} {\widetilde S}^+ e^{- y J^-_0} = S^+ (z) + y S^- (z)~.
 \eea
 
 Now, the spin fields and the fermions have
 \bea
 \chi(y_1; z){\widetilde S}(y_2;w) = {\sqrt{k\over 2} \over{(z-w)^{1/2}}}
 \Big([(y_1 -y_2)^2 - 2y_1 y_2]\pd_2 + (2y_1 -y_2)\Big){\widetilde S}(w; y_2)~,
 \nonumber
 \eea
 \bea
 {\widetilde S}^{\al}(y_1; z) {\widetilde S}^{\bt} (y_2; w)
 = {{(y_1 - y_2)}\over{(z-w)^{3/8}}} + {1\over 2} \left([(1-y_1 y_2)
 \pd_2 + 2y_1]\chi (x_2; w)\right)(z-w)^{1/8}~.
 \nonumber
 \eea

 % OOOOOOOOOOOOOOOOOOOOOOOOOOOOOOOOOO    Correlators  OOOOOOOOOOOOOOOOOOOOOOOOOOOOOOOOOOOOOOOO

 % OOOOOOOOOOOOOOOOOOOOOOOOOOOOOOOOOO    Correlators  OOOOOOOOOOOOOOOOOOOOOOOOOOOOOOOOOOOOOOOO

 % OOOOOOOOOOOOOOOOOOOOOOOOOOOOOOOOOO    Correlators  OOOOOOOOOOOOOOOOOOOOOOOOOOOOOOOOOOOOOOOO

  \section{Amplitudes}
  \noindent

  Maybe the most important feature of superstrings on AdS$_3$ spaces
  is that the theory can be described exactly, without relying on the
  low-energy supergravity approximation. This motivates us to
  propose explicit expressions for two, three and four
  point superstring amplitudes.

  \subsection{2-pt correlators}
  \noindent

  By ghost
  charge violation, the only non trivial correlators in the NS sector are
  $$
  \langle\,\mathcal{W}_{h_1}(x_1,y_1)\,\mathcal{W}_{h_2}(x_2,y_2)\,\rangle=
  \langle \,\Omega_{j_1}(y_1)\Omega_{j_2}
  (y_2)\,\rangle~ \times\hspace{6cm}
  $$
  \be\label{<WW>}
  \times \lim_{x_{12}'\rightarrow x_{12}}\,P_{h_1-1}(x_1,x_1')\,P_{h_2-1}(x_2,x_2')
  \,\langle \,\psi(x_1')\, \psi(x_2') \,\rangle
  \,\langle \,\Phi_{h_1}(x_1)\,\Phi_{h_2}(x_2) \,\rangle~,
  \ee
  and
  $$
  \langle\,\mathcal{X}_{h_1}(x_1,y_1)\,\mathcal{X}_{h_2}(x_2,y_2)\,\rangle=
  \langle \,\Phi_{h_1}(x_1)\,\Phi_{h_2}(x_2) \,\rangle~ \times\hspace{6cm}
  $$
  \be\label{<XX>}
  \times \lim_{y_{12}'\rightarrow  y_{12}} \, P_{j_1+1}(y_1,y_1')\,P_{j_2+1}(y_2,y_2')
  \,\langle \,\chi(y_1')\, \chi(y_2')\, \rangle
  \,\langle \, \Omega_{j_1}(y_1)\,\Omega_{j_2}(y_2)\,\rangle~.
  \ee
  Two point functions involving the mixed operator $\mathcal V_h$
  and $\widetilde{\mathcal V}_h$ are vanishing.

  The above expressions already expose the power and elegance of the method:
  we are reducing the computation of superstring amplitudes to the action of some
  differential  operators on the
  already known bosonic amplitudes\footnote{See previous chapters for
  explicit expressions.} combined with
  simple if not trivial free fermionic amplitudes.\\

  \subsection{3-pt correlators}
  \noindent

  Examples of non-vanishing three-point functions are
  \bea \label{3points}
  \langle\mathcal{W}_{h_1}^-\,\mathcal{W}_{h_2}^0\,\mathcal{W}_{h_3}^-\rangle~,
  \qquad\qquad
  \langle\mathcal{Y}_{h_1}^{-1/2}\,\mathcal{W}_{h_2}^-\,
  \mathcal{Y}_{h_3}^{-1/2}\rangle~,
  \eea
  where we introduced an upper index in the vertex operators to
  keep track of the ghost number. We have analogous correlators for
  $\mathcal{W}^-\to \mathcal{X}^-$. Other correlators involve
  the mixed operators  $\mathcal V_h$
  and $\widetilde{\mathcal V}_h$. \\

  We start with the
  $\langle\mathcal{W}_{h_1}^-\,\mathcal{W}_{h_2}^0\,
  \mathcal{W}_{h_3}^-\rangle$ correlator:\\
  \\
  The picture $0$ primary is obtained as usual by application of the
  picture changing operator on $\mathcal{W}^-$, \emph{i.e.},  $\mathcal{W}^0 =
  \Gamma_{+1} \mathcal{W}^- \equiv [ Q_{BRS}, \xi \mathcal{W}^-]$.
  The only non-vanishing
  contribution comes from the supercurrent term $e^\varphi G$ in $\Gamma_{+1}$.
  This yields
 {\colsep
 \bea
 \mathcal{W}^0&=&\lim_{\begin{subarray}{c} x'\rightarrow x \\
 x''\rightarrow x',x\end{subarray}}
 \oint \frac{d w}{2\pi i}\,
 G(w,x'')\,||P_{h-1}(x,x')||^2\,\psi(z,x')\Phi_h(z,x,\bx)\,\Omega_h(z,y,\by)\nonumber\\
 &\sim & 
 \lim_{\begin{subarray}{c} x'\rightarrow x \\ x''\rightarrow
 x',x\end{subarray}}\,||P_{h-1}(x,x')||^2\oint \frac{d w}{2\pi i}\,
 \Big[ -(\pd_{x''}\psi(w,x''))\,(\pd_{x''} K(w,x''))+\nn \\ &+&
(\pd_{x''}^2\psi(w,x''))\, K(w,x'')+ \psi(w,x'')
(\pd_{x''}^2K(w,x''))+
 \nn\\
 &+& \half \psi(w,x'')(\pd_{x''}\psi(w,x''))(\pd_{x''}^2\psi(w,x''))\Big]~ 
\psi(z,x') \Phi_h(z,x,\bx)\Omega_h(z,y,\by).
  \eea}
  We have changed notation, from $\Omega_j$ to $\Omega_h$ since physical fields have $h=j+1$. 
 In the previous formula we have introduced an auxiliary variable $x''$ in order to
 allow for 
 non trivial OPE between fermions. Using the expressions given above for fermionic
OPEs and the action 
 of the current $K(x)$ on the primary fields $\Phi(y)$, we find
 {\colsep
 \bea \label{wo}
 \mathcal{W}^0&=&
 \lim_{\begin{subarray}{c} x'\rightarrow x \\ x''\rightarrow
 x',x\end{subarray}}\,||P_{h-1}(x,x')||^2
 \Big[ 2h(\pd_{x'}\psi(z,x'))\,+2K(z,x)+\nn \\ &+& 2 \,\psi(z,x')\pd_{x}+
  \psi(z,x')(\pd_{x'}\psi(z,x'))\Big]~  \psi(z,x') \Phi_h(z,x,\bx)\Omega_h(z,y,\by). 
 \eea}
 We used that 
 \be
 \lim_{x''\rightarrow x'} \psi(w,x'')\psi(z,x')\sim  \lim_{x''\rightarrow
 x'}\frac{(x'-x'')^2}{w-z}~,
 \ee
 hence, the only contribution coming from the fermions to $\mathcal{W}^0$ is 
 \be
 \lim_{x''\rightarrow x'} \pd_{x''}^2 \left(\psi(w,x'')\psi(z,x')\right) \sim
\frac{2}{w-z}~.
 \ee
 Notice that we have taken the limit $x''\to x'$ and $x''\to x$ for the fermionic
 and bosonic 
 variables   respectively. 
 
 With this form of $\mathcal{W}^0$ the three-point function consists only in the
 second term of (\ref{wo}) since 
 the fermionic contribution $\langle
 \psi(z_1,x'_1)\psi(z_2,x'_2)\psi(z_3,x'_3)\psi(z_3,x'_3) \rangle$ vanishes.

In this form, we are lead to compute 

 {\colsep\bea
\langle \, \mathcal{W}^-(x_1,\bx_1) \mathcal{W}^-(x_2,\bx_2)
\mathcal{W}^0(x_3,\bx_3)\, \rangle&=&(\mathrm{const.})  
~\lim_{x_{123}'\to x_{123}} ||P_{h_1-1}\, P_{h_2-1}\,P_{h_3-1}||^2\times\nn\\ &\times& \langle
\Phi_{h_1}(x_1)\Phi_{h_2}(x_2)K(x_3)\Phi_{h_3}(x_3) \rangle \,
 \langle \psi(x_1)\psi(x_2)\rangle \times\nn\\
 &\times&\,\langle \Omega_{h_1}(y_1)\Omega_{h_2}(y_2)\Omega_{h_3}(y_3)\rangle~.
\eea}
 For simplicity we dropped the superghost contribution
  $z_{13}^{-1}$, that anyway cancels with the ghost contribution
  $z_{12}z_{23}z_{13}$. \\

  The second correlator in (\ref{3points}) reads
  {\colsep
  \bea
  \,&&
  \langle\,\mathcal{Y}^{-1/2}_{h_1}(x_1,y_1)\,\mathcal{W}^{-1}_{h_2}(x_2,y_2)\,
  \mathcal{Y}^{-1/2}_{h_3}(x_3,y_3)\,\rangle=\nn\\
  \,&=&\lim_{x'_{123}\rightarrow x_{123}}
  || P_{h_1-{1\over2}}\,P_{h_2-1}\,P_{h_3-{1\over2}}||^2
  \, \langle \,S_{\alpha} (x'_1)\,\psi(x'_2)\, S_{\beta} (x'_3)\, \rangle \,
  \langle \,\Phi_{h_1}(x_1)\,\Phi_{h_2}(x_2)\,\Phi_{h_3}(x_3) \,\rangle
  \nn\\
  &\times&
  \lim_{y'_{13}\rightarrow
  y_{13}}||P_{j_1+{1\over2}}\,P_{j_3+{1\over2}}||^2\,
  \langle \,\widetilde S_{\alpha}(y'_1)\,\widetilde S_{\beta}(y'_3)\,\rangle
  \,\langle\,
  \Omega_{j_1}(y_1)\,\Omega_{j_2}(y_2)\,\Omega_{j_3}(y_3)\,
  \rangle.
  \eea}
  where we have dropped the charge variables dependence of the projectors
  in order to simplify the notation.

  We get easily the amplitudes integrating over the $z$'s. As usual,
  the three points can be fixed at 0,1, and  $\infty$.

 % OOOOOOOOOOOOOOOOOOOOOOOOOOOOOOOOOO    Conclusions  OOOOOOOOOOOOOOOOOOOOOOOOOOOOOOOOOOOOOOOO

 % OOOOOOOOOOOOOOOOOOOOOOOOOOOOOOOOOO    Conclusions  OOOOOOOOOOOOOOOOOOOOOOOOOOOOOOOOOOOOOOOO

 % OOOOOOOOOOOOOOOOOOOOOOOOOOOOOOOOOO    Conclusions  OOOOOOOOOOOOOOOOOOOOOOOOOOOOOOOOOOOOOOOO

 \section{Discussion}
 \noindent

 We expect the twist  operators  $\sigma_n^{\pm\pm}$ to be in correspondence  with the
  worldsheet vertex operators $\mathcal X_h$ and $\mathcal W_h$
  {\colsep
  \bea
  \sigma_n^{++}~\longleftrightarrow~
  \mathcal{X}_h(x,\bx,y,\by)~,\\
  \sigma_n^{--}~\longleftrightarrow~
  \mathcal{W}_h(x,\bx,y,\by)~,
  \eea}
  while the twists $\sigma_n^{\mp\pm}$ should correspond to the mixed vertex operators
  {\colsep
  \bea
  \sigma_n^{-+}~\longleftrightarrow~
  \mathcal{V}_h(x,\bx,y,\by)~,\\
  \sigma_n^{+-}~\longleftrightarrow~
  \mathcal{\widetilde V}_h(x,\bx,y,\by)~.
  \eea}
  From these identifications, it would be interesting to check if the three point function of the form
  $\langle\,\sigma_n^+ \,\sigma_m^+ \,(\sigma_{m+n-1}^+)^{\dagger} \,\rangle$
  agree with the
  worldsheet result of $\langle\, \mathcal X^-\mathcal X^0\mathcal X^-  \,\rangle$.
  In the same way the correlator
  $\langle\,\sigma_n^- \,\sigma_m^- \,(\sigma_{m+n-1}^-)^{\dagger} \,\rangle$
  should correspond to $\langle\, \mathcal W^-\mathcal W^0\mathcal W^- \,\rangle$.

 % OOOOOOOOOOOOOOOOOOOOOOOOOOOOOOOO  OOOOOOOOOOOOOOOOOOOOOOOOOOOOOOOOOO

 % OOOOOOOOOOOOOOOOOOOOOOOOOOOOOOOO  OOOOOOOOOOOOOOOOOOOOOOOOOOOOOOOOOO

 % OOOOOOOOOOOOOOOOOOOOOOOOOOOOOOOO  OOOOOOOOOOOOOOOOOOOOOOOOOOOOOOOOOO

 % OOOOOOOOOOOOOOOOOOOOOOOOOOOOOOOO  OOOOOOOOOOOOOOOOOOOOOOOOOOOOOOOOOO

 % OOOOOOOOOOOOOOOOOOOOOOOOOOOOOOOOOOOOO     WAKIMOTO    OOOOOOOOOOOOOOOOOOOOOOOOOO

 % OOOOOOOOOOOOOOOOOOOOOOOOOOOOOOOOOOOOO     WAKIMOTO    OOOOOOOOOOOOOOOOOOOOOOOOOO

 % OOOOOOOOOOOOOOOOOOOOOOOOOOOOOOOOOOOOO     WAKIMOTO    OOOOOOOOOOOOOOOOOOOOOOOOOO

 % OOOOOOOOOOOOOOOOOOOOOOOOOOOOOOOOOOOOO     WAKIMOTO    OOOOOOOOOOOOOOOOOOOOOOOOOO

 \appendix
 \renewcommand{\theequation}{A.\arabic{equation}}
 \chapter{Amplitudes \`a la Wakimoto}
 \label{app A}

In this section we construct a free field representation for the
$\widehat {\mathcal H}_6$ algebra starting from the standard
Wakimoto realization for $\widehat{SL}(2,\mathbb{R})$ and
$\widehat{SU}(2)$ \cite{wak,dot} and contracting the currents of both
CFTs as indicated in section 2. Then we use this approach to
compute two, three and four-point correlators that only involve
$\F^\pm$ vertex operators and reproduce the results obtained in
the previous sections. This free field representation was
introduced by Cheung, Freidel and Savvidy \cite{cfs} and used to
evaluate correlation functions for $\widehat {\mathcal H}_4$.

\section{$\widehat {\mathcal H}_6$ free field representation}
\noindent

 The \emph{Wakimoto representation}\index{Wakimoto
 representation|textbf} of the $\widehat{SL}(2,\mathbb{R})$ current
 algebra requires  a pair of commuting ghost fields $\beta_1(z)$
 and $\gamma^1(z)$ (the index 1 here is a label) with propagator
 $\langle\beta_1(z)\gamma^1(w)\rangle=1/(z-w)$, and a free boson
 $\phi(z)$ with
  $\langle\phi(z)\phi(w)\rangle=-\log (z-w) $. The
 $\widehat{SL}(2,\mathbb{R})$  currents can then be written as
 {\setlength\arraycolsep{2pt} \beqa \label{sl2w} K^+(z) &=&
 \displaystyle-\beta_1 \  , \nonumber\\
 K^-(z)&=&-\beta_1\gamma^1\gamma^1+\alpha_{+}\gamma^1\pd
 \phi-k_1\pd\gamma^1 \  , \\
 K^3(z) &=& \displaystyle -\beta_1\gamma^1+\frac{\alpha_{+}}{2}\pd
 \phi  \  , \nonumber \eeqa} where $\alpha_+^2\equiv 2(k_1-2)$.
 Similarly for $\widehat{SU}(2)$ we introduce a second pair of
 ghost fields $\beta_2(z)$ and $\gamma^2(z)$ (here the index 2 is a
 label) with world-sheet propagator
 $\langle\beta_2(z)\gamma^2(w)\rangle=1/(z-w)$, and a free boson
 $\varphi(z)$ with $\langle\varphi(z)\varphi(w)\rangle=-\log
 (z-w)$. The currents are then given by
 {\setlength\arraycolsep{2pt}\beqa \label{su2w} J^+(z) &=&
 \displaystyle-\beta_2 \  , \nonumber\\
 J^-(z)&=&\beta_2\gamma^2\gamma^2-i \alpha_{-}\gamma^2\pd
 \varphi-k_2\pd\gamma^2 \  , \\
 J^3(z) &=&-\beta_2\gamma^2+\frac{i \alpha_{-}}{2}  \pd \varphi \ ,
 \nonumber \eeqa} where  $\alpha_-^2\equiv 2(k_2+2)$. In order to
 obtain a Wakimoto realization for the $\widehat {\mathcal H}_6$
 algebra, we rescale the two ghost systems \beq \label{scalw}
 \beta_{\alpha}\to\sqrt{\frac{k_\a}{2}} \, \beta_{\alpha} \ ,
 \hspace{1cm}
 \gamma^{\alpha}\to\sqrt{\frac{2}{k_\a}}\,\gamma^{\alpha} \ , \eeq
 and introduce the light-cone fields $u$ and $v$ \be \phi = -i
 \sqrt{\frac{k_1}{2}} \m_1 u - \frac{i}{\sqrt{2 k_1}}
 \frac{v}{\m_1} \ , \hspace{1cm} \varphi = \sqrt{\frac{k_2}{2}}
 \m_2 u - \frac{1}{\sqrt{2 k_2}} \frac{v}{\m_2} \ , \ee with
 $u(z)v(w)\sim \, \ln (z-w)$. We then perform the current
 contraction as prescribed in $(\ref{contr})$, with the result
 {\colsep
 \ba
 \label{h6w}
 P^+_{\alpha}(z)&=&-\beta_{\alpha} \ , \nonumber\\
 P^{-\alpha}(z)&=&-2 \pd\gamma^{\alpha} -2 i \m_\a \pd
 u\,\gamma^{\alpha} \ , \\
 J(z)&=&i\, \m_\a \beta_{\alpha}\gamma^{\alpha}-\pd v
 +\frac{\m_1^2+\m_2^2}{2}\p u \ ,
 \nonumber\\
 K(z)&=&-\pd u  . \nonumber
 \ea} The $\widehat {\cal H}_6$
 stress-energy tensor follows from the limit of
 $T_{SL(2,\mathbb{R})}(z)+T_{SU(2)}(z)$ and is given by \be T(z) =
 \sum_{\a=1}^2 :\beta_{\alpha}(z)\pd\gamma^{\alpha}(z):+:\pd
 u(z)\,\pd v(z): - \frac{i}{2} (\m_1+\m_2) \p^2 u\ , \ee where the
 last term appears when expressing the normal ordered product of
 the currents in terms of the Wakimoto fields.

 The $\F^+_{p,\jh}$ primary vertex operators similarly follow from
 the $SL(2,\mathbb{R}) \times SU(2)$ primary vertex operators in
 the ${\cal D}^-_l \times V_{\tilde{l}}$ representation \be
 V_{l,m;\tilde l, \tilde m} =
 (-\g^1)^{-l-m}(-\g^2)^{\tilde{l}-\tilde{m}} e^{\frac{2l}{\a_+}
 \phi+\frac{2i\tilde{l}}{\a_-} \varphi} \ , \ee where $m$ is the
 eigenvalue of $K^3$. Introducing the charge variables we can
 collect all the components in a single field \be
 \Psi^-_{l}(z,x_1)\Omega_{\tilde l}(z,x_2) = \left ( x_1 + \g^1
 \right )^{-2l} \left ( x_2- \g^2 \right )^{2\tilde{l}} e^{\frac{2
 l}{\a_+} \phi+\frac{2\tilde{l}}{\a_-} i \varphi} \ , \ee that in
 the large $k_1$, $k_2$  limit becomes, using (\ref{pp+}) and
 introducing a normalization factor $ N(p,\jh)$ \be
 \F^+_{p,\jh}(z,x_\a) = N(p,\jh) e^{-\sqrt{2} \m_\a p x_\a \g^\a -
 i p v - i \left( \jh + \frac{\m_1^2+\m_2^2}{2} p \right ) u }  \ .
 \label{wakp} \ee It is easy to verify that this field satisfies
 the correct OPEs with the $\widehat{\cal{ H}}_6$ currents and that its
 conformal dimension is $h(p,\jh) = - p \jh +\frac{\m_1
 p}{2}(1-\m_1 p)+\frac{\m_2 p}{2}(1-\m_2 p)$. If we choose the
 normalization factor $N(p,\jh) = (\g(\m_1 p)\g(\m_2 p))^{-1/2}$
 the vertex operators $(\ref{wakp})$ precisely reproduce the
 results obtained in the previous sections.

 The  $\F^-_{p,\jh}$ vertex operators can be represented using an
 integral transform \cite{cfs} \beq \label{p<0w}
 \Phi_{p,\hj}^-(z,x^{\alpha})= \prod_{\a=1}^2  \int\,
 d^2x_{\alpha}\, \g(\m_a p)\frac{\m_\a^2 p^2}{2 \pi^2} \ e^{-\m_\a
 p x_{\alpha}x^{\alpha}} \Phi_{-p,\jh + \m_1 +
 \m_2}^+(z,x_{\alpha}) \ . \eeq

 The Wakimoto representation can also be derived from the
 $\s$-model action written in the following form
 \beq
 S= \int
 \frac{d^2 z}{2 \pi} \left\{ -\pd u \bar{\pd}v + \sum_{\a=1}^2
 \left [ \beta_{\alpha}\bar{\pd}\gamma^{\alpha}
 +\bar{\beta}^{\alpha}\pd\bar{\gamma}_{\alpha}-
 \beta_{\alpha}\bar{\beta}^{\alpha} e^{-i \m_\a u}\right ] \right\}
 \, \label{action} \ .
 \eeq
 The non-chiral $SU(2)_I$ currents are
 \beq
 \mathcal
 J^a(z,\bz)=i\,\gamma^{\alpha}(\sigma^a)_{\alpha}^{\phantom{\alpha}\beta}\,\beta_{\beta}
 \ , \hspace{1cm} \qquad \bar{\mathcal J}^a(z,\bz)=-i\,\bar
 \beta^{\alpha}(\sigma^a)_{\alpha}^{\phantom{\alpha}\beta}\,
 \bar\gamma_{\beta} \ .
 \eeq
 Using the equations of motion  \beq \beta_{\alpha}=e^{i\m_\a u}\pd
 \bar{\gamma}_{\alpha} \ , \hspace{2cm} \bar{\pd}\beta_{\alpha}=0 \
 , \eeq one can verify that they satisfy  $\bar\pd\mathcal
 J^a+\pd\bar{\mathcal J}^a=0$. Moreover their OPEs with the
 Wakimoto free fields are
 {\colsep
 \ba
  \mathcal
 J^a(z,\bz)\gamma^{\alpha}(z,\bz) &\sim&
 i\,\frac{\gamma^{\beta}(\sigma^a)_{\beta}^{\phantom{\beta}\alpha}}{z-w}
 \ , \hspace{1cm} \bar{\mathcal J}^a(z,\bz)\bar
 \gamma_{\alpha}(z,\bz) \sim -i\,
 \frac{(\sigma^a)_{\alpha}^{\phantom{\alpha}\beta}\bar
 \gamma_{\beta}}{\bz-\bar w} \ ,
 \nonumber \\
 \mathcal J^a(z,\bz)\beta_{\alpha}(z) &\sim&
 -i\,\frac{(\sigma^a)_{\alpha}^{\phantom{\alpha}\beta}
 \beta_{\beta}}{z-w} \ , \hspace{1cm} \bar{\mathcal J}^a(z,\bz)\bar
 \beta^{\alpha}(\bz) \sim i\,\frac{\bar
 \beta^{\beta}(\sigma^a)_{\beta}^{\phantom{\beta}\alpha}}{\bz-\bar
 w} \ .
 \ea}

 \section{ The $\sigma$-model view point}
 \noindent

Elements of the ${\bf H}_6$ Heisenberg group can be parametrized as \cite{cfs}
  \beq
g(u,v,\gamma^{\alpha},\bar \gamma_{\alpha})=e^{\frac{\gamma^{\alpha}}{\sqrt
2}P^+_{\alpha}}e^{uJ-vK}e^{\frac{\bar \gamma_{\alpha}}{\sqrt
2}P^{-\alpha}}\ .
  \eeq
As usual the $\sigma$-model action can be written in terms of the Maurer-Cartan forms and reads
\beq \label{wakact1} S = \frac{1}{2 \pi}\,\int d^2\sigma \left( -\pd u
\bar{\pd} v + \sum_{\a=1}^2 e^{i\mu_\a u} \pd\bar\gamma_\a
\bar\pd\gamma^\a \right), \eeq
where we have used $\langle J,K\rangle=1$ and $\langle
P^{+}_{\alpha},P^{-\alpha}\rangle=2$.
The metric and $B$ field are then given by
{\setlength\arraycolsep{2pt} \beqa
ds^2&=&-2dudv+2\sum_\a e^{i\mu_\a u} d\gamma^{\alpha}d\bar{\gamma}_{\alpha}\ ,\\
B&=&-du\wedge dv+\sum_\a e^{i\mu_\a u}d\gamma^{\alpha}\wedge
d\bar{\gamma}_{\alpha}\ . \eeqa}

Two auxiliary fields $\beta_{\alpha}$ and
$\bar{\beta}^{\alpha}$, defined by the OPE's
  \beq
\beta_{\alpha}(z)\gamma^{\beta}(w)\sim\frac{\delta_{\alpha}^{\beta}}{z-w}
\ , \eeq
complete the ghost-like systems that appear in the Wakimoto
representation.

With the help of $\beta_{\alpha}$ and
$\bar{\beta}^{\alpha}$, the action can be written as
  \beq \label{wakact2} S =
\frac{1}{2\pi} \int \, d^2 z \left( -\pd u \bar{\pd} v + \sum_{\a=1}^2 [
\bar\beta^{\alpha}  \pd \bar\gamma_{\alpha} + \beta_{\alpha}
\bar{\pd} \gamma^{\alpha} - e^{-i\mu_\a u} \beta_{\alpha}
\bar\beta^{\alpha} ] \right)\ , \eeq
that gives us back (\ref{wakact1}) upon using the equations of motion for $\beta_{\alpha}$ and
$\bar{\beta}^{\alpha}$.

In the Wakimoto representation, the currents can be written  as \cite{cfs}
{\setlength\arraycolsep{2pt}\beq \begin{array}{rcl}
P^+_{\alpha}(z) &=&
-\beta_{\alpha}(z), \\
P^{-\alpha}(z)&=& -2(\partial \gamma^{\alpha} + i
\partial u \gamma^{\alpha})(z), \\
J(z) &=& -(\partial v - i \sum_\a \mu_a \beta_{\alpha} \gamma^{\alpha})(z),  \\
K(z) &=& - \partial u(z)\ .
\end{array}\eeq}

A simple identification of the $\bf H_6$ group parameters and
the string coordinates, recast the metric in the more standard form
of (\ref{pp2trans}).
Generalizing the results of \cite{cfs}, it is easy to show that string coordinates
and Wakimoto fields are related as follows
{\setlength\arraycolsep{2pt}\beq \begin{array}{rcl}
u(z,\bar{z}) &=& \displaystyle{ u(z) + \bar u(\bar{z})}, \\
v(z,\bz)  &=& v(z) + \bar
v(\bar{z})+2i\bar{\gamma}_{L\alpha}(z) \gamma_R^{\alpha}(\bar{z}),\\
w^{\alpha}(z,\bz) &=&\displaystyle{ e^{-{i}\mu_\a u(z)}
[e^{i\mu_\a u(z)}\gamma_L^{\alpha}(z)
+ \gamma_R^{\alpha}(\bar{z})] },\\
\bar{w}_{\alpha}(z,\bar{z})&=&\displaystyle{e^{+{i}\mu_\a u(z)}[\bar{\gamma}_{L\alpha}(z)
+e^{i\mu_\a\bar u(\bar{z})} \bar{\gamma}_{R\alpha}(\bar{z})]} \ .
\end{array}\eeq}

 \section{Correlators}
 \noindent

 In order to evaluate the correlation functions in this free-field
 approach, we first integrate over the zero modes of the Wakimoto
 fields using the invariant measure \be \int \, du_0 dv_0
 \prod_{\a=1}^2 d\gamma^{\alpha}_0\,d\bar{\gamma}_0^{\alpha}e^{i
 \m_\a u_0} \ . \ee The presence of the interaction term \be S_I=
 \sum_{\a=1}^2 S_{I\a} = - \sum_{\a=1}^2 \int \frac{d^2w}{2 \pi}
 ~\beta_\a(w)\bar{\beta}^\a(\bar{w}) e^{-i \m_\a u(w,\bar{w})} \ ,
 \ee in the action $(\ref{action})$ leads to the insertion in the
 free field correlators of the screening operators \be
 \sum_{q_1,q_2=0}^\infty \prod_{\a=1}^2 \frac{1}{q_\a!}\,\left (
 \int \frac{d^2w_\a}{2\pi} \b_\a \bar{\b}^\a e^{-i \m_\a u} \right
 )^{q_\a} \ . \ee

 Negative powers of the screening operator are needed in order to
 get sensible results for $n$-point correlation functions other
 than the `extremal' ones, that only involve one $\F_{p_n,\jh_n}^-$
 vertex operator and $n-1$ $\F_{p_i,\jh_i}^+$ vertex operators.
 This means that the sum over $q_\a$ should effectively runs over
 all integers, $q_\a \in \mathbb{Z}$, not only the positive ones.
 An `extremal' $n$-point function can be written as
 {\colsep
 \ba & &
 \sum_{q_1,q_2=0}^\infty \prod_{\a=1}^2 \frac{1}{q_\a !}
 \left\langle
 \prod_{i=1}^{n-1}{\Phi}^+_{p_i,\hj_i}(z_i,\bz_i,x_{i\alpha},\bx^{\alpha}_i)
 \Phi^+_{-p_4,\hj_4+\m_1+\m_2}(z_n,\bz_n,x_{n
 \alpha},\bx^{\alpha}_n)\,S_{I \a}^{q_\a} \,
 \right\rangle \ \label{genwak} \\
 &=&  \delta\left(\sum_i^{n-1} p_i-p_n\right) \prod_{i < j \ne 4}
 |z_i -z_j|^{-2p_i(\hj_j+\h \, p_j)-2p_j(\jh_i+\h \, p_i)} \prod_{i
 \ne n} |z_i-z_n|^{2p_n(\hj_i+\h \, p_i) - 2p_i(\jh_n-\h \,
 p_n+\m_1+\m_2)}
 \nb \\
 &&  \sum^\infty_{q_1,q_2=0} \delta\left(\, L -\m_1 q_1 -\m_2 q_2
 \,\right)~\prod_{\a=1}^2 R(\m_\a) \left | \,e^{-\m_\a
 x_n^{\alpha}\sum_{i=1}^{n-1}  p_i x_{i\alpha}}\, \right |^2
 \frac{1}{q_\a!}\, \left ( - 2 \m_\a^2 \mathcal{I}_{\a, n} \right
 )^{q_\a}  \ , \nonumber
 \ea}
 where $L = - \sum_{i=1}^n \jh_i$, $\h
 = \frac{\m^2_1+\m_2^2}{2}$  and \be {\cal I}_{\a, n} = \int
 \frac{d^2 w}{2\pi} \prod_{i=1}^{n-1}|z_i-w|^{-2 \m_\a p_i}
 |z_n-w|^{2\m_\a p_n} \left | \sum_{i=1}^{n-1} \frac{p_i
 x_{i\a}}{w-z_i}- \frac{p_n x_{n \a}}{w-z_n} \right |^2 \ ,
 \label{intwak} \ee with the constraint $p_n x_{n\a} =
 \sum_{i=1}^{n-1} p_i x_{i\a}$. Finally the constant $R(\m)$,
 related to the normalization of the operators in $(\ref{wakp})$,
 is given by \be R^2(\m) = \frac{\g(\m p_n)}{\prod_{i=1}^{n-1}
 \g(\m p_i)} \ . \ee In  $(\ref{genwak})$ the two $\d$-functions
 arise from the integration over $u_0$ and $v_0$. Similarly the
 integration over the $\g_{0\a}$ leads to four other $\d$-functions
 that constrain the integration over the $x_{n\a}$ variables and
 give the exponential term. The other terms in $(\ref{genwak})$
 follow from the contraction of the free Wakimoto fields. Note that
 due to the second $\d$-function in $(\ref{genwak})$ the correlator
 is non vanishing only when $L = \m_1 q_1 + \m_2 q_2$ where $q_\a
 \in \mathbb{N}$. Therefore the same structure we found before
 using current algebra techniques appears: for the generic
 background $\m_1 \ne \m_2$ only one term from the double sum in
 $(\ref{genwak})$ contributes while for the $SU(2)_I$ invariant
 wave we have to add several terms. Let us consider some examples.
 We will need the following integral \cite{dotfat}
 {\colsep
 \ba & &\int  d^2
 t |t-z|^{2(c-b-1)}|t|^{2(b-1)}|t-1|^{-2a} =
 \frac{\pi \g(b)\g(c-b)}{\g(c)}|z|^{2(c-1)}|F(a,b,c;z)|^2  \nb \\
 &-& \frac{\pi
 \g(c)\g(1+a-c)}{(1-c)^2\g(a)}|F(1+a-c,1+b-c,2-c;z)|^2 \ .
 \label{dfintegral}
 \ea}
 It follows from the general expression
 $(\ref{genwak})$ that the two-point function $\langle +- \rangle$
 coincides with $(\ref{pm})$, since only the $q_\a=0$ terms are
 non-vanishing. For the $\langle ++- \rangle$ three-point coupling
 the integral $(\ref{intwak})$ gives \be - 2 \, \m_\a^2 \, {\cal
 I}_{\a,3} = |z_{12}|^{-2\m_\a p_3}|z_{23}|^{2\m_\a p_1}
 |z_{13}|^{2\m_\a p_2} \frac{\g(\m_\a p_3)}{\g(\m_\a p_1) \g(\m_\a
 p_2)} \left | x_{1\a}-x_{2\a} \right |^2 \ , \ee and the result
 precisely agrees with $(\ref{cgppm})$, $(\ref{qpppm})$. When
 $\m_1=\m_2$ the sum over $q_\a$ reconstructs the $SU(2)_I$
 invariant coupling $(\ref{su2ppm})$.

 The four-point function $\langle+++-\rangle$ can be evaluated in a
 similar way. In this case
 {\colsep
 \ba
 - 2 \, \m_\a^2 \, {\cal I}_{\a, 4}
 &=&  |z_{12}|^{-2\m_\a p_4}|z_{14}|^{-2\m_\a (p_1-p_4)}
 |z_{34}|^{-2\m_\a (p_3-p_4)}|z_{24}|^{2\m_\a (p_2-p_4)} \nb \\
 & & \left [ C_{12}(\m_\a) \left | x_{31\a} f(\m_\a,x_\a,z) \right
 |^2 + C_{34}(\m_\a) \left | x_{31\a} g(\m_\a,x_\a,z) \right |^2
 \right ] \ , \label{wpppm}
 \ea}
 where the functions $f$ and $g$ are
 as defined in $(\ref{ex7})$ and \be C_{12}(\m) =
 \frac{\g(\m(p_1+p_2))}{\g(\m p_1)\g(\m p_2)} \ , \hspace{1cm}
 C_{34}(\m) = \frac{\g(\m p_4)}{\g(\m p_3)\g(\m (p_4-p_3))} \ . \ee
 We find again complete agreement with $(\ref{cor-pppm})$.

 Finally the correlator  $\langle+-+-\rangle$ can be obtained from
 the $\langle+++-\rangle$ correlator performing the integral
 transform $(\ref{p<0w})$ of the vertex operator inserted in $z_2$
 \cite{cfs}, that is we send $(p_2,\jh_2) \rightarrow
 (-p_2,\jh_2+\m_1+\m_2)$ and evaluate the $x_{2\a}$ integral. We
 first rewrite
 {\colsep
 \ba {\cal T} &\equiv& \int d^2 x_{2 \a} \frac{\left
 | e^{-\m_\a p_2 x^\a_{24}} \right |^2}{\G(q_\a+1)} \left [
 C_{12}(\m_\a) \left | x_{31\a} f(\m_\a,x_\a,z) \right |^2 +
 C_{34}(\m_\a) \left | x_{31\a} g(\m_\a,x_\a,z) \right |^2 \right
 ]^{q_\a}
 \nb \\
 &=& \int d^2 x_{2 \a} \frac{\left | e^{-\m_\a p_2 x^\a_{24}}
 \right |^2}{\G(q_\a+1)} \left [ A x_{2\a}\bar{x}_{2\a} + B
 \bar{x}_{2\a}+\bar{B}x_{2\a}+E \right ]^{q_\a} \ ,
 \ea}
 and then
 evaluate the integral using
 \be \int d^2 u  \left | e^{-u} u^t
 \right |^2 = \pi (-1)^{-1-t} \g(1+t) \ ,
 \ee
 which is a limit of
 $(\ref{dfintegral})$. The result is
 \be
 {\cal T} = \left |
 e^{\m_\a p_2  x^\a_{24} \frac{B_\a}{A}} \right |^2
 \frac{|x^\a_{24}|^{-q_\a}}{2A} \left ( \frac{B\bar{B}-EA}{\m_\a^2
 p_2^2} \right )^{\frac{q_\a}{2}} I_{q_\a}\left ( 2\m_\a p_2  \left
 | x^\a_{24} \right | \sqrt{\frac{B\bar{B}-EA}{A^2}} \right ) \ ,
 \label{wakt}
 \ee
 where $I_{q_\a}$ is a modified Bessel function of
 integer order and
 {\colsep
 \ba
 && \frac{\m_\a p_2  x^\a_{24} B}{A} = -\m_\a
 p_2 x_{1\a} x^\a_{24} +\m_\a p_3 x_{13\a}  x^\a_{24}z -\m_\a p_2
 x_{13\a}  x^\a_{24}z(1-z) \p_z \ln S(\m_\a,z,\bar{z}) \ , \nb \\
 && A = -\frac{\m^2_\a p^2_2}{\tilde{C}_{12}} |z|^{-2
 \m_\a(p_1-p_2)} S(\m_\a,z,\bar{z})\ , \hspace{1cm} 2 \m_\a p_2
 \left |x^\a_{24}\right |\sqrt{\frac{B\bar{B}-EA}{A^2}} = \z_\a \ .
 \ea}
 The functions and constants that appear on the left-hand side
 of the previous equations were defined in $(\ref{pm8}-\ref{pm9})$
 and $(\ref{pm11})$. Combining $(\ref{wakt})$ with the rest of the
 $\langle +++-  \rangle$ correlator we obtain the $\langle +-+-
 \rangle$ correlator and also in this case the result coincides
 with $(\ref{pm12})$ when $\m_1 \ne \m_2$ and with
 $(\ref{pmsu2inv})$ when $\m_1=\m_2$.

 \printindex\newpage\thispagestyle{empty}

 % OOOOOOOOOOOOOOOOOOOOOOOOOOOOOOOOOOOOOOO  BIBLIOGRAPHY   OOOOOOOOOOOOOOOOOOOOOOOOOOOOOO

 % OOOOOOOOOOOOOOOOOOOOOOOOOOOOOOOOOOOOOOO  BIBLIOGRAPHY   OOOOOOOOOOOOOOOOOOOOOOOOOOOOOO

 % OOOOOOOOOOOOOOOOOOOOOOOOOOOOOOOOOOOOOOO  BIBLIOGRAPHY   OOOOOOOOOOOOOOOOOOOOOOOOOOOOOO

 % OOOOOOOOOOOOOOOOOOOOOOOOOOOOOOOOOOOOOOO  BIBLIOGRAPHY   OOOOOOOOOOOOOOOOOOOOOOOOOOOOOO

 \end{document}